\documentclass[preprint,1p,12pt]{elsarticle}

\usepackage{amssymb}
\usepackage{graphicx}
\usepackage{dcolumn}
\usepackage{bm}
\usepackage{array}
\usepackage{color}

\journal{Comptes Rendus Physique}

\begin{document}

\begin{frontmatter}



\title{The ampere and the electrical units in the quantum era}


\author{W. Poirier, S. Djordjevic, F. Schopfer, O. Th\'evenot}

\address{Laboratoire national de m\'etrologie et d'essais, 29 avenue Roger Hennequin, 78197 Trappes, France}

\begin{abstract}
By fixing two fundamental constants from quantum mechanics, the Planck constant $h$ and the elementary charge $e$, the revised Syst\`eme International (SI) of units endorses explicitly quantum mechanics. This evolution also highlights the importance of this theory which underpins the most accurate realization of the units. From 20 May 2019, the new definitions of the kilogram and of the ampere, based on fixed values of $h$ and $e$ respectively, will particularly impact the electrical metrology. The Josephson effect (JE) and the quantum Hall effect (QHE), used to maintain voltage and resistance standards with unprecedented reproducibility since 1990, will henceforth provide realizations of the volt and the ohm without the uncertainties inherited from the older electromechanical definitions. More broadly, the revised SI will sustain the exploitation of quantum effects to realize electrical units, to the benefit of end-users. Here, we review the state-of-the-art of these standards and discuss further applications and perspectives.
\end{abstract}

\end{frontmatter}


\section{Ampere definition: from electromechanics to quantum mechanics}
\subsection{Ampere and the hierarchy of electrical units}
\begin{figure*}[!h]
\includegraphics[width=5.2in]{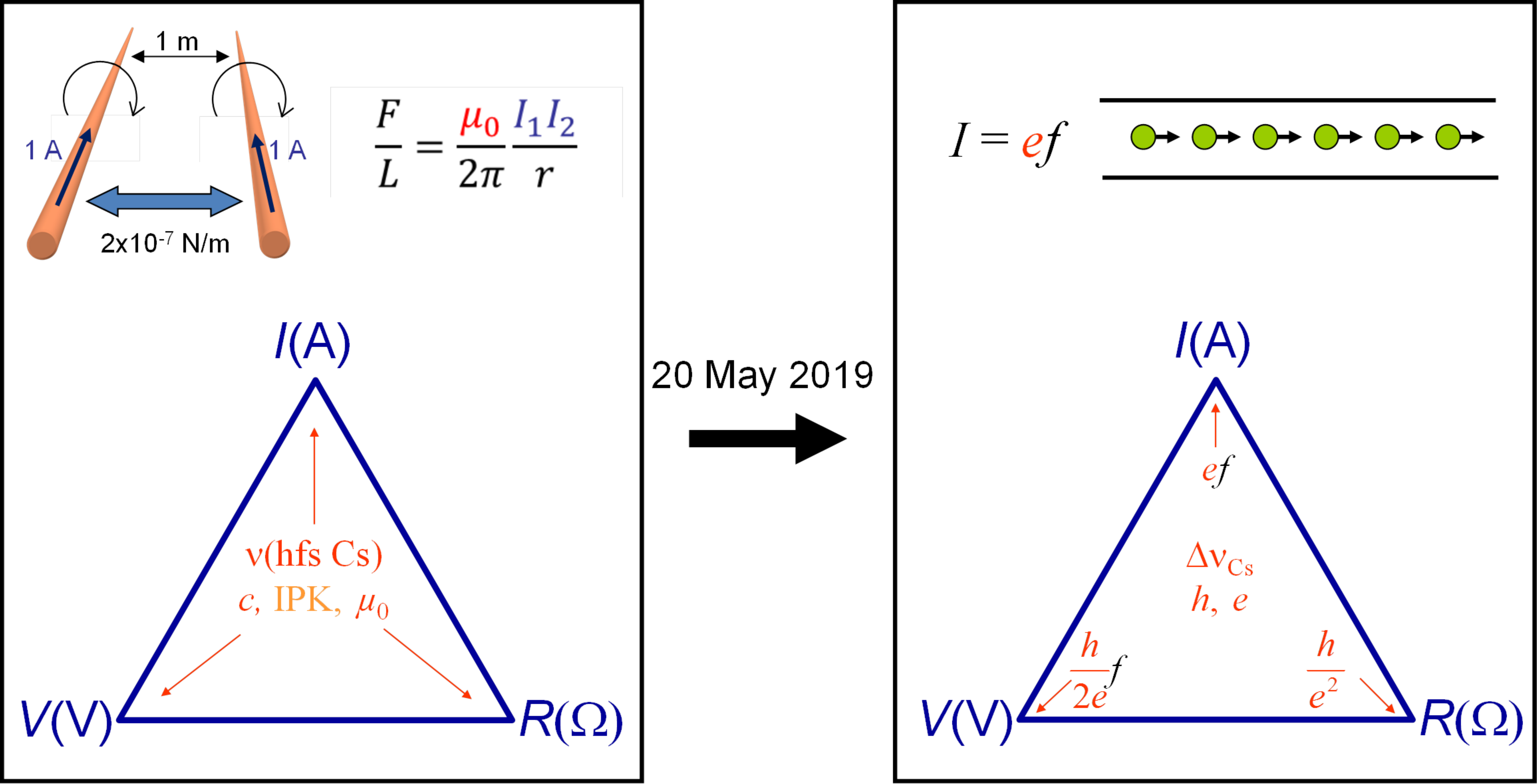}
\caption{Left-top: Schematic of the ampere definition based on Ampere's force law and $\mu_0$. Left-bottom: Illustration of the link of electrical units to the unperturbed ground state hyperfine transition frequency of the caesium 133 atom (noted $\nu(\mathrm{hfs}~\mathrm{Cs})$ before 20 May 2019), IPK (International Prototype of the Kilogram), $c$ (light velocity in vacuum) and $\mu_0$ (magnetic constant of vacuum). Right-top: Schematic of the ampere definition based on the elementary charge $e$. Right-bottom: Illustration of the link of electrical units to $\Delta \nu_\mathrm{{Cs}}$ (notation which replaces $\nu(\mathrm{hfs}~\mathrm{Cs})$ after 20 May 2019), $h$ (Planck constant) and $e$ (the elementary charge), where $f$ is the frequency.}\label{fig:Fig-Intro}
\end{figure*}
In 1948, a new definition of the unit of electrical current, based on Ampere's force law, was established on the occasion of the 9th General Conference of Weight and Measurements (CGPM). Funded on the theory of electromagnetism, this definition, reported in table~\ref{table:AncienSI}, fixes the exact value of the attractive force experienced by two current carrying wires in an ideal situation (fig.\ref{fig:Fig-Intro}-left). Doing so, the value of the magnetic constant of vacuum $\mu_0=4\pi\times10^{-7}~\mathrm{N/A^{2}}$ is fixed.
\begin{table}[h]
\newcolumntype{M}[1]{>{\centering\arraybackslash}m{#1}}
\begin{center}
\begin{tabular}{|c|M{10cm}|}
  \hline
  \textbf{Units} & \textbf{Definitions} \tabularnewline \hline
  kilogram (kg) & The kilogram is the unit of mass; it is equal to the mass of the international prototype of the kilogram. \tabularnewline \hline
  ampere (A) & The ampere is that constant current which, if maintained in two straight parallel conductors of infinite length, of negligible circular cross-section, and placed 1 m apart in vacuum, would produce between these conductors a force equal to $2\times10^{-7}$ newton per metre of length.\tabularnewline \hline
  volt (V) & The volt is the potential difference between two points of a conducting wire carrying a constant current of 1 ampere, when the power dissipated between these points is equal to 1 watt. \tabularnewline \hline
  ohm ($\Omega$) & The ohm is the electric resistance between two points of a conductor when a constant potential difference of 1 volt, applied to these points, produces in the conductor a current of 1 ampere, the conductor not being the seat of any electromotive force. \tabularnewline \hline
  farad (F) & The farad is the capacitance of a capacitor between the plates of which there appears a potential difference of 1 volt when it is charged by a quantity of electricity of 1 coulomb. \\
  \hline
\end{tabular}
\caption{Definitions of the kilogram, the ampere, the volt, the ohm and the farad before 20 May 2019.}
\label{table:AncienSI}
\end{center}
\end{table}
It was confirmed in the Syst\`eme International d'unit\'es\cite{BIPM}, adopted at the 11th CGPM, and maintained since then\cite{BrochureSI}. Let us remark that this definition therefore bounds the unit ampere to the unit newton, hence to the kilogram, meter and second. It results that all electrical units depend on the mechanical units, as highlighted by definitions in table \ref{table:AncienSI} and illustrated by fig.\ref{fig:Fig-Intro}-left.

The ampere definition describes a though experiment, and the closest implementation of this experiment is the ampere balance\cite{Vigoureux1965}. It consists in comparing the weight of a mass in the gravitational field with the magnetic force that is exerted between two coils supplied by a current. The accuracy of the ampere achieved using the ampere balance was limited by the measurement of mechanical dimensions from which the electromagnetic force is computed. Relative measurement uncertainties\cite{Vigoureux1965} were not better than a few parts in $10^{6}$.

An alternative route\cite{Awan2011Book,Trapon1998,Trapon03,Schurr2009} for improving the accuracy of realization of electrical units was to implement the farad instead of the ampere by exploiting the link of the electric constant of vacuum $\epsilon_0$ to $\mu_0$ and the velocity of light in vacuum $c^{2}=1/\mu_0\epsilon_0$. The farad can indeed be accurately realized from $\epsilon_0$ and the meter using a calculable standard of capacitance\cite{Clothier64,Awan2011Book,Delahaye1987,Small1997}. This device is based on a robust theorem\cite{Thompson56} derived by A. Thompson and D. Lampard which stipulates that, in a cylindrical system made of four electrodes of infinite lengths, the two linear cross-capacitances are linked by a universal relationship only dependent on the exact electric constant $\epsilon_0$. For a device based on four cylindrical electrodes, the linear cross-capacitance is equal to $\gamma=\frac{\epsilon_0\ln(2)}{\pi}\simeq1.95$ pF/m. The interest of this route was reinforced when the velocity of light $c$ was fixed at the value of 299792458 $\mathrm{ms^{-1}}$ to define the meter from the second $s$ in 1983. Doing so, $\epsilon_0$ was also fixed at an exact value. Realizations of the farad using Thompson-Lampard calculable capacitance standards have been achieved with uncertainties\cite{Trapon03,Small1997,Jeffrey1998} of a few parts in $10^8$. The ohm was then realized from impedance comparisons. To complete the chain of electrical units, the volt was also realized from $\epsilon_0$ and mechanical units using the volt balance\cite{Clothier89,Funk1991}. This experiment compares the weight of a mass in the gravitational field with the electrostatic force occurring between the two electrodes of a capacitance between which a voltage is applied. So, the volt was realized with uncertainties of a few parts in $10^7$. Until today, the uncertainty of the realizations of the electrical units have hardly changed. Although the Josephson effect and the quantum Hall effect have revolutionized the traceability of voltage and resistance measurements, the realizations of the volt and the ohm have remained limited by the electromagnetic definition of the ampere. The adoption at the 26th CGPM of the revised SI\cite{Mills05,Mills2006,Milton2007,Draft2015} based on constants of nature, and particularly the new definition of the ampere based on the elementary charge will disconnect the electrical units from the mechanical ones (fig.\ref{fig:Fig-Intro}-right). This evolution will rule out the previous limits and this will have a direct impact on the accuracy of the realizations of the electrical units from 20 May 2019, the date of implementation of the revised SI.
\subsection{The quantum revolution}
In the 20th century quantum mechanics brings a new description of the reality, \emph{\emph{i.e.}} the physics of particles, fields and solids. Relying on the indistinguishability of particles in quantum mechanics, the formalism of the second quantization was developed to describe many-body systems\cite{Mahan1981}, in particular crystalline solids where electrons occupy Bloch states satisfying the crystal periodicity\cite{Kittel1966,Kittel1987}. Beyond the description of energy bands, one famous success of the quantum theory of solids is the BCS (Bardeen-Cooper-Schrieffer) theory\cite{Bardeen1957} of the superconductivity which is explained by the condensation of Cooper pairs. This has opened the way to the discovery of the Josephson effect\cite{Josephson62} a few years after. In the 80's, the solid-state quantum physics is then used to describe electronic transport properties in small devices at low temperatures such as the quantization of the conductance in electronic conductors\cite{Landauer1957,Wees1988}, the wave function localization by disorder\cite{Anderson1958,Abrahams1979} and the Coulomb blockade\cite{Averin1986}. The first two are essential for the description of the quantum Hall effect\cite{Klitzing1980} while the last underpins the single electron tunneling\cite{Likharev1985}.
\subsection{the Josephson effect}
\begin{figure*}[!h]
\includegraphics[width=5.2in]{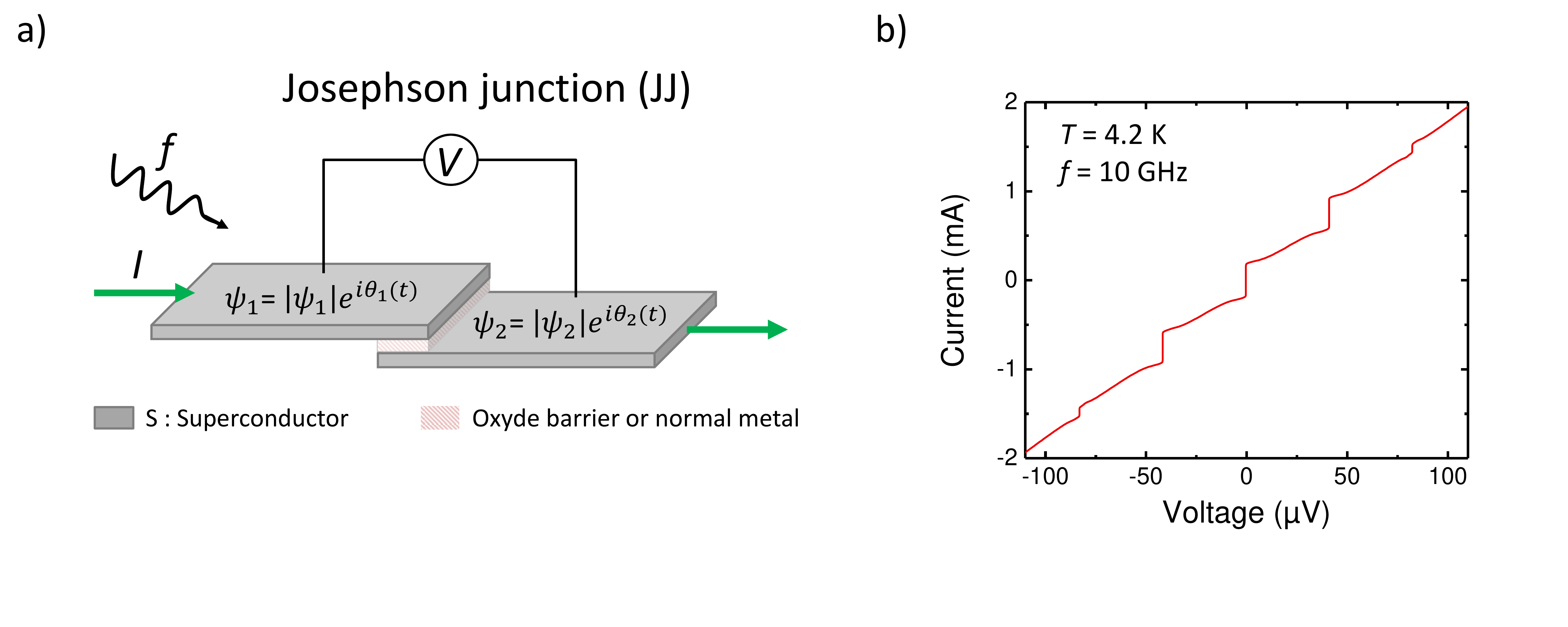}
\caption{a) Schematic of a Josephson junction formed by two superconducting electrodes (S) seperated by an oxyde barrier (I) or a thin layer of normal metal (N), irradiated by an external microwave source of frequency $f$. b) $I-V$ characteristic of two SINIS Josephson junctions in series at 4.2 K and submitted to a 10 GHz external frequency. Shapiro steps appear in the $I-V$ characteristic at multiple integer of $2f/K_{\mathrm{J}}\sim 40~\mu$V.}\label{fig:Fig_Jos_1}
\end{figure*}
The ac Josephson effect has been predicted by Brian Josephson in 1962 \cite{Josepshon1962}. It manifests itself as quantized voltage steps in the dc current-voltage ($I-V$) characteristic of two weakly coupled superconductors submitted to a microwave irradiation of frequency $f$ (fig.\ref{fig:Fig_Jos_1}a). First demonstrated by Shapiro in 1963 \cite{Shapiro63}, the quantized steps appear at $V_{n} = nf/K_{J}$, where $n$ is an integer and $K_{J}\equiv 2e/h$ is the Josephson constant (fig.\ref{fig:Fig_Jos_1}b).

\paragraph{Josephson equations} The Josephson effects are a consequence of the existence of a macroscopic coherent quantum state in the superconductors. The BCS theory \cite{Bardeen1957} states that, due to a weak attractive interaction, the electrons near the Fermi surface bind into Cooper pairs, and form a condensate sharing a macroscopic wave function $\psi = |\psi|e^{i\theta}$. The macroscopic properties of the superconducting state like the Meissner effect or the quantization of flux are related to the existence of the phase $\theta$ of the macroscopic wave function which is maintained over macroscopic distances and hence is responsible for the long range order. The BCS ground state is a phase coherent linear combination of states with different number of pairs, in which the phase and the number of pairs are related by an uncertainty relation. The macroscopic number of pairs participating in the superconducting state explains the well defined phase. The Josephson effects appear when the phase locking of the pairs is weakened, \emph{i.e.} when Cooper pairs can be transferred between the two superconducting regions called a Josephson junction (JJ).

Considering a tunnel junction between two superconducting electrodes, Josephson showed, by using second-order perturbation theory in the tunneling Hamiltonian, that a current of Cooper pairs flows in the junction and is related to the phase difference $\varphi = \theta_2-\theta_1$ of the superconducting wavefunctions on each side of the tunnel barrier (fig.\ref{fig:Fig_Jos_1}a) :
 \begin{equation}\label{Josephson1}
   I_{s} = I_{c}\sin \varphi,
\end{equation}
where $I_{c}$ is the critical current.
This equation states that a dc supercurrent flows with no voltage drop when the phases are time independent.

The ac Josephson effect relates the time dependence of the phase $\varphi$ to the voltage drop $V$ between the two superconductors :

\begin{equation}\label{ac-Josephson}
  \frac{\mathrm{d}\varphi}{\mathrm{d}t}=\frac{2eV}{\hbar}.
\end{equation}
Combining the two equations (\ref{Josephson1}) and (\ref{ac-Josephson}), the supercurrent oscillates at a frequency $f_{J} = \frac{2eV}{h}$ in the presence of a voltage difference. It can be interpreted as the emission of a photon of energy $hf_{J}$ when the pair undergoes the energy change of $2eV = 2\Delta\mu$ (where $\Delta\mu$ is the electrochemical potential difference between the two superconductors).

\paragraph{Observation of the ac Josephson effect}\label{Bessel} In order to observe this effect, Josephson proposed to modulate the Josephson oscillation frequency by biasing the junction with a dc voltage and an external microwave voltage of frequency $f$, such that $V(t) = V_{dc}+V_{ac}\cos(2 \pi ft)$. In that case, the Josephson current can be analyzed in terms of Bessel functions. At the condition, $V_{dc}=n\frac{h}{2e}f$, the current has a dc component which extends over an amplitude $\Delta I_{n}= 2I_{c}|J_{n}(2eV_{ac}/hf)|$, where $J_{n}$ is the $n$th order Bessel function. Hence the synchronization of the Josephson oscillation with the external frequency gives rise to constant voltage steps in the dc $I-V$ characteristic as illustrated in fig.\ref{fig:Fig_Jos_1}b.

Moreover, it is important to note that the time integral of a voltage pulse $V(t)$ across a Josephson junction, $\int V dt = n\frac{h}{2e}$, is quantized in multiples of the quantum of flux in the superconductor $\Phi_{0} = h/2e = 2.067\times10^{-15}$ V.$\mathrm{s}^{-1}$, so that the Shapiro steps, $V_{n} \equiv n \Phi_{0} f$, can be interpreted as $n$ quantized voltage pulses per period of the external signal. Furthermore, when biased by a current pulse of appropriate amplitude and width, a voltage pulse of quantized area can be generated. We will see in Section \ref{Section_JAWS} that the precise control of the timing of individual pulses is another way to synthesize quantized voltages.

\begin{figure*}[!h]
\includegraphics[width=5in]{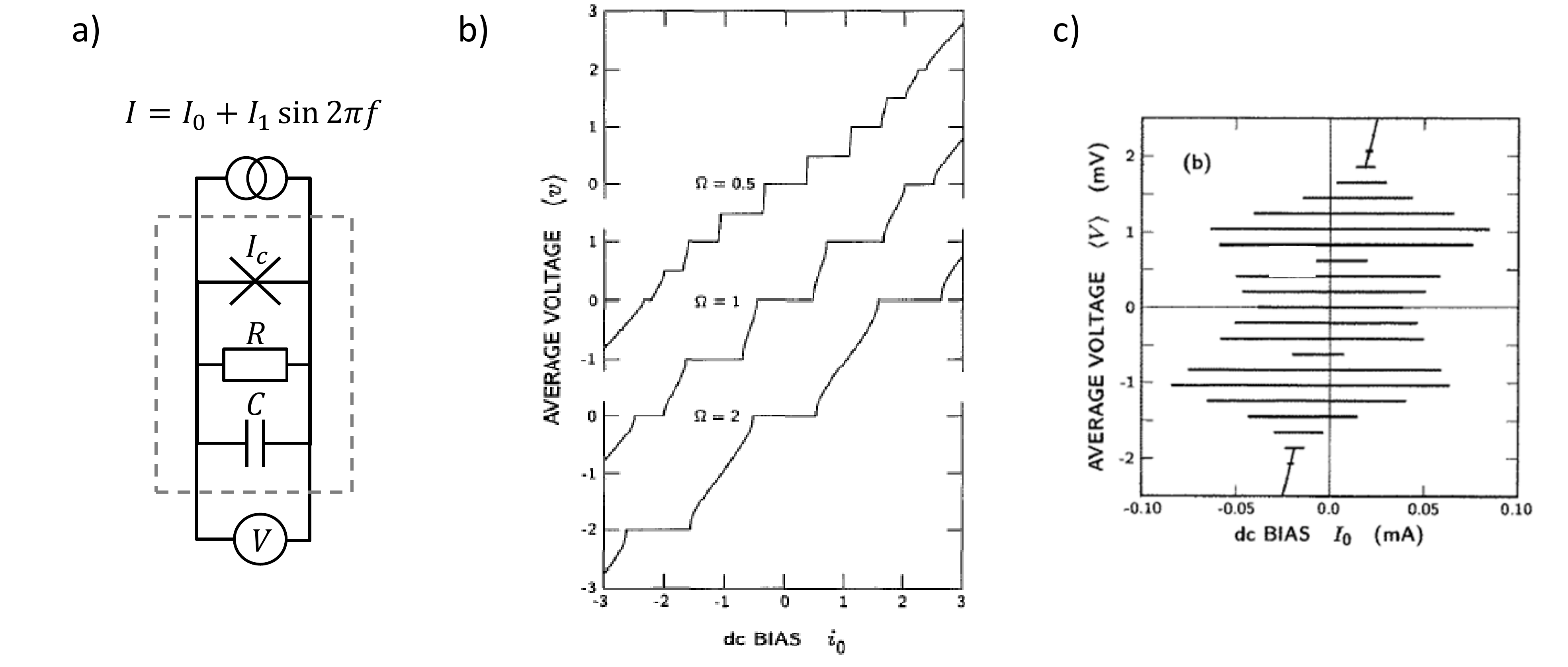}
\caption{$I-V$ characteristics calculated within the Stewart-McCumber model: a) Schematic diagram of the current driven, resistively and capacitively shunted junction (RCSJ) model. b) Three $I-V$ characteristics calculated in the limit of the resistively shunted junction model (RSJ) ($C=0$ and $\beta = 0$). The y-axis is the average voltage in units of the characteristic voltage $I_c R$, and the x-axis is the dc current relative to $I_c$ ($i_0=I_0/I_c$). The three different curves correspond to three values of $\Omega=f/f_c=0.5, 1$ and $2$ for three values $i_1=I_1/I_0=1.19, 1.70$ and $3.00$ respectively chosen to simultaneously maximize the widths of the $n=0$ and $n =1$ steps. From ref.\cite{Kautz1994}. And c) $I-V$ characteristic calculated in the limit of highly hysteretic JJ for $\beta = 200$ ($I_c$ = 0.2 mA, $R$ = 100 $\Omega$, $C$ = 20 pF, $I_1$ = 16 mA, $f_1$ = 100 GHz). Zero-current crossing steps are visible. Adapted from ref.\cite{Kautz1996}.} \label{fig:Fig_Jos_1b}
\end{figure*}

\paragraph{Dynamics of the Josephson junctions}\label{RCSJ} To describe the dynamics of a realistic Josephson junction (JJ), current components other than the supercurrent must be taken into account. It is usually done in the frame of the resistively and capacitively shunted junction (RCSJ) model or Stewart McCumber model \cite{Stewart1968,McCumber1968}, where the JJ is represented by an ideal Josephson element, obeying equations (\ref{Josephson1}) and (\ref{ac-Josephson}), which is shunted by a resistance $R$ and a capacitance $C$, as depicted in fig.\ref{fig:Fig_Jos_1b}a. In the presence of an external current source, the bias current, $I$, is equal to the sum of the currents in the three parallel channels. It results that the behaviour of the JJ is governed by the following second order non-linear equation for $\varphi$:
\begin{equation}
I=\frac{\hbar C}{2e}\frac{\mathrm{d}^{2}\varphi}{\mathrm{d}t^{2}}+\frac{\hbar }{2eR}\frac{\mathrm{d}\varphi}{\mathrm{d}t}+I_{c}\sin\varphi.
\label{RCSJ_equation}
\end{equation}

For small phase differences ($\varphi \ll 1$), $\sin\varphi \backsim \varphi$; the problem becomes linear and similar to a parallel $RLC$ resonator, where the Josephson element can be identified with the kinetic inductance $L_{J} = \hbar/2 e I_{c}$. The resonant angular frequency of the circuit is $\omega_{p}=2\pi f_p=(L_{J}C)^{-1/2}$, where $f_{p}$ is the plasma frequency. Another important parameter is the characteristic angular frequency $\omega_{c}=2\pi f_{c}=R/L_{J}=\frac{2e}{\hbar}I_{c}R$. The quality factor $Q$ is given by $Q=\omega_{p}RC$. The latter is related to the well known McCumber parameter $\beta$ by: $Q^{2} = \beta$. By noting that the equation (\ref{RCSJ_equation}) for the phase is similar to the equation of the damped motion of a particle  in a tilted washboard potential (the capacitance playing de role of the mass and the resistance the role of the damping term), $\beta$ is often used to characterize the damping of the JJ: $\beta \leq 1$ corresponds to the case of overdamped JJ and $\beta \gg 1$ to the case of underdamped JJ.

The $I-V$ characteristics under microwave irradiation can be calculated by assuming that the junction is driven by a current source with dc and rf components: $I=I_{0}+I_{1}\sin2\pi f$. The amplitude of the constant-voltage steps, $\Delta I_{n}$, can be expressed in terms of Bessel functions, as in the case of a voltage-biased JJ, if the rf voltage across the JJ is approximately sinusoidal, i.e. when most of the rf current flows in the linear elements rather than in the Josephson element. These limiting cases are useful for the design of the different Josephson voltage standards (see Section \ref{JVS}) \cite{Kautz1992,Kautz1996,Kautz1995}; however in most of the realistic cases, numerical calculations are needed to reproduce the wide variety of experimental $I-V$ characteristics under microwave irradiation illustrated in fig.\ref{fig:Fig_Jos_1b}b and fig.\ref{fig:Fig_Jos_1b}c. For overdamped JJs, the displacement current in the capacitance can be neglected, the $I-V$ characteristics are non-hysteretic as demonstrated in the simulations done by Kautz \cite{Kautz1994} (fig.\ref{fig:Fig_Jos_1b}b). There, the rf components are adjusted to optimize simultaneously the amplitude of the $n = 0$ and $n = 1$ steps. For underdamped JJ, the $I-V$ characteristics can be highly hysteretic with the first few constant-voltage steps crossing the zero-current axis if the current through the capacitance is the dominant one, as illustrated in fig.\ref{fig:Fig_Jos_1b}c \cite{Kautz1996}.

\paragraph{Universality of the Josephson effect}
Although the prediction of the Josephson effects was done for tunnel junctions, they can be observed for a very wide range of $'$weak links$'$. The universality of the relation has been tested early after the discovery of the effect at some parts in $10^{8}$ \cite{Clarke1968}. These measurements were improved in the 80's by Tsai and coworkers \cite{Tsai1983} and reached a  relative uncertainty of 2 parts in $10^{16}$ by comparing different types of junction (Nb-Cu-Nb junction to an In microbridge). The lowest uncertainty has been achieved with two similar junctions to 3 parts in $10^{19}$ \cite{Jain1987}. This gives a very high confidence that the correction to the frequency-to-voltage relation might be very small if any. In parallel, several theoretical works justified the universal character of the relation in a superconducting ring interrupted by a barrier \cite{Bloch1968,Bloch1970,Fulton1973} and the absence of corrections to a level of $10^{-20}$ \cite{Penin2010}.

\paragraph{Towards the Josephson voltage standards}
Since a Josephson junction acts as a perfect frequency-to-voltage converter based on fundamental constants, it was soon proposed to use these steps to improve the voltage standards \cite{Taylor1967,Field1973,Witt2011} taking advantage of the high accuracy of time references. Today, microwave sources can be referenced and locked to atomic clocks to a level of a few parts in $10^{11}$. However, the very small value of the flux quantum fixes the scale of the Shapiro steps to 20 $\mu$V at 10 GHz for a single junction. This low value is an obstacle to the development of practical voltage standards for which outputs of 1 V to 10 V are desirable. This challenge has been addressed by the successful development of highly-integrated series arrays of underdamped or overdamped Josephson junctions, which will be described in section \ref{TheQuantumvoltagestandard}.

\subsection{The quantum Hall effect}
\subsubsection{The effect and its physics}
\begin{figure*}[!h]
\includegraphics[width=5.2in]{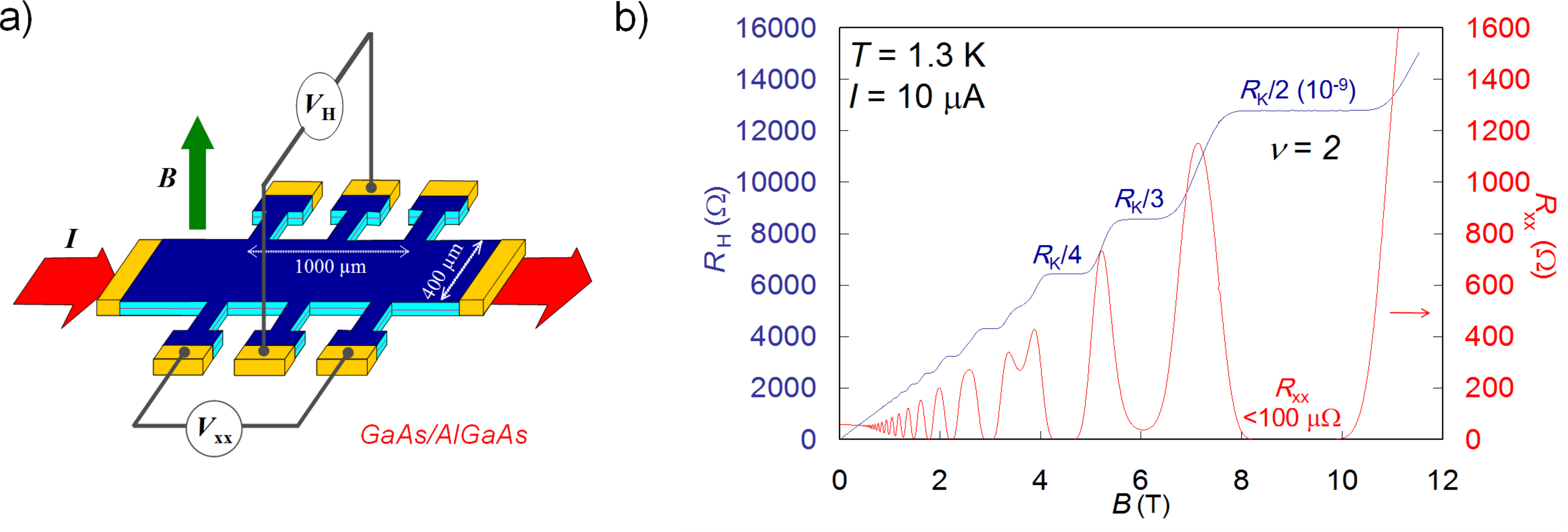}
\caption{a) Schematic of eight-terminals Hall bar based on a GaAs/AlGaAs heterostructure. b) Hall resistance, $R_\mathrm{H}$, and longitudinal resistance, $R_\mathrm{xx}$, as a function of the magnetic induction $B$ at $T=1.3$ K.}\label{fig:Fig-QHEManifestation}
\end{figure*}
The quantum Hall effect\cite{Klitzing1980}, discovered by K. von Klitzing in 1980, occurs in a two-dimensional electron gas (2DEG), like a Hall bar (see fig.\ref{fig:Fig-QHEManifestation}a), under a perpendicular magnetic field. As reported in fig.\ref{fig:Fig-QHEManifestation}b, it manifests itself by a quantization of the transverse resistance $R_\mathrm{H}$ at values $R_\mathrm{K}/i$, where $i$ is an integer and $R_\mathrm{K}\equiv h/e^2$ is the von Klitzing constant. Simultaneously, the longitudinal resistance forms minima, $R_\mathrm{xx}\sim 0$, revealing that the 2DEG is in a dissipation-less state.

In a perpendicular magnetic field, the classical motion of an electron of charge ($-e$) moving in a two dimension (2D) space is a cyclotron motion that drifts under the application of an electric field. The resistivity tensor is given by:
\begin{equation}
\rho_{xx}=\frac{m^*}{n_se^2\tau} \mathrm{~~~and~~~~} \rho_{xy}=\frac{B}{n_se},
\end{equation}
where $n_s$ is the carrier density, $m^*$ is the effective mass and $\tau$ is the scattering time. This classical model explains the Hall effect observable at low magnetic field or high temperature. Moreover, it emphasizes that the Hall resistance of 2D conductors is independent of dimensions, i.e. scale invariant. On the other hand, nor the quantization of the Hall resistance neither the cancellation of the longitudinal resistivity are predicted.
To explain these two features of the QHE, a quantum mechanics description must be considered.
\begin{figure*}[!h]
\includegraphics[width=5.2in]{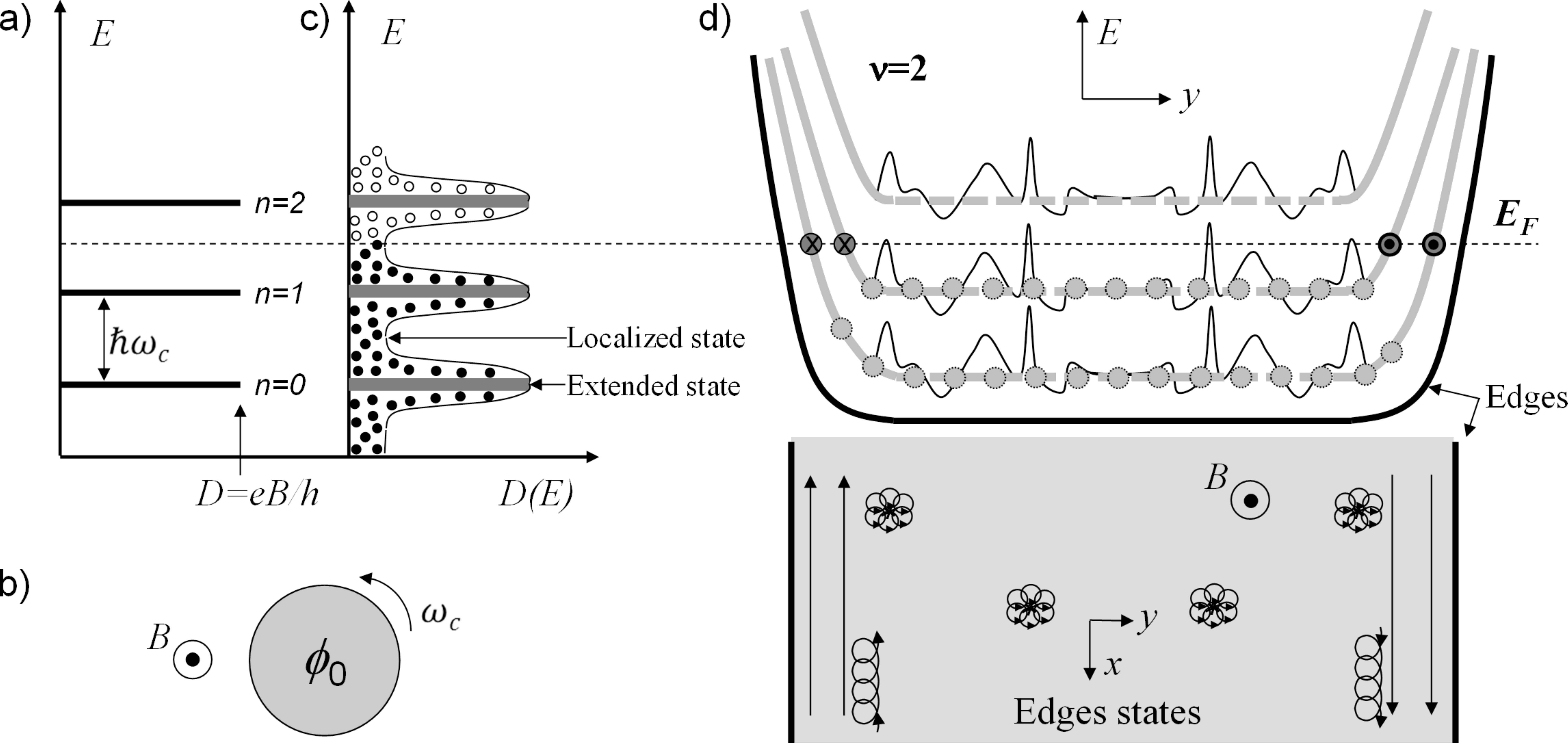}
\caption{a) Quantization of the density of states D(E) in Landau levels in a ballistic 2D conductor. b) Illustration of the quantization of the cyclotron motion. c) Density of states in a disordered 2D conductor. d) Device with edges in the QHE regime (at $\nu=2$). Landau levels bent by the confining potential (top). Schematic of two edge states and localized states in real space (bottom).}\label{fig:Fig-QHEEdges}
\end{figure*}

Several review papers\cite{Beenakker1991,Klitzing2004,Doucot2004,Buttiker2009} or books\cite{Janben1994,Girvin1999,Yoshioka1998,Goerbig2006} about the QHE physics can be consulted.
The hamiltonian of a free electron in two dimensions (2D) in presence of a magnetic field (potential vector A) is given by:
\begin{equation}
H=\frac{1}{2m^*}(P+eA)^2=\hbar\omega_c(a^+a+1/2),
\end{equation}
where $\omega_c=eB/h$ is the cyclotron pulsation and $a$ is a scaling operator obeying $[a^+,a]=1$. This hamiltonian, gauge invariant, is that of an harmonic oscillator whose energy spectrum is quantized in Landau levels (LL)(fig.\ref{fig:Fig-QHEEdges}a) at values given by:
\begin{equation}
\epsilon_n=\hbar\omega_c(n+1/2),
\end{equation}
where $n$ is an integer. The cyclotron motion is quantized (fig.\ref{fig:Fig-QHEEdges}b) and the energy spectrum is highly degenerate with regards to the center of guidance of the cyclotron orbit. It results that the density of states for each LL is $eB/h$ (one spin value) and therefore is equal to the density of flux quanta $n_B=B/\phi_0$. Calculations show that each electron occupies a surface $2\pi l_B^2=1/n_B$ in real space, \emph{i.e.} the area crossed by a flux quantum where $l_B=\sqrt{\hbar/eB}$ is the magnetic length. This explains the relationship $n_s=\nu n_B$ in quantized Hall states, where the LL filling factor $\nu$ is an integer. The electronic fluid is therefore incompressible and a high energy $\hbar\omega_c$ is required to add an electron.

To explain both the existence of a Hall resistance plateau and the drop to zero of the longitudinal resistance, disorder must be considered. It introduces a spatially varying potential which lifts the Landau level degeneracy (fig.\ref{fig:Fig-QHEEdges}c). This leads to extended states in narrow energy bands centered around $\epsilon_n$, and localized states at energies in between Landau levels. In the high magnetic field limit (or smooth potential), localized states correspond to closed equipotential lines around peaks or deeps of the potential while delocalized states spread along equipotential lines in valleys of the potential, as illustrated in fig.\ref{fig:Fig-QHEEdges}d. Only delocalized states can carry current. By varying the LL filling factor $\nu$ (variation of $B$ or $n_s$), the Fermi energy $E_\mathrm{F}$ can be continuously changed. While it is located at energies corresponding to localized states, the total net current, and thus the Hall resistance remains constant. Moreover, excitations towards extended states are blocked by the energy gap which prevents dissipation and leads to $R_\mathrm{xx}\sim 0$. Let us note that residual dissipation exists at finite low temperature due to conduction through localized states. As the filling factor $\nu$ moves closer to a LL energy, electrons experience a localisation/delocalisation\cite{Huckestein1995,Pruisken1988} transition (divergence of the localization length) which is considered to be a quantum phase transition\cite{Evers2008}.

To explain the values, $h/\nu e^2$, at which the Hall resistance is quantized, let us consider a real device geometry with edges, as illustrated in fig.\ref{fig:Fig-QHEEdges}d. The confining potential introduced by edges bends the Landau levels. At integer $\nu$ value, the Fermi energy $E_\mathrm{F}$ in the bulk intercept only localized states that do not carry any current. On the other hand, it intercepts LL extended states at edges which defines an integer number ($\nu=2$ in fig.\ref{fig:Fig-QHEEdges}d) of one-dimensional (1D) states. The velocity group of these states is reversed from one edge to the other. Thus, states with opposite momentum are spatially separated which forbids electron backscattering given the large width of the device compared to the magnetic length (note that conduction by hopping between localized states leads to a residual backscattering at finite low temperature). It results that these 1D-edges states are ballistic (their transmission is unity). From the scattering theory\cite{Landauer1957,Landauer1970} of the electronic transport, the conductance of a 1D ballistic state (one spin) is known to be $h/e^2$. This is a direct consequence of the Pauli principle combined with the Heisenberg time-energy uncertainty principle\cite{Glattli2009}. The two-terminal conductance of a Hall conductor is then obtained by simply counting the number of edge-states. Besides, backscattering being cancelled the dissipation can only occurs in contacts. More generally, the conductance properties in the QHE regime of a system with contacts at given chemical potentials can be obtained from the occupation of edge states\cite{Beenakker1991}. In this framework, the Landauer-Buttiker theory\cite{Buttiker1985,Buttiker1988} describes the conductance of multi-terminals conductors and notably the Hall bar. It notably predicts that the QHE requires phase coherency of the wave-function only at the scale of $l_B$.
\subsubsection{A universal and robust quantum effect}
The QHE is a universal quantum effect, which means that the quantized Hall resistance is linked to $h/e^2$ independently of the two-dimensional conductor considered. The first explanation of the QHE, proposed by Laughlin in 1981, showed that the universal character of the QHE originates from the gauge invariance of the hamiltonian\cite{Laughlin1981}. More precisely, let us consider a closed 2D ribbon submitted to a perpendicular magnetic field and a transverse electric field. The system being invariant by application of a flux quantum through the ribbon (filled states are simply shifted by one unity if the Fermi energy is in between two Landau levels), the variation of energy ($\Delta U=eV$) has a purely electrostatic origin and is caused by the transfer of one electron from one edge of the ribbon to the other. The expression of the current, $I=\Delta U/\phi_0=(e^2/h)V$, gave the first explanation of the universal quantized value\cite{Doucot2004}. This argument was then generalized by Thouless \emph{et al} showing that the Hall conductance is an topological invariant\cite{Niu1985,Thouless1994,Watson1996}. It was also demonstrated that neither electron-electron interaction\cite{Yennie1987} nor the gravitational field\cite{Hehl2004} lead to any correction. However, one work using quantum electrodynamic calculations reports on a correction to $h/e^2$ caused by a renormalization of the electric charge by the magnetic field. A tiny relative correction, $\alpha B^2$, amounting to $10^{-21}$ for $B=20$ T is predicted\cite{Penin2009}.

The universality and the reproducibility\cite{Schopfer2007} of the von Klitzing constant $R_\mathrm{K}$ were proved by showing the agreement of the quantized Hall resistance measured in several two-dimensional semiconductors with relative uncertainties down to a few $10^{-11}$. Different devices by the nature of the 2DEG (Silicon-MOSFET\cite{Hartland1991,Jeckelmann1995}, GaAs/AlGAs\cite{Jeckelmann2001}, InGaAs/InP\cite{Delahaye1986}, graphene\cite{Janssen2011,Ribeiro2015}), by their electronic properties (carrier density, electronic mobility, filling factor)\cite{Jeckelmann1995,Jeckelmann1997,JeckelmannIEEE2001} and their geometry (Hall bar size)\cite{Jeanneret1995} were tested.
\subsection{Single electron tunneling current sources}
\paragraph{Single electron pumps}
\begin{figure*}[!h]
\includegraphics[width=5.2in]{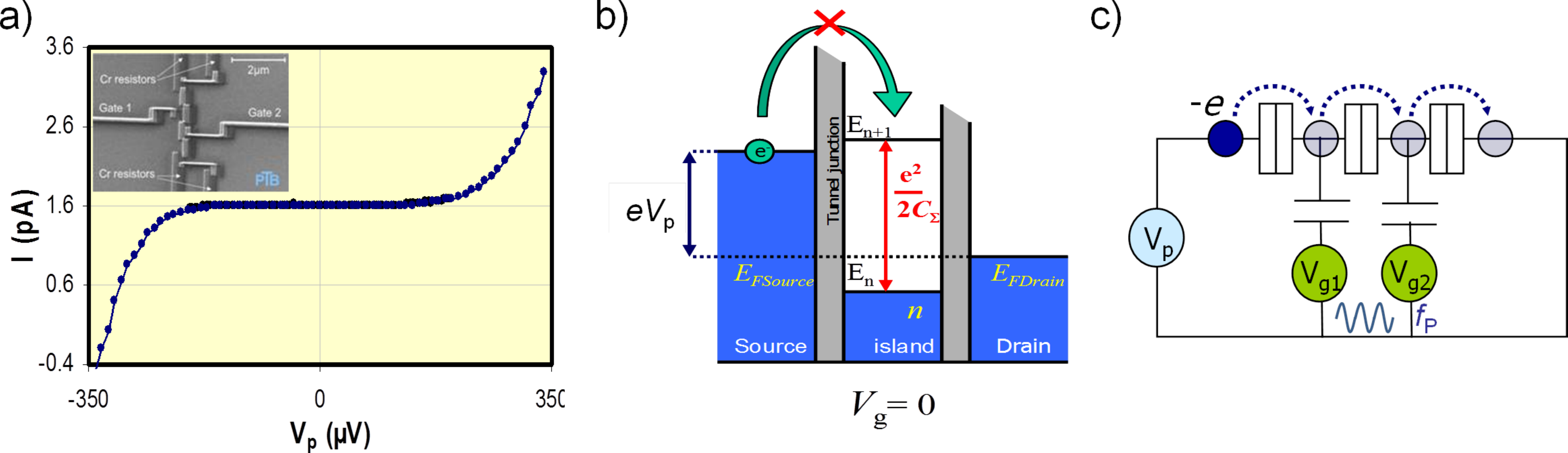}
\caption{a) Example of a current \emph{versus} voltage ($I-V$) curve showing a 1.6 pA current step obtained with a pumping frequency of $f_\mathrm{P}=10$ MHz. SEM picture of a metallic pump fabricated at PTB. b) Principle of Coulomb blockade in a single electron transistor (SET): the charging energy $e^2/2C_\Sigma$ opposes to the addition of an extra electron. c) Principle of the transfer one by one of electron in a pump device based on three tunnel-junctions and two islands. Energy states in islands are controlled by AC gate voltages synchronized at frequency $f_\mathrm{P}$. Adapted from ref.\cite{Feltin2009}.}\label{fig:Fig-Pump}
\end{figure*}
In the 90's, the possibility to manipulate a single electron charge has been demonstrated \cite{Lafarge1991} in mesoscopic conductors. Some of these devices, called single-electron pumps, have enabled the control of the transfer of electrons one by one at a rate fixed by an external frequency $f_\mathrm{P}$\cite{Pothier1992,Pothier,Kouwenhoven1991}, resulting in a current $I = Qf_\mathrm{P}$, where $Q\equiv e$. Fig.\ref{fig:Fig-Pump}a shows a current plateau at a value of 1.6 pA that is observed in the $I-V$ characteristic of such a device operating at a frequency of 10 MHz.

The operation of single-electron pumps relies on the charge quantization in a small metallic island, isolated by tunnel barriers of capacitance $C$ and resistance $R$, where Coulomb blockade manifests itself\cite{Grabert1991}. The transfer of electrons one by one occurs if 
: 1) the charging energy $e^2/C$ prohibiting the addition of a second electron, as illustrated in fig.\ref{fig:Fig-Pump}b, is larger than the thermal energy $k_BT$. This requires very low operating temperatures and small tunnel barrier capacitances and  2) the electron state has an energy thickness $\delta E$ much smaller than $e^2/C$, \emph{i.e.} $\delta E\ll e^2/C$. Considering the charging time of the island of the order of $RC$ and the energy-time Heisenberg uncertainty principle, one deduces $\delta E\sim h/(RC)$ which leads to a tunnel resistance $R\gg h/e^2$.

First single-electron tunneling (SET) pumps \cite{Pothier1992} consisted of several (at least two) small metallic islands in series isolated by tunnel junctions. Each island is capacitively coupled to a voltage generator (gate voltage) synchronized to $f_\mathrm{P}$ which is used to control its charge state. By adjusting carefully the amplitude and the phase of the gate voltages, $n_{Q}e$ charges can be transferred from island to island at each pumping cycle, where $n_{Q}$ is the number of charges. This principle is described in fig.\ref{fig:Fig-Pump}c. The output current delivered by these devices is therefore ideally equal to $I_{\mathrm{P}} = n_{Q}ef_\mathrm{P}$. A current plateau forms by varying the polarization voltage of the pump $V_\mathrm{P}$, as reported in fig.\ref{fig:Fig-Pump}a.
\paragraph{Accuracy of metallic single-electron pumps}
In a 7-junction device, quantized currents of a few pA have been generated for frequencies in the MHz range and an error rate of charge transfer per cycle as low as $1.5\times10^{-8}$ was measured\cite{Keller1996}. Given the low current values, the accuracy of such devices was determined by measuring the voltage at the terminals of a calibrated (in terms of $\mu_0$ or $R_\mathrm{K}$) cryogenic capacitor charged with a precise number of electrons in terms of $K_\mathrm{J}$\cite{Keller1999}. The quantization of the current was demonstrated with a relative uncertainty of $9.2\times10^{-7}$ for currents below 1 pA\cite{Keller2007}. In a similar experiment, a relative uncertainty of $1.66\times10^{-6}$ was reached with a 5-junction R-pump\cite{Camarota2012}. The limit in the uncertainty achieved comes from the small currents that theses metallic pumps can accurately generate. Large $RC$ values and serialization of several junctions used to reduce co-tunneling events indeed result in a strong frequency dependence which prevents generating larger currents with accuracy\cite{Grabert1991,Pekola2013}. As a trade-off between accuracy and increased current, several alternative quantum current sources have then been proposed\cite{Pekola2013,Kaestner2015,Okazaki2016}. They will be discussed in section \ref{SectionAmpere}
\subsection{Quantum standards in the SI based on the electromechanical definition of the ampere}
The high reproducibility and universality of the JE and the QHE motivated their use to realize the units of voltage and resistance respectively. The development of high-quality 10 V Josephson arrays and GaAs/AlGaAs Hall bars on one side and accurate comparisons bridges on the other side have allowed metrologists, in the late eighties, performing accurate calibrations of voltage references and resistors from $K_\mathrm{J}$ and $R_\mathrm{K}$ respectively with relative uncertainties around $10^{-9}$. A prerequisite before using the JE and the QHE in metrology was to link the Josephson voltage and the quantum Hall resistance to the volt and the ohm as defined in the SI. This means calibrating $K_\mathrm{J}$ and $R_\mathrm{K}$ in terms of SI units. In 1990, the Josephson effect and the quantum Hall effect were recommended by the CIPM to maintain the units of voltage and resistance in national metrology institutes (NMIs) and values for the two quantum constants were adopted: $K_\mathrm{J}=K_\mathrm{J-90}\times(1\pm4\times10^{-7})~\mathrm{GHz/V}$\cite{KJ} and $R_\mathrm{K}=R_\mathrm{K-90}(1\pm1\times10^{-7})~\Omega$\cite{RK} where $K_\mathrm{J-90}=483597.9~\mathrm{GHz/V}$ and $R_\mathrm{K-90}=25812.807~\Omega$. The uncertainties of determination of these two constants are much larger than the reproducibility of the quantum phenomena. This comes from the definition of the current unit based on Ampere's force law, which imposes the implementation of complex electromechanical experiments to measure $K_\mathrm{J}$ (the volt balance) and $R_\mathrm{K}$ (the Thompson-Lampard calculable capacitor).

To benefit even so from the high-reproducibility of the Josephson and quantum Hall effects for the traceability of voltage and resistance measurements, the conventional exact values (without uncertainties), $K_\mathrm{J-90}$ and $R_\mathrm{K-90}$, were recommended by the Comit\'e International des Poids et Mesures (CIPM)\cite{Quinn89} as the reference values in calibration certificates based on the implementation of these quantum effects. The voltage and the resistance measurements traceable to $K_\mathrm{J-90}$ and $R_\mathrm{K-90}$ give representations, and not realizations in the SI, of the volt and the ohm. It results that the current realized from $(K_\mathrm{J-90}R_\mathrm{K-90})^{-1}$ by application of Ohm's law from the representations of the volt and the ohm gives a representation of the ampere (not a realization). These decisions resulted in a major improvement of the reproducibility of the units of voltage and resistance, as realised by national metrology institutes (NMIs).
\begin{figure*}[!h]
\includegraphics[width=5.2in]{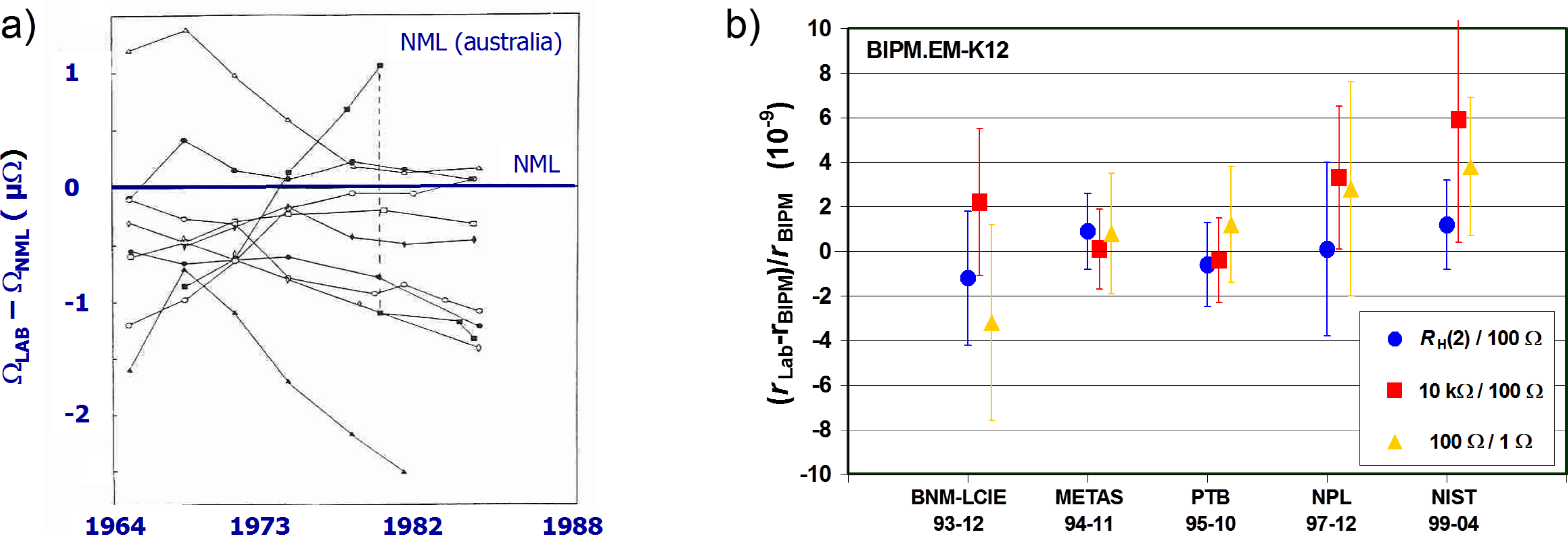}
\caption{a) Deviation in $\mu\Omega$ between realizations of the ohm in several NMIs and NML (australia) till 1988. b) Relative deviations of resistance ratio measurements performed in different NMIs and in BIPM by using the QHE and modern resistance bridges.}\label{fig:Fig-QHEBeforeAfter}
\end{figure*}
This is highlighted by fig.\ref{fig:Fig-QHEBeforeAfter} which shows a reduction of the relative deviations in resistance measurements between NMIs, from about $10^{-6}$ before the use of the QHE a) down to $10^{-9}$ after its recommendation by the CIPM. The exploitation of the JE led to a similar strong improvement.

This is highlighted by fig.\ref{fig:Fig-QHEBeforeAfter} which reports the deviations between ohm realizations performed by different NMIs, before a) and after b) the use of the QHE. It shows a reduction from about $10^{-6}$ down to $10^{-9}$ of the relative deviations. The exploitation of the JE led to a similar strong improvement.

However, this artifice, which makes the traceability of the electrical measurements advantageous for end-users, is not applicable to experiments where realizations of units, and not representations, are required. This is notably the case of high-precision experiments involving both mechanical measurements and electrical measurements traced to quantum effects: it would indeed be incoherent to drop out the uncertainties of $K_\mathrm{J}$ and $R_\mathrm{K}$, the origin of which is mechanical. To illustrate this difficulty, one can evoke the realisation of the farad either from $\mu_0$ using the Thompson-Lampard calculable capacitor or from $R_\mathrm{K-90}$ using the QHE. Another example, discussed in subsection \ref{Kibble balance}, is the Kibble balance experiment that links the kilogram to electrical units. A main motivation and issue of the revised SI was to get rid of this artifice.
\subsection{A new definition of the ampere from the elementary charge}
\subsubsection{The revised SI}
The article "Redefinition of the kilogram: a decision whose time has come" by Mills \emph{et al}\cite{Mills05} in 2005 has not only initiated a cogitation about a redefinition of the kilogram without an artefact, but has also crystallized a consensus around a major evolution of the SI to overpass the limits imposed by the definitions of others units: the ampere, the kelvin, the mole. Concerning the ampere, the goal was to find a definition to fully benefit from the high reproducibility and universality of the quantum electrical standards. More generally, the ambition of the revised SI was to take into account modern physics, \emph{i.e.} quantum physics and statistical physics and avoid definitions closely linked to given practical realizations. The revised SI based on fixed fundamental constants ($\Delta \nu_{Cs}$, $c$, $h$, $e$, $k$, $K_\mathrm{cd}$) fulfils these requirements\cite{REVSIBIPM}.
\begin{table}[h]
\newcolumntype{M}[1]{>{\centering\arraybackslash}m{#1}}
\begin{center}
\begin{tabular}{|c|M{10cm}|}
  \hline
  \textbf{Units} & \textbf{Definitions} \tabularnewline \hline
  kilogram (kg) & The kilogram, symbol  kg, is the SI unit of  mass. It is defined by taking the  fixed numerical value of the Planck constant $h$ to be $6.626 070 15\times10^{-34}$ when expressed in  the  unit  J s,  which  is  equal  to  $\mathrm{kg~m^{2}~s^{-1}}$, where the  metre  and  the  second  are defined in terms of $c$ and $\Delta \nu_\mathrm{{Cs}}$.\tabularnewline \hline
  ampere (A) & The ampere, symbol A, is the SI unit of electric current. It is defined by taking the fixed numerical value of the elementary charge $e$ to be $1.602 176 634\times10^{-19}$ when expressed in the unit C, which is equal to A s, where the second is defined in terms of $\Delta \nu_\mathrm{{Cs}}$\\  \hline
\end{tabular}
\caption{Definitions of the kilogram and the ampere associated to the fixing at exact values of $h$ and $e$ from $20$ May 2019. The unperturbed ground state hyperfine transition frequency of the caesium 133 atom $\Delta \nu_\mathrm{{Cs}}$ is 9 192 631 770 Hz. Values\cite{Newell2018} of $h$ and $e$ were established from the CODATA 2017 adjustment\cite{Mohr2018}.}
\label{table:nouveauSI}
\end{center}
\end{table}
Electrical metrology is directly concerned by the definitions of the kilogram and the ampere based on the constants $h$ and $e$ respectively. They are presented in table~\ref{table:nouveauSI}.

The setting of this revised SI follows many works aiming at improving the knowledge of fundamental constants before fixing their exact values. The goal was to reduce the measurement uncertainties not only of $c$, $h$, $e$ and $k$ but also of the constants $K_\mathrm{J}$, $R_\mathrm{K}$ and $Q$ in order to check the solid-state theory on which quantum electrical standards rely.
\begin{figure*}[!h]
\includegraphics[width=5.2in]{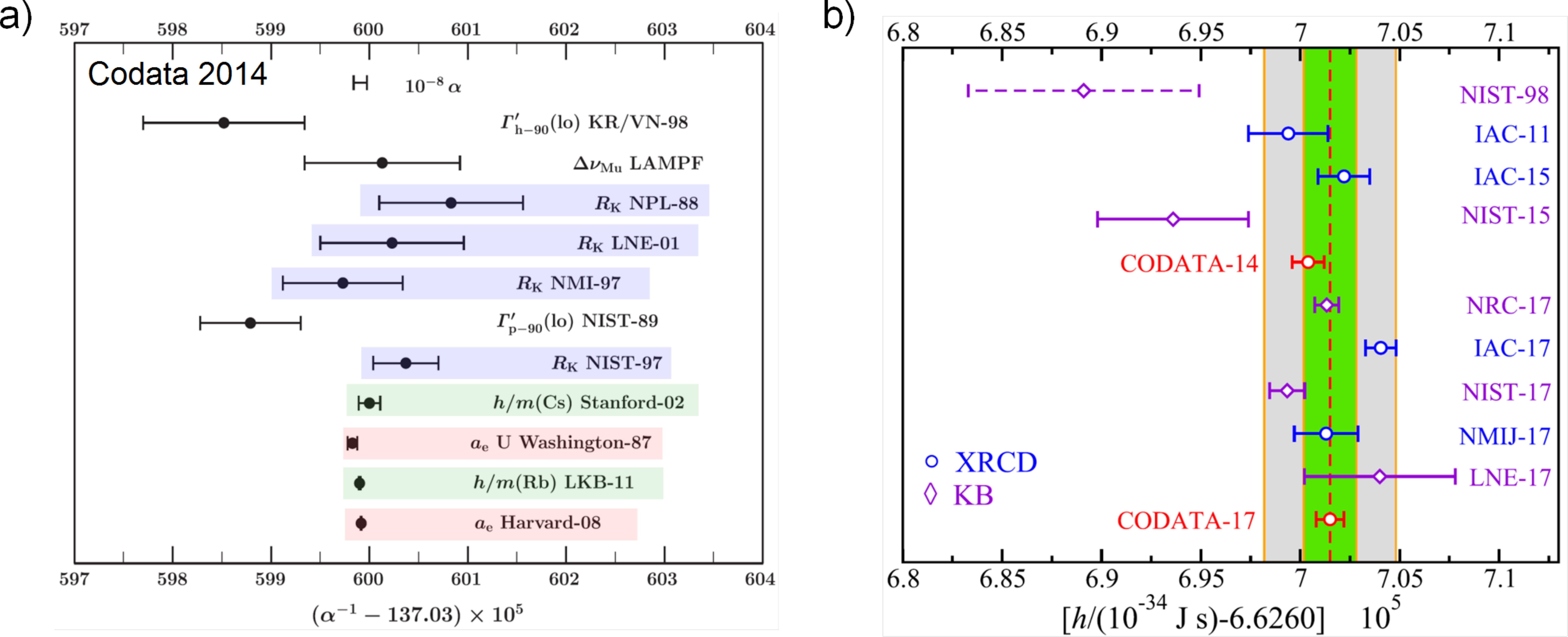}
\caption{a) Determinations of the fine structure constant $\alpha$: from $R_\mathrm{K}$, from $h/m$ by atomic interferometry and from the abnormal magnetic moment of the electron $a_e$ using quantum electrodynamic calculations. b) Determinations of the Planck constant $h$ using a Kibble balance (KB) or a silicon sphere (XRCD).}\label{fig:Fig-RKKJ}
\end{figure*}
\subsubsection{Determinations of $R_\mathrm{K}$ and $\alpha$}
 The von Klitzing constant $R_\mathrm{K}$ can be measured through a comparison with the impedance $1/(2\pi fC)$ using a quadrature bridge, where $f$ is the operation frequency and $C$ is a capacitance calibrated from the Thompson-Lampard calculable capacitor that was described before. It is interesting to compare the determinations of $R_\mathrm{K}$ with measurements of $h/e^2$ in order to test the QHE theory. It is equivalent to compare the determinations of the fine structure constant $\alpha=\frac{\mu_0c}{2(h/e^2)}$ with the estimations $\alpha=\frac{\mu_0c}{2R_\mathrm{K}}$. Determinations of $\alpha$ can be obtained from measurements of $h/m_{at}$ [$m_{at}$ is an atomic mass (cesium or rubidium atoms)] by atomic interferometry, or measurements of the abnormal magnetic moment of the electron combined with quantum electrodynamic calculations. Results\cite{codata16} reported in fig.\ref{fig:Fig-RKKJ}a, shows that they are in agreement with the estimations from $R_\mathrm{K}$, including that of LNE measured with relative uncertainty of $5.3\times10^{-8}$ using a specific five electrodes calculable capacitor\cite{Trapon03}. It results that $R_\mathrm{K}=\frac{h}{e^2}(1+\epsilon_\mathrm{K})$ with $\epsilon_\mathrm{K}=(2.2\pm1.8)\times10^{-8}$. This confirms the QHE theory, futhermore supported by universality tests which shows that $R_\mathrm{K}$ is independent on 2D material with uncertainties down to a few $10^{-11}$. From all data, an accurate value of $\alpha$ is deduced\cite{Mohr2018}: $\alpha^{-1}=137.035999139\times(1\pm2.3\times10^{-10})$.
\subsubsection{Determinations of $K_\mathrm{J}$ and $h$}
 $K_\mathrm{J}$ constant was initially determined using the volt balance. The uncertainty of measurement of $K_\mathrm{J}$ was improved using the Kibble balance (a watt balance) and the $R_\mathrm{K}$ value. This experiment consists in measuring the mechanical power of a mass $m$ moving at velocity $v$ under the gravitational field $g$ in terms of an electric power in a coil calibrated from $1/R_\mathrm{K}K_\mathrm{J}^2$, which is an estimate of the Planck constant $h$. The Kibble balance therefore establishes a relationship between the kilogram and $h$. Its advantage is that the watt does not depend on the ampere definition, contrary to the volt. This allows to overcome some technical difficulties, for example the geometric calibration of the coil. Comparing $K_\mathrm{J}$ value to its theoretical expectation requires the knowledge of $2e/h$. The latter constant can be obtained from $h/e^2$ and the determination of $h$ from the Avogadro constant $N_\mathrm{A}$. The Planck constant can indeed be deduced from the relationship $h=\frac{cA_r(e)M_u\alpha^2}{2R_\infty N_\mathrm{A}}$, where $A_r(e)$ is relative atomic mass of the electron, $M_u$ is the molar mass constant and $R_\infty$ is the Rydberg constant. The Avogadro constant $N_\mathrm{A}$ can be determined from the number of atoms in a silicon sphere of volume $V_{\mathrm{sphere}}$ and mass $m_{\mathrm{sphere}}$, according to $N_\mathrm{A}=\frac{M_\mathrm{Si}V_{\mathrm{sphere}}}{\sqrt{8}d_{220}^3 m_{\mathrm{sphere}}}$, where $M_\mathrm{Si}$ is the silicon molar mass and $d_{220}$ is the inter-atomic distance measured by X-ray diffraction. Here, one can note that the mass can be realised from $N_\mathrm{A}$. Testing the agreement of $K_\mathrm{J}$ with $2e/h$ is as comparing determinations of $h$ using the Kibble balance (assuming QHE and JE theories are valid) and determinations from $N_\mathrm{A}$. Analysis from CODATA 2014 group demonstrated the absence of significant disagreement between the two determinations of $h$, reported in fig.\ref{fig:Fig-RKKJ}b. It was deduced that $K_\mathrm{J}=\frac{2e}{h}(1+\epsilon_\mathrm{J})$ with $\epsilon_\mathrm{J}=(-0.9\pm1.5)\times10^{-8}$. This result and the universality tests of the JE show that the JE theory could be adopted. The last adjustment of constants carried out by CODATA 2017 group\cite{Mohr2018}, that considered new results, determined an accurate value of the Planck constant equal to $h=6.626070150\times10^{-34}(1\pm1.0\times10^{-8})$ J.s. From this value and that of $\alpha$, the value of the elementary charge obtained is $e=1.6021766341\times10^{-19}(1\pm5.2\times10^{-9})$ C.
\subsubsection{Closure of the metrological triangle}
\begin{figure*}[!h]
\includegraphics[width=5.2in]{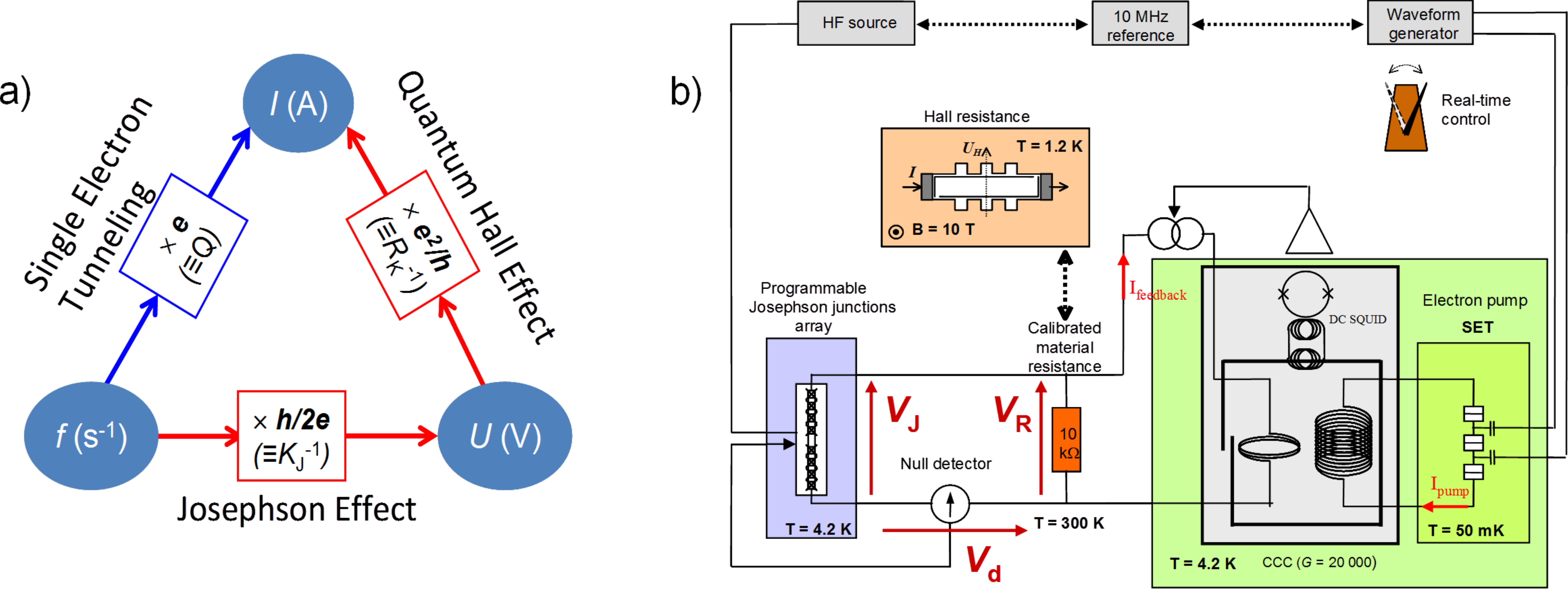}
\caption{a) Illustration of the metrological triangle \cite{Likharev1985} and realization of the current $I$ from frequency $f$, either by using the relation $I=Qf$ in the single electron tunnelling effect, or by using Ohm's law associated with the Josephson and the quantum Hall effects, $I=n(R_\mathrm{K}K_\mathrm{J})^{-1}f$, where $n$ is an integer. Adapted from ref.\cite{Brun-Picard2016}. b) Schematic of LNE experiment to close the metrological triangle \cite{Feltin2009,Devoille2012}. The current generated by a metallic pump $I_\mathrm{pump}$ is amplified by a factor 20000 using a cryogenic current comparator (CCC). This amplified current feeds a calibrated resistor from the QHE. The voltage drop $V_\mathrm{R}$ at resistance terminals is measured by comparison with a Josephson reference voltage $V_\mathrm{J}$. From ref.\cite{Devoille2012}.}\label{fig:Fig-CloseTriangle}
\end{figure*}
 This experience, illustrated in fig.\ref{fig:Fig-CloseTriangle}a, consists in comparing the realization of the ampere from the frequency by implementing SET devices on one side and by applying Ohm's law to quantum voltage and resistance standards on the other\cite{Likharev1985,Feltin2009}. It leads to the determination of the constants product $R_\mathrm{K}K_\mathrm{J}Q$ which is theoretically equal to 2 if $R_\mathrm{K}=h/e^2$, $K_\mathrm{J}=2e/h$ and $Q=e$. This is a fundamental test of consistency of the quantum solid state physics, where one of the issues it to check that the quasi-particles either handled in the SET quantum dot, or flowing along the sample QHE edge or in the Cooper pair have the same elementary charge. Any discard to the expected value would question a part of quantum mechanics. In practice, this direct comparison implementing the three quantum standards together has never been performed. Instead, the current generated by SET devices was measured by using a secondary resistor (or capacitor) calibrated from $R_\mathrm{K}$, whose voltage at its terminals was compared to a Josephson voltage reference. Among the many works that have reported measurements of the output current of SET devices, only three of them have claimed the closure of the metrological triangle\cite{Keller2007,Devoille2012,Scherer2012}. The main reason is that determining $R_\mathrm{K}K_\mathrm{J}Q$ requires that the expected number of charges $n_Q$ is really transferred with accuracy at each cycle. Some authors consider that there is no proof of that for recent non-adiabatic SET devices, described in section \ref{Newmonoelectronicdevices}, and that an independent measurement of $n_Q$ is required to make a determination of $Q$. Thus, the three determinations were based on the use of metallic SET pumps, the physics of which was well understood and, above all, it was demonstrated in a 7-junction device that the error rate of charge transfer per cycle was as low as $1.5\times10^{-8}$\cite{Keller1996}. Moreover, some quantization criteria, such as observation of a linear frequency dependence of the current and of a minimization of co-tunneling events by setting the biasing voltage, could be used to confirm quantization state. Using a measurement method based on the charging of a calibrated cryogenic capacitor, research works \cite{Keller2007} and \cite{Scherer2012} achieved closure of the metrological triangle with relative uncertainties of $9.2\times10^{-7}$ and $1.66\times10^{-6}$ respectively. In work\cite{Devoille2012} from LNE, the SET device current is amplified using a CCC\cite{Sassine2010} before measurement (fig.\ref{fig:Fig-CloseTriangle}b) and the uncertainty achieved is of $1.3\times10^{-5}$. Although confirming solid state physics, these works have not contributed to an estimation of the elementary charge $e$ from $R_\mathrm{K}$ and $K_\mathrm{J}$ in the CODATA calculations because of the too large uncertainties achieved.
\subsubsection{Impact of the new ampere definition}
Exact values of $h$ and $e$ chosen to establish the new SI definitions and reported in table~\ref{table:nouveauSI} were obtained by truncating the digits of the values determined in the previous SI. One advantage of the new definitions is that they do no specify any given realization. The value of the elementary charge expressed in coulomb, \emph{i.e.} in ampere.seconde, simply means that the electrical current corresponds to a fixed flux value of elementary charges per time unit. Thus, any experiment based on the handling of elementary charges can, in principle, constitute a realization of the ampere. Besides, let us note that the definitions of all others electrical units have not changed in the revised SI. Following the verification of the quantum theories of the JE and the QHE with lower uncertainties, and to some extend of the single electron tunneling effect, the relationships $K_\mathrm{J}=\frac{2e}{h}$, $R_\mathrm{K}=\frac{h}{e^2}$ and $Q=e$ are adopted in the revised SI. It results that the JE, the QHE  and the SET effect are recommended experiments to realize the volt, the ohm and the ampere. Fig.\ref{fig:Fig-Intro}-right illustrates the link of these three units to $h$, $e$ and $\Delta \nu_\mathrm{Cs}$. Constants being exact, the uncertainty of realization of units comes from the implementation of the quantum phenomena, and no more from the definition of the ampere itself. Recommendations for the \emph{mise en pratique} of the electric units were written (Draft\cite{SIAnnexe2} for Appendix 2 of the SI Brochure for the “Revised SI”). Here, are reported those concerning the volt, the ohm and the ampere.
\paragraph{Practical realization of the volt, V}~\\
The volt, V, can be realized from $K_\mathrm{J}$ using the Josephson effect. Although the $2e/h$ value can be used, a truncated value with 15 significant digits is recommended: $K_\mathrm{J} = 483 597.848 416 984~\mathrm{GHz V^{-1}}$. This value is lower than the value $K_\mathrm{J-90}$ by a relative amount of $106.665\times10^{-9}$. As a consequence, the numerical value of a voltage measured in terms of the new SI volt is larger than the value measured in terms of $K_\mathrm{J-90}$ by the same amount.
\paragraph{Practical realization of the ohm, $\Omega$}~\\
a) The ohm, $\Omega$, can be realized from $R_\mathrm{K}$ by using the quantum Hall effect in a manner consistent with the CCEM Guidelines\cite{Delahaye2003}. Although the $h/e^2$ value can be used, a truncated value with 15 significant digits is recommended: $R_\mathrm{K} = 25 812.807 459 3045~\Omega$. This value is larger than the value $R_\mathrm{K-90}$ by a relative amount of $17.793\times10^{-9}$. As a consequence, the numerical value of a resistance measured in terms of the new SI ohm is larger than the value measured in terms of $R_\mathrm{K-90}$ by the same amount; or\\
b) by comparing an unknown resistance to the impedance of a known capacitance using, for example, a quadrature bridge, where, for example, the capacitance has been determined by means of a calculable capacitor and the value of the electric constant of vacuum $\epsilon_0=1/\mu_0c^2$. In the revised SI, the the magnetic constant of vacuum has no longer the exact value $4\pi\times10^{-7}\mathrm{N/A^2}$. It is obtained from the fine structure constant value $\mu_0=2\alpha/ce^2$. The value determined from the CODATA 2017 adjustment is $\mu_0=4\pi[1+2.0(2.3)\times10^{-10}]\times10^{-7}\mathrm{N/A^2}=12.566 370 6169(29)\times10^{-7}\mathrm{N/A^2}$. It results the value of $\epsilon_0$ is no longer exact either. Its relative uncertainty is equal to that of $\mu_0$ since $c$ is fixed.
\paragraph{Practical realization of the ampere, A}~\\
a) The ampere can be realized by using Ohm's law, the unit relation A = V/$\Omega$, and using practical realizations of the SI derived units the volt V and the ohm $Ω\Omega$, based on the Josephson and quantum Hall effects, respectively; or\\
b) by using the relation $I = C·dU/dt$, the unit relation A = F·V/s, and practical realizations of the SI derived units the volt V and the farad F and of the SI base unit second s; or\\
c) by using a single electron transport (SET) or similar device, the unit relation A = C/s, the value of $e$ given in the definition of the ampere and a practical realization of the SI base unit second s. However, SET implementations still have technical limitations and often larger relative uncertainties than some other competitive techniques.

\emph{Mise en pratique} based on quantum electrical standards were also recommended for the units coulomb, farad and watt\cite{SIAnnexe2}. Generally, the revised SI promotes quantum solid-sate physics. This is to sustain further development of the quantum electrical standards and their applications.
\section{The volt and the ohm from quantum standards}
\subsection{The Quantum voltage standard}
\label{TheQuantumvoltagestandard}\label{JVS}
\paragraph{Introduction} Nowadays, in NMIs, coexist three generations of state-of-the-art Josephson voltage standards (JVS): the conventionnal and programmable Josephson voltage standards (CJVS and PJVS), and the Josephson arbitrary waveform synthesizers (JAWS or ACJVS). These JVS are very complex superconductive circuits with thousands of junctions in series. The development of JVS had to overcome several difficulties concerning the quality of the JJ (homogeneity of the junction parameters over large areas of the order of cm$^{2}$, stability of the material, high yield), the optimization of the microwave design and the design of the bias electronics. Not only, at each generation, the number of junctions and the complexity of the circuit have increased but also the domain of applications.

The first generation was dedicated to dc voltage metrology. The main achievement of these standards is the increase of the voltage output from a few millivolts\cite{Field1973,Endo83} to 10 V\cite{Lloyd1987,Popel1992} and the establishment of the basis of the present standard volt conservation to a few parts in $10^{10}$. The PJVS have opened the way to the rapid dc voltage selection and to low frequency ac applications. Finally, JAWS are achieving the mutation towards programmability and higher frequencies by allowing the generation of arbitrary waveforms with fundamental accuracy up to the MHz range. Several review papers on JVS have been published covering all the aspects of the Josephson voltage standards \cite{Popel1992,Kautz1992,Kautz1996,Hamilton2000,Kohlmann2003,Jeanneret2009,Behr2012}. Here we will present the state-of-the-art of the three generations of JVS.
\subsubsection{Conventional devices}
\begin{figure*}[!h]
\includegraphics[width=5.2in]{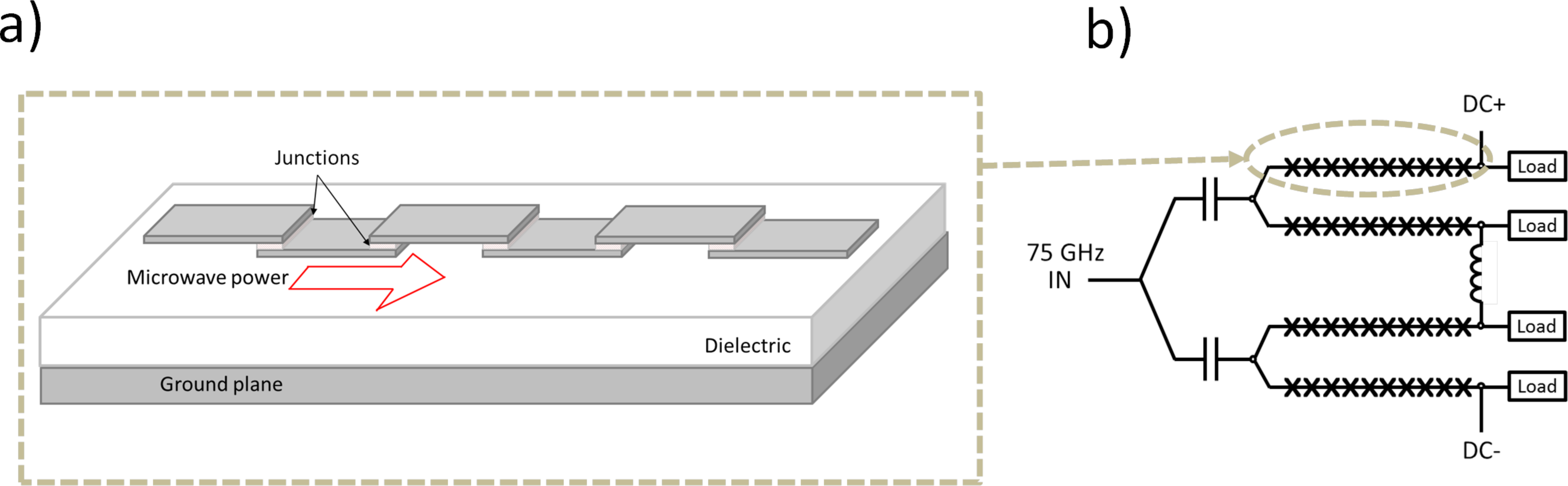}
\caption{Array of Josephson junctions: a) Series of Josephson junctions introduced in a microstrip transmission line. The junctions are distributed in the microstrip over a superconducting ground plane, separated from the strip by a dielectric layer. b) Schematic of the circuit of a typical Josephson voltage standard: to increase the number of JJ while limiting the effect of rf power attenuation in the transmission line, the JVS is formed of parallel series arrays terminated by matched loads that absorb the rf power and avoid reflections. A network of low and high-pass filters splits the microwave power into parallel paths (here four), and allows all junctions to be connected in series at dc. Inspired from \cite{Hamilton2000}.} \label{fig:Fig_Jos_2}
\end{figure*}
\paragraph*{Principle}
The CJVS are based on the idea proposed by Levinsen in 1977 \cite{Levinsen1977} to use zero-crossing steps shown in fig.\ref{fig:Fig_Jos_1b}c corresponding to $\beta\geq 100$. First, there are no stable regions between the first steps, this ensures the quantization of the voltage. Second, all the steps can be selected with the same bias current ($I\sim0$), and the array can be disconnected from the bias source during the measurements. This relaxes, to some extent, the need of perfectly identical JJ parameters in the arrays, and for this reason it allowed the fabrication of the first 10 V arrays of JJ\cite{Lloyd1987,Popel1992}.

\paragraph*{Junction}
SIS (Superconducting/Insulator/Superconductor) junctions are fabricated with Nb/Al$_{2}$O$_{3}$/Nb thin film structures \cite{Gurvitch1983}, which ensure clean interfaces and thin insulating junction barrier. Niobium (Nb) is mechanically and chemically stable, preventing the Josephson arrays from aging problems. Moreover, the critical temperature of Nb of 9 K allows working in liquid helium at 4.2 K \cite{Jeanneret2009}. The junctions are planar junctions of width $w$ and length $l$ made by the superposition of two superconducting films separated by the insulating barrier. They are imbedded in the microstrip of a superconducting microstrip-transmission line as illustrated in fig.\ref{fig:Fig_Jos_2}a. The choice of the Josephson junction parameters $w$, $l$ and $I_{c}$ and the operating frequency $f$ can be determined within the RSCJ model (Section \ref{RCSJ} and fig.\ref{fig:Fig_Jos_2}) with the aim to increase the stability of phase-lock against thermal noise and chaos, and to avoid any spatial dependence of the junction phase over the junction area. For a detailed discussion on the subject, see ref.\cite{Kautz1992,Kautz1996}. Typical parameters for the junctions are $w$ = 30 $\mu$m, $l$ = 18 $\mu$m corresponding to a critical current of 110 $\mu$A and working frequencies are around 75 GHz \cite{Hamilton2000}.

\paragraph*{Microwave circuit}
Typical 10 V Josephson voltage standards are composed of about 14000 to 20000 Josephson junctions \cite{Hamilton1989,Muller1997}. To reach 10 V, high order constant voltage steps are exploited (typically, at 10 V, a JVS of 14000 JJ works in average on the voltage step $n=5$).
A homogeneous rf power distribution for all the junctions is a key element for proper operation of the JVS, hence, the problems of rf power attenuation and reflections should be circumvented. The rf power attenuation 
along a series array limits the number of junctions in one array to about few thousands. To reach more than 10000 JJ, several arrays are needed. This is realized by connecting series arrays in a series/parallel circuit as shown in fig.\ref{fig:Fig_Jos_2}b. This circuit allows splitting the microwave power into parallel paths while maintaining a dc path. The reflections are avoided by a matched load at the end of each series array.
\begin{figure*}[!h]
\includegraphics[width=5in]{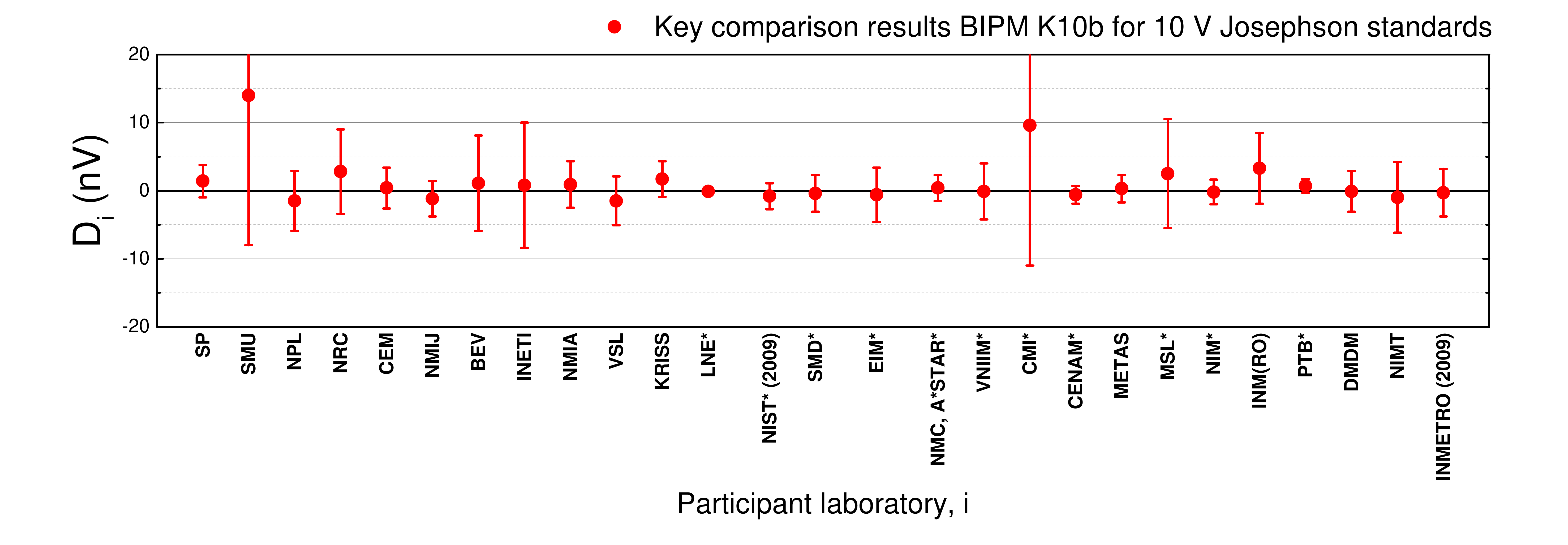}
\caption{
Results of the Key comparison BIPM.EM-K10.b, SIM.EM.BIPM-K10.b and K10.b.1, COOMET.EM.BIPM-K10.b for 10 V Josephson standards, expressed in terms of degrees of equivalence. The degree of equivalence is given by a pair of terms, $[D_i, U_i]$, where $D_i$ is the result of the measurement carried out by laboratory $i$ expressed as the difference from the BIPM value, and $U_i$ is the expanded uncertainty, $U_i = k u_i$, where $k = 2$ is the coverage factor and $u_i$ the combined uncertainty. Adapted from ref.\cite{BIPMK10b}.}\label{fig:Fig_Jos_2b}
\end{figure*}
\paragraph*{Measurement system and applications}
The measurement system is composed of the microwave source phase locked to a 10 MHz frequency referenced to an atomic clock through a GPS receiver, a bias electronics which allows selecting the zero-crossing steps, an oscilloscope to visualize the steps and to optimize the microwave power and frequency. Once the step is selected, the array is disconnected from the bias source. The CJVS are used for the calibration of the 1.018 V and 10 V outputs of Zener-diode-based dc reference standards used in the traceability chain and for the calibration of the gain and linearity of high precision digital voltmeters. Fig.\ref{fig:Fig_Jos_2b} presents the results of the international Key comparison BIPM EM K10b \cite{BIPMK10b}. It shows that for most of the participants, the degree of equivalence, which is defined as the voltage difference with respect to the BIPM value associated with the expanded uncertainty $U_i$ corresponding to a coverage factor $k=2$, is below 5 parts in $10^{10}$ \cite{Wood2009} and some are at a few parts in $10^{11}$ \cite{Solve2009}. The wide dissemination of CJVS is ensured by the commercialization of these standards by two hightech companies specialized in superconducting electronics \cite{Hypres,Supracon}.
\subsubsection{PJVS at 10 V or more}
\begin{figure*}[!h]
\includegraphics[width=5.2in]{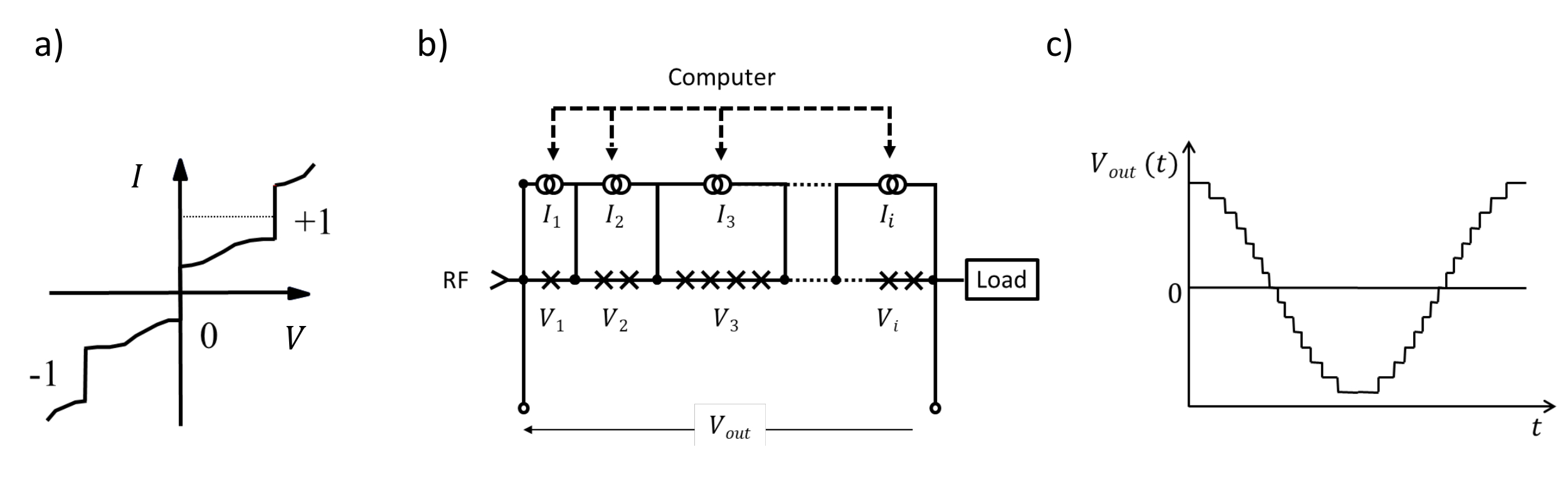}
\caption{Programmable Josephson voltage standards (PJVS) principle : a) $I-V$ characteristic of a non hysteretic Josephson junction ($\beta\leq1$) under microwave irradiation, showing the three constant voltage steps $n=\pm1, 0$ used in PJVS. b) Principle of the digital to analog converter based on a Josephson array \cite{Hamilton1995}. Here the array is divided in $i$ segments containing a number of JJ following a binary sequence. Each of the segments is biased by a computer controlled current source. c) AC voltage generation: the output voltage $V_{out}(t)$ of the PJVS is a stepwise approximated waveform.}\label{fig:Fig-Jos-PJVS}
\end{figure*}
\paragraph*{Principle}
Josephson voltage standards based on zero-crossing steps are not satisfying for programmable applications due to the difficulty to select a precise voltage step number $n$, and because of the susceptibility to noise-induced-spontaneous transitions. To circumvent the problem, in 1995, Hamilton \emph{et al.}\cite{Hamilton1995} proposed to use overdamped junctions for which each bias current value corresponds to a single voltage value (fig.\ref{fig:Fig-Jos-PJVS}a). He also suggested to transform the array into a {"}digital-to-analogue converter (DAC) of fundamental accuracy{"} by dividing a series array of $N_{\mathrm{J}}$ junctions into segments containing different numbers of JJ as depicted in fig.\ref{fig:Fig-Jos-PJVS}b. Each segment is current biased on one of the three constant voltage steps $n = -1, 0$ or $+1$, such that the output voltage can take any value between $\pm N_{J} \Phi_{0} f$ by increment of the voltage of the smallest segment. Most of the 1 V arrays were subdivided in segments with a number of JJ following a binary sequence \cite{Benz1997}. On the other hand, 10 V arrays, depending on the junction technology, can have very different sequences\cite{Yamamori2006,Muller2009,Burroughs2011}. The programmability is possible thanks to a computer controlled bias source. An ac voltage $V_{out}(t)$ can be generated by biasing sequentially different segments so that the output signal is a stepwise approximated waveform as sketched in fig.\ref{fig:Fig-Jos-PJVS}c.
\begin{table}[h]
\newcolumntype{M}[1]{>{\centering\arraybackslash}m{#1}}
\small

\begin{center}
\begin{tabular}{|l|c|c|c|}
  \hline
  Characteristic & PTB \cite{Muller2009} & NIST \cite{Dresselhaus2011,Burroughs2011} & NMIJ \cite{Yamamori2008,Yamamori2010} \\ \hline
  Voltage & 10 V & 10 V & 17 V \\
  Frequency & 70 GHz & 18.3 GHz & 16 GHz \\
  Junctions & 69 632 & 268 800 & 524 288 \\
  Material & Nb/Nb$_{x}$Si$_{1-x}$/Nb & Nb/Nb$_{x}$Si$_{1-x}$/Nb & NbN/TiN$_{x}$/NbN \\
  Temperature & 4.2 K & 4.2 K & 9.8 K \\
  Stacks & 1 & 3 & 2 \\
  Transmission line & Microstrip & Coplanar waveguide & Coplanar waveguide \\
  Parallel arrays & 128 & 32 & 64 \\
  \hline
\end{tabular}
\caption{Main characteristics of the 10 V PJVS arrays developed by PTB, NIST and NMIJ.}
\label{table:10V_PJVS}
\end{center}
\end{table}
\paragraph*{Junctions and microwave circuit}
The PJVS technology has evolved during about 15 years. For a historical overview of the development, the reader can consult recent reviews \cite{Jeanneret2009, Behr2012}. The current technology is based on SNS (superconducting-normal metal-superconducting) junctions, which are intrinsically overdamped junctions ($\beta \leq 1$). Kautz \cite{Kautz1995} showed that the high critical currents (in the mA range) of SNS junctions provide good immunity to thermal and electrical noise and that the frequency of operation should be very close to the characteristic frequency, $f_{c}=\frac{2e}{h}I_{c}R$, in order to obtain simultaneously the maximum amplitude for the voltage steps $n=\pm1,0$.

Among the three Institutes that fabricate 10 V PJVS, (NIST, PTB, and NMIJ/AIST), NIST and PTB are using Nb/Nb$_{x}$Si$_{1-x}$/Nb junctions first developed at NIST \cite{Baek2006}. Despite the same material, the arrays operate at very different frequencies, 18 GHz at NIST \cite{Dresselhaus2011} and 70 GHz at PTB \cite{Muller2009}. The operating frequency is adjusted by tuning the $I_{c}R_{n}$ product by changing both the thickness of the barrier and by varying the Nb content of the amorphous Nb$_{x}$Si$_{1-x}$ by a few percent. The choice of NMIJ is to use NbN/TiN$_{x}$/NbN junctions benefiting of the higher critical temperature of NbN at 16 K in order to fabricate PJVS able to operate at a temperature above 4.2 K in cryocoolers.

The use of Nb$_{x}$Si$_{1-x}$ and TiN$_{x}$ enables the vertical stacking of the JJ. Double and triple stacked JJ are currently used in the 10 V PJVS from NIST and NMIJ, to limit the size of the array while increasing the number of JJ to compensate for the reduction of the operating frequency. PTB could achieve 20 V \cite{Muller2013} with a double stack.
The two different domains of operating frequency lead to different microwave designs: at 70 GHz, the microstrip transmission line of the CJVS has been adapted, while in the 20 GHz  range, coplanar waveguides (CPW) are used. Table \ref{table:10V_PJVS} summarizes the parameters of the different 10V PJVS.


\begin{figure*}[!h]
\includegraphics[width=5.2in]{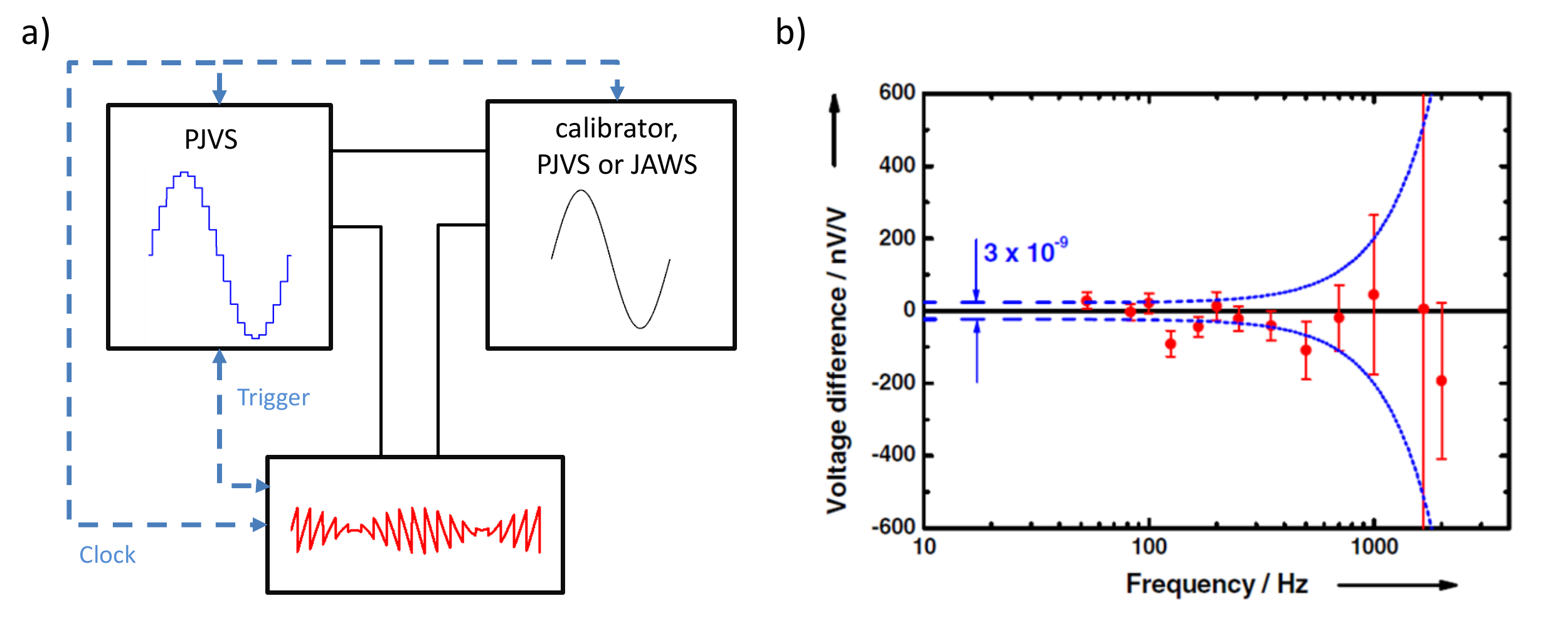}
\caption{a) Principle of the differential sampling technique with a PJVS. A stepwise sinusoidal waveform is compared with an unknown ac signal using a fast analogue to digital converter (ADC) replacing the null detector used in dc voltage comparisons. b) Comparison of two 4-sample waveforms ($[0, +V_{max}, 0, -V_{max}]$) generated by two 10 V PJVS. Adapted from ref.\cite{Behr2012}. }\label{fig:Fig-Jos-3-d}
\end{figure*}

\paragraph*{Measurement set-up and applications}
The accuracy of 10 V PJVS has been demonstrated by comparison to 10 V CJVS; no significant difference between the voltage standards were measured within 1.2 part in $10^{10}$ \cite{Djordjevic2008} and 2.6 parts in $10^{10}$ \cite{Tang2012} ($k$=2). The recent comparison of two cryocooled 10 V PJVS \cite{Rufenacht2018b} illustrates the advantages of the PJVS over the CJVS. The complete automation and synchronisation of both systems allow voltage reversals over very long measurement time (28 h) and enable the use a very sensitive null detector. The authors have measured the voltage difference at $10$ V between the two systems with a relative combined uncertainty of 2.9 parts in $10^{11}$ ($k$ = 2). Today PJVS tend to replace CJVS not only for the calibration of Zener dc references but also for the calibration of the gain and linearity of high precision digital voltmeters through automated measurements\cite{VandenBrom2007,Kaneko2015}.

For low-frequency ac applications ($<$1 kHz), the generation of stepwise approximated waveforms has been used to calibrate ac-dc thermal converters, however the transients, \emph{i.e.} the unquantized parts of the signal between two quantised voltage levels of the waveform, contribute to the uncertainty and are difficult to handle \cite{Burroughs2009}. Despite this, uncertainties of the order of 1 part in $10^{6}$ have been reported \cite{Seron2012,Budovsky2012}.

Today, the PJVS are mainly used with sampling techniques (for review \cite{Behr2012,Rufenacht2018}): the stepwise waveform is compared with an unknown ac signal using a fast analogue to digital converter (ADC) replacing the null detector used in dc voltage comparisons (differential sampling or ac quantum voltmeter \cite{Behr2007,Rufenacht2008,Behr2012,Rufenacht2018}) as depicted in fig.\ref{fig:Fig-Jos-3-d}a or by alternately measuring both signals with the same ADC (\cite{Jeanneret2012,Palafox2009,Rufenacht2018}). This comparison is done only when the voltage of the array is on a quantized plateau of the stepwize approximated waveform. However as the frequency is increased the length of each plateau is reduced (for a given number of samples) and this limits the accuracy above few kHz. Fig.\ref{fig:Fig-Jos-3-d}b shows the comparison of two 4-sample waveforms generated by two 10 V PJVS showing an agreement of the voltage standards, below 400 Hz, within the type-A uncertainty of 3 part in $10^{9}$ \cite{Behr2012}. Comparing two waveforms by differential sampling rather than by sampling them successively by the same sampler reduces the errors due to the gain and the non linearity of the ADC at the expense of the necessity to lock the PJVS, the ac-source and the ADC to a common frequency reference. Indeed, any phase jitter from the ac-source or the ADC is detrimental in terms of uncertainties. The performance of the differential sampling systems (at 7 or 10 VRMS) have been studied in different laboratories \cite{Rufenacht2008,Lee2013,Rufenacht2008,Ihlenfeld2015,Amagai2016,Kaneko2015}. Today, liquid cryogen-free PJVS systems have been demonstrated \cite{Yamada2010,Rufenacht2015,Schubert2016}, and fully automated systems are available from NIST and Supracon.
A very interesting application of PJVS was suggested for impedance ratio measurements based on two PJVS systems generating square-waves: the Josephson two-terminal-pair impedance bridge \cite{Lee2010}. The recent variants of impedance bridges (see section \ref{QIS}) are set-up with pulse driven arrays, which generate pure sinusoidal waveforms (see section \ref{Section_JAWS}).

\subsubsection{JAWS: Pulse driven arrays} \label{Section_JAWS}

\begin{figure*}[!h]
\includegraphics[width=5.2in]{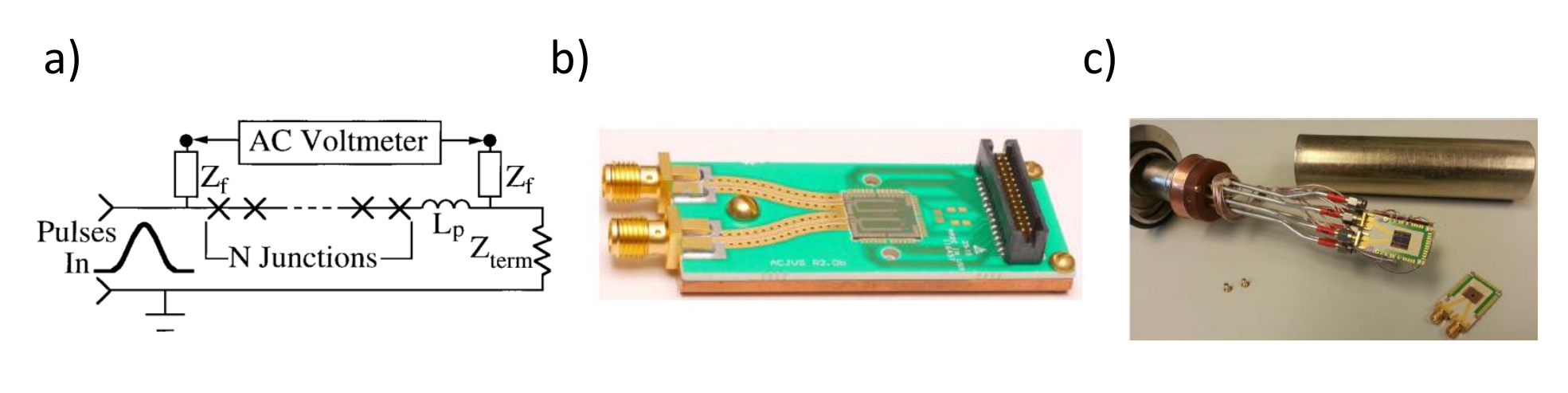}
\caption{a) Principle of pulse-driven arrays: A series array of $N$ Josephson junctions distributed along a wide bandwidth transmission line is biased by current pulses that generate quantized voltage pulses with a time integrated area $\Phi_0$ at each junction. A pulse train of frequency $f$ generates an average voltage $N\Phi_0f$ across the array, which is measured at the low-pass-filtered output of the array. An arbitrary waveform can be generated by gating the input pulse train with a long digital word generator. From ref.\cite{Benz1998}. b) Photograph of the NISTS's new 1 V JAWS package. From ref.\cite{Flowers-Jacobs2016a}.
c) Photograph of the PTB's 8-channel cryoprobe. From ref.\cite{Kieler2015}.}\label{fig:Fig-Jos-5-a}
\end{figure*}

\paragraph*{Principle}
To resolve the problem of transients in the generation of ac signals based on PJVS, Benz \emph{et al.} proposed in 1996 \cite{Benz1996} to bias the array by a train of short current pulses generated by a pulse generator as depicted in fig.\ref{fig:Fig-Jos-5-a}a. 
For a given pulse area, each JJ generates a quantized voltage pulse \ref{ac-Josephson}. The array then acts as a pulse quantizer transferring a single flux quantum $\Phi_0$ for each input pulse (see section \ref{RCSJ}). The voltage across the array is determined by the repetition rate $f$ (fig.\ref{fig:Fig-Jos-5-a}a), which can be modulated to generate arbitrary waveforms. As the pulses can be generated at a very high speed ($\sim$ 15 GHz) compared to the frequency of the desired signal (in the MHz range), generating arbitrary waveforms can be dealt with oversampling techniques. In particular, noise shaping techniques using delta-sigma modulation algorithms \cite{Pavan2017} can be applied to push most of the quantization noise to high frequencies. By this way, pulse sequences corresponding to pure sine-wave can be determined with extremely low distortion \cite{Flowers-Jacobs2016a}.
Most of the difficulties lie in the way to generate the bipolar pulses at a rate of $\sim$ $15\times10^{9}$ pulses per second and to ensure that the pulses propagate with low distortion in the transmission line, such that each bias pulse generates one quantized voltage output pulse for all the junctions. For more details on JAWS, the reader can consult recent reviews \cite{Behr2012,Rufenacht2018}.

The denomination of the pulse-driven technique depends on the domain of application (ACJVS for pure sinewaves generation or QVNS for quantum voltage noise source for a pseudo-random noise source used in electrical based thermometry, see section \ref{QVNS}). Today, due to the high complexity of these systems, only NIST and PTB are developing JAWS systems, however, they have established close collaborations with several groups \cite{Filipski2012,VandenBrom2012,Hagen2012,Sosso2017}.

\begin{figure*}[!h]
\includegraphics[width=5.2in]{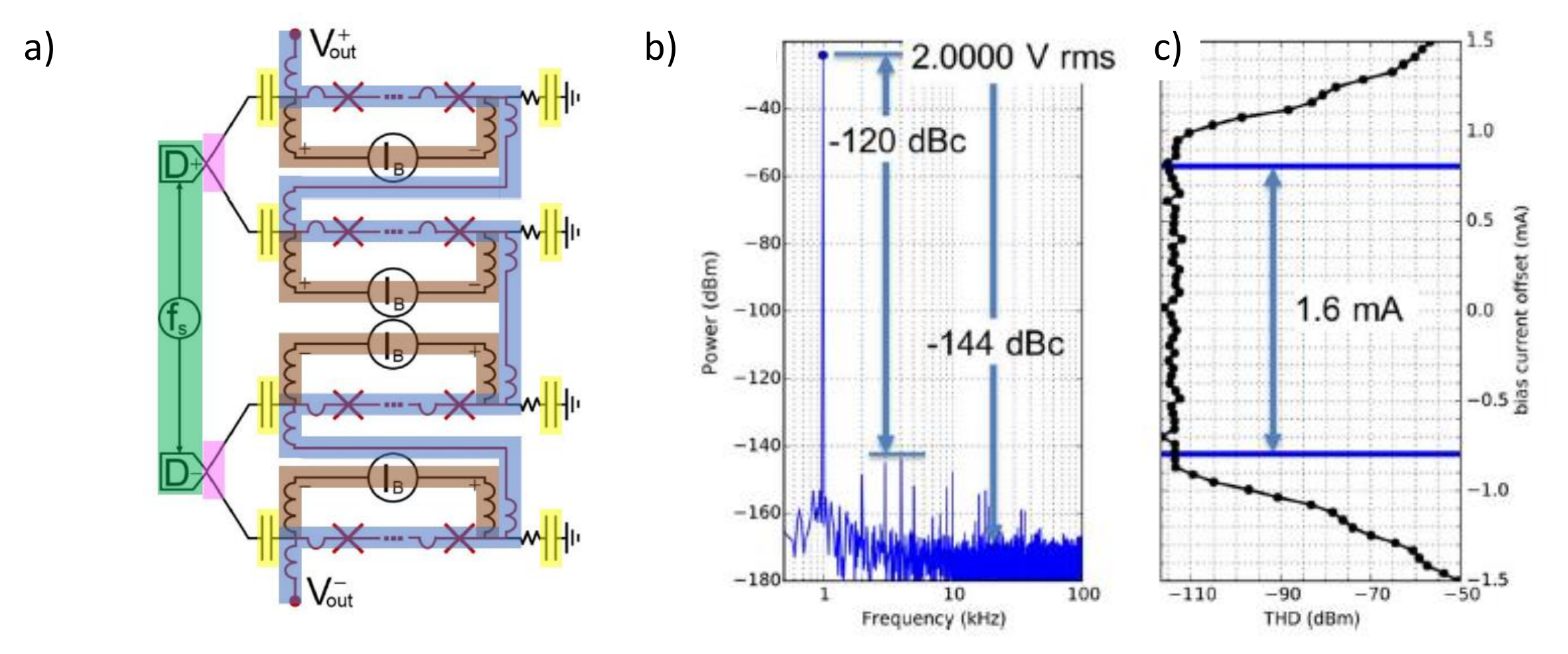}
\caption{a) Schematic of the 1 V JAWS chip from NIST showing the series parallel connection of four JJ arrays fed with only two pulse generator channels (green), thanks to on-chip power splitters (Wilkinson dividers) (pink) \cite{Flowers-Jacobs2016b}. Arrays are capacitively coupled to the pulse generator channels via inside-outside dc blocks (yellow). This filters the low frequency components of the pulse train and avoids common mode voltage on the load. In order to restore the complete pulse spectrum in the array and therefore improve the operating margins, floating low frequency current sources (in brown) bias each array via inductive taps. The total quantized low-frequency voltage is obtained by connecting the arrays in series via inductive taps. From ref.\cite{Rufenacht2018}. b) Digitally sampled spectral measurement showing a low-distortion JAWS output voltage with an rms magnitude of 2 V. Adapted from ref.\cite{Flowers-Jacobs2016a}. c) Total harmonic distortion (THD) versus dither offset current showing the 1.6 mA operating current range. Adapted from ref.\cite{Flowers-Jacobs2016a}.}\label{fig:Fig-Jos-5-b}
\end{figure*}

\paragraph*{Junctions, microwave circuit, bias techniques and applications}
The junctions used for JAWS are SNS junctions based on the same technology used for PJVS (Nb/Nb$_{x}$Si$_{1-x}$/Nb) optimized around 20 GHz \cite{Baek2006,Kohlmann2016}. Double or triple stacked \cite{Flowers-Jacobs2016b,Kieler2015} junctions are used. The maximum output is lower than for PJVS, but recently several breakthroughs have been reported \cite{Kieler2015,Behr2015,Benz2015,Flowers-Jacobs2016a,Flowers-Jacobs2016b}. The main difficulty lies in the broadband nature of the pulses, which have significant power at frequencies up to 30 GHz \cite{Flowers-Jacobs2016b}, and which are very sensitive to non-linearities in the coplanar waveguide. Many techniques have been adapted from the PJVS arrays to improve the propagation of the pulses in the transmission lines \cite{Flowers-Jacobs2016b,Dresselhaus2009}. In addition, pulse generation methods have been optimized over more than 15 years \cite{Benz2014,Zhou2015,Brevik2017}.

Recently, the two groups of NIST and PTB have demonstrated rms amplitude up to 3 V. Kieler \emph{et al.} could reach an rms voltage of 1 V by summing the voltages of 8 arrays (on 4 separate chips) for a total of 63 000 JJs \cite{Kieler2015,Behr2015}. Each array is connected to a separate channel of an 8-channel ternary pulse pattern generator in order to minimize the pulse distortion. Fig.\ref{fig:Fig-Jos-5-a}c shows the cryoprobe with the 4 chips. A direct comparison with a PJVS has demonstrated an agreement better than 1 part in $10^{8}$ ($k = 1$) at 250 Hz \cite{Behr2015}. Flower-Jacobs \emph{et al.} \cite{Flowers-Jacobs2016a} have demonstrated rms amplitudes of 2 V (fig.\ref{fig:Fig-Jos-5-b}a and fig.\ref{fig:Fig-Jos-5-b}b) \cite{Flowers-Jacobs2016a} and recently even 3 V \cite{Flowers-Jacobs2018} in a cryocooler, by connecting 2 chips of 4 arrays (fig.\ref{fig:Fig-Jos-5-b}a) and 2 chips of 8 arrays respectively for a total of 102 480 JJs and 204 960 JJs respectively.

JAWS systems are mainly used for the calibration of thermal transfer standards with high input impedance in order to avoid loading the array output. However, direct calibration of thermal converters with lower impedance might be improved by using buffer \cite{Karlsen2016} or transconductance amplifiers \cite{Budovsky2016,Palafox2016} similar to the ones developed for implementation with a PJVS \cite{Seron2012,Budovsky2012}. Another challenge is to limit the major systematic error due to the voltage leads when measuring at frequencies up to 1 MHz \cite{VandenBrom2012,VandenBrom2016}. JAWS systems allow to test the non-linear behavior of electronic component by generating multi-tone waveforms \cite{Toonen2009,Behr2017}. Other applications concern Johnson noise thermometry (see section \ref{QVNS}) and impedance bridges (see section \ref{QIS}).
\subsection{The quantum Hall resistance standard}
\subsubsection{Usual QHR}
\begin{figure*}[!h]
\includegraphics[width=5.2in]{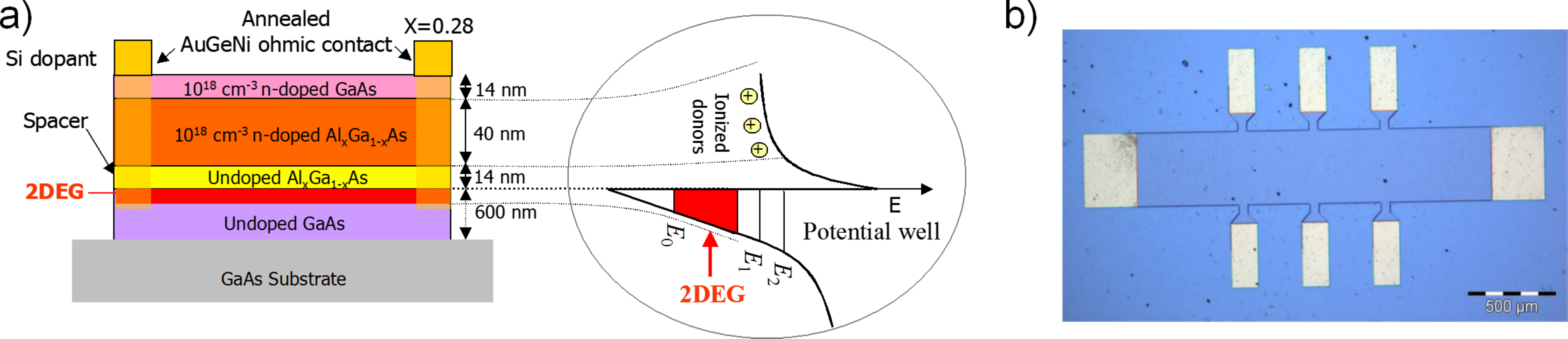}
\caption{a) Typical GaAs/AlGaAs heterostructure used to form a two-dimensional electron gas (2DEG): layer stacking (left), energy bands, 2D subbands and potential well (right). b) Optical picture of a typical device based on a 2DEG having a Hall bar geometry and eight terminals. Electrical contacts are made from annealed AuGeNi deposits (C2N/LNE).}\label{fig:Fig-UsualQHR}
\end{figure*}
Quantum Hall resistance standards\cite{Jeckelmann2001,Poirier2009,Poirier2011} are usually based on Hall bars made of GaAs/AlGaAs heterostructures, the electronic properties of which are well adapted to the metrological application. The two-dimensional electron gas forms at the interface between two semiconductors having different electron doping and energy gap (fig.\ref{fig:Fig-UsualQHR}a). 
These heterostructures can be fabricated by molecular beam epitaxy (MBE) or metal organic chemical vapor deposition (MOCVD).
For resistance metrology application, carrier densities and electron mobilities in the ranges from $3\times10^{15}$ to $5\times10^{15}\mathrm{cm^2}$ and from 10 to 80 $\mathrm{T}^{-1}$ respectively are optimal. Samples are fabricated using usual lithography and etching techniques with a wide Hall bar geometry (typically 400 $\mu$m) characterized by two current contacts and usually three pairs of voltage terminals (fig.\ref{fig:Fig-UsualQHR}b). The Hall bar design aims at optimizing contact surface, breakdown currents and edge state equilibrium. Contacts to the 2DEG are realized by annealing a AuGeNi deposit (diffusion of germanium).

The $\nu=2$ Hall resistance plateau of usual QHRS is quantized to within one part in $10^{9}$ at a magnetic induction of about 10 T, a temperature below 2 K and a measurement current below 100 $\mu$A (the breakdown current reaches a few hundreds of $\mu$A for best devices). As a result of many studies performed by metrologists, technical guidelines\cite{Delahaye2003}, concerning samples properties and characterizations, have been recommended to check the quality of a QHR. The verification of some technical criteria ensures that the Hall resistance is accurately quantized: contact resistances below 10 $\Omega$, longitudinal resistances below 100 $\mu\Omega$, spatial homogeneity, insensitivity to the direction of the magnetic field...
\begin{figure*}[!h]
\includegraphics[width=5.2in]{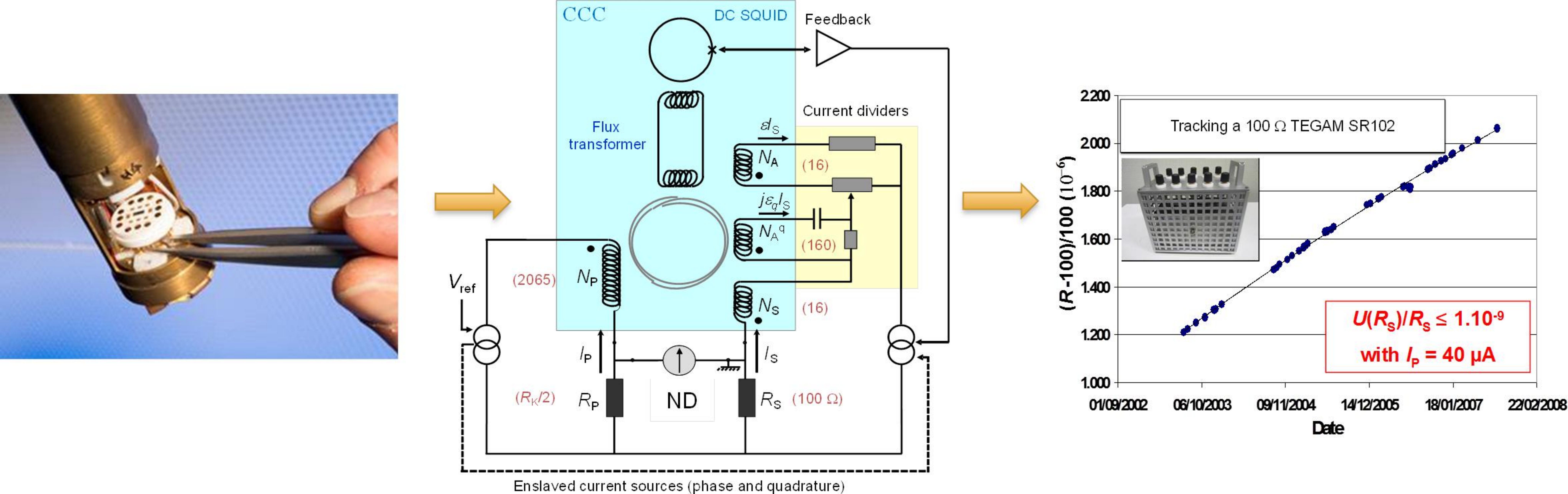}
\caption{Calibration of a resistor from the QHE: the resistance of a resistor (right) is compared to the quantized Hall resistance of a GaAs/AlGas device (left) using a CCC-based resistance bridge (center).}\label{fig:Fig-CalibrationR}
\end{figure*}

The most accurate way to calibrate a resistance from $R_\mathrm{K}/2$, described in fig\ref{fig:Fig-CalibrationR}, relies on a resistance bridge\cite{Jeckelmann2001,Poirier2009} based on a cryogenic current comparator (CCC)\cite{Harvey1972}. Briefly, the method consists in measuring the ratio of the two currents circulating through the resistors and generating at their terminals the same drop voltage. The current ratio is determined using the CCC which is a perfect transformer operating in direct current. More precisely, this device can measure the ratio of two currents in terms of the ratio of the numbers of turns of the two windings through which circulate the two currents with a relative uncertainty below $10^{-10}$. Its accuracy relies on a flux density conservation property of the superconductive toroidal shield (Meissner effect), in which superconducting windings are embedded. Owing to a flux detector based on a dc superconducting quantum interference device (SQUID), the current noise resolution of the CCC can be as low as 80 pA.turn/Hz$^{1/2}$\cite{Lafont2015}. Recent resistance bridges\cite{Drung2009,Sanchez2009,Williams2010,PoirierCPEM2018,Ribeiro2015} can calibrate a 100 $\Omega$ resistor from $R_\mathrm{K}/2$ with a relative uncertainty of a few $10^{-10}$. They are more sensitive, accurate, versatile and automated than the first generation developed in the eighties. In the revised SI, the reference value that must be used in resistance calibration certificate is $R_\mathrm{K}=h/e^2$. It differs from the value $R_\mathrm{K-90}$ so the numerical value of a resistance measured in terms of the new SI ohm is larger than the value measured in terms of $R_\mathrm{K-90}$ by a relative amount of $1.7793\times10^{-8}$.
\subsubsection{Arrays}
\begin{figure*}[!h]
\includegraphics[width=5.2in]{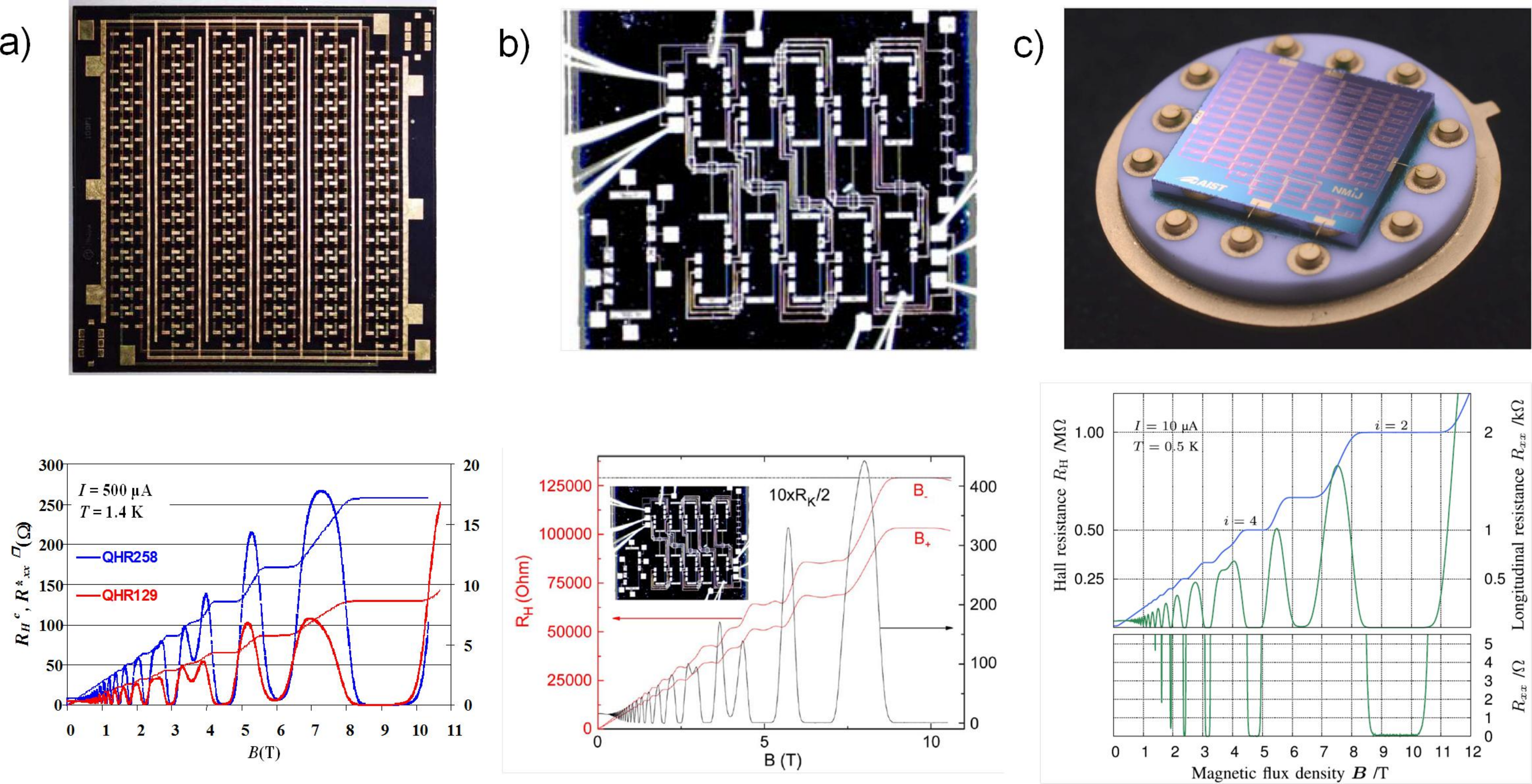}
\caption{a) Top: picture of an array developed by LNE in collaboration with the Laboratoire d'Electronique de Philips: it is based on 100 Hall bars connected in parallel. Bottom: magneto-resistance measurements carried out for arrays of $R_\mathrm{K}/200$ (QHR129) and $R_\mathrm{K}/100$ (QHR258) nominal values on the $\nu=2$ plateau. From ref.\cite{Poirier2002}. b) Top: picture of an array developed by PTB made of 10 Hall bars connected in series. Bottom:  magneto-resistance measurements carried out for two magnetic field directions. From ref.\cite{KonemannIEEE2011}. c) Top: picture of an array developed by NMIJ of nominal value close to 1 M$\Omega$ made of 88 Hall bars. Bottom: magneto-resistance measurements. From ref.\cite{Oe2016}.}\label{fig:Fig-Arrays}
\end{figure*}

The perfect equipotentiality along edges and the quantization of any two-terminal resistance at $R_\mathrm{H}$ value are two fundamental properties of the QHE that can be exploited together to get rid of the resistance of the connections between multiply-connected Hall bars. More precisely, let us consider a Hall bar with a resistance $r_1$ connected in series with a current terminal. The two-terminal resistance, $R_\mathrm{2T}$, equal to $R_\mathrm{H}+r_1$, becomes $R_\mathrm{H}+r_1r_2/R_\mathrm{H}$ if adding a connection to a second terminal at same potential as that of the first terminal. The relative effect of series resistances, $r$, is reduced according to $(r/R_\mathrm{H})^n$, where $n$ is the number of connections. The so-called multiple connection technique\cite{Delahaye1993} can be used to realize arrays of QHRs extending the range of quantized resistance values. It was successfully used to realize array resistance standards having values\cite{Piquemal1999,Poirier2002,Bounouh2003,Poirier2004}, in the range from 100 $\Omega$ to 1.29 M$\Omega$, quantized in terms of $R_\mathrm{K}$ to within a few parts in $10^9$ (\ref{fig:Fig-Arrays}a).
Devices were based on several tenths of GaAs/AlGaAs Hall bars combined in series and/or in parallel using a triple or quadruple connection technique. The achievement of quantized arrays relies on GaAs/AlGaAs heterostructures having very homogeneous electronic density (to within a few percents) so that all Hall bars are quantized at same magnetic induction. Multiple interconnections require good ohmic contacts and perfect insulating layers (\emph{i.e.} without pinholes) to electrically isolate different levels of connections. After these founding results by LNE, others NMIs undertook research to develop Hall bar arrays. PTB realized standards (fig.\ref{fig:Fig-Arrays}b) made of ten Hall bars connected in series or in parallel\cite{KonemannIEEE2011,Konemann2011}. NMIJ pursues the development of arrays to achieve not only resistance standards of 10 k$\Omega$\cite{Oe2013} and 1 M$\Omega$\cite{Oe2016} (fig.\ref{fig:Fig-Arrays}c) resistance values but also voltage dividers\cite{Domae2017}.
\begin{figure*}[!h]
\includegraphics[width=5.2in]{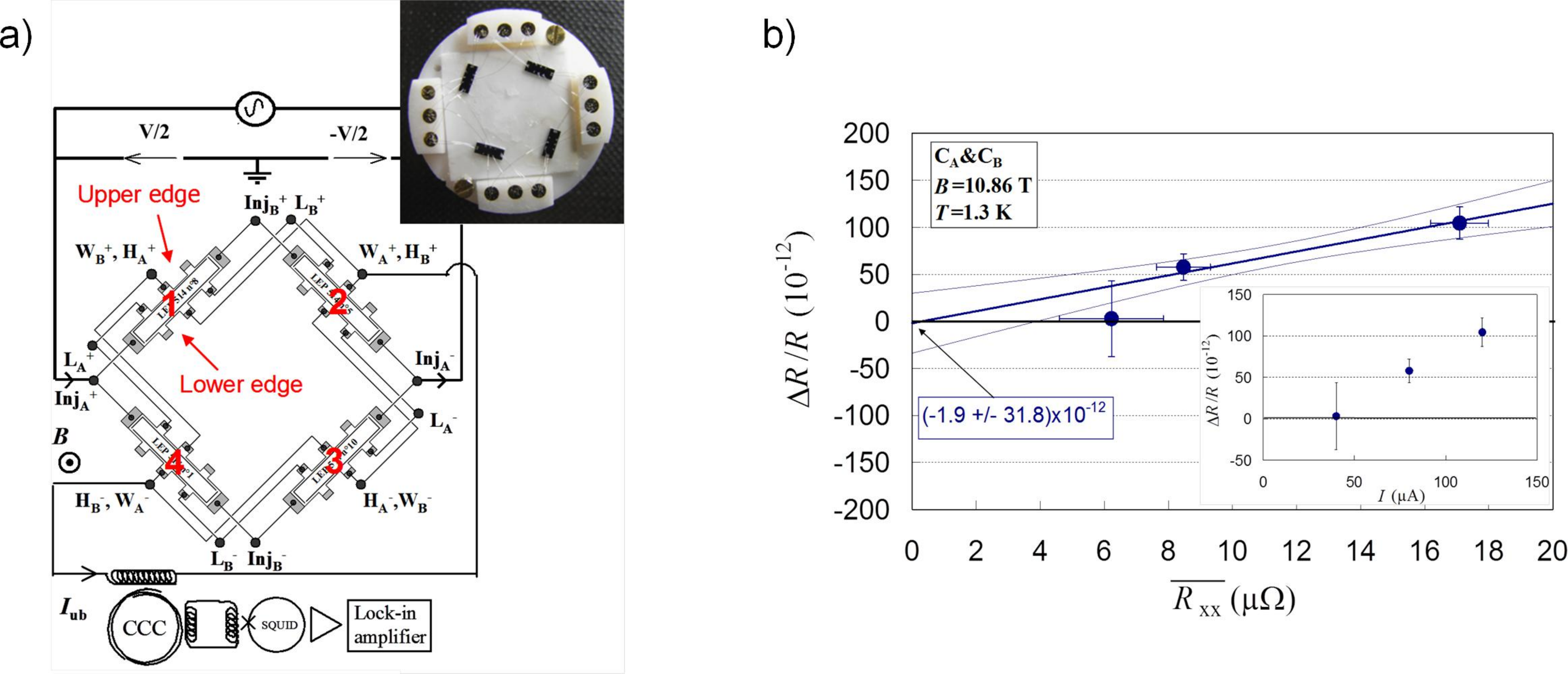}
\caption{a) Scheme of the Wheatstone bridge based on the triple connection of four GaAs/AlGaAs (LEP514) Hall bars mounted on a single sample holder (picture). The unbalance current $I_\mathrm{ub}$ is measured using a CCC winding. b) Extrapolation at zero dissipation ($\bar{R}_\mathrm{xx}=0$) of the relative deviation, $\Delta R/R$, of the quantized resistance of one standard from the others. One finds $\Delta R/R(\bar{R}_\mathrm{xx}=0)=(-1.9\pm31.8)\times10^{-12}$. From ref.\cite{Schopfer2013}.}\label{fig:Fig-Wheatstone}
\end{figure*}

One particular array that can be implemented is the Wheatstone bridge which is made of four Hall bars. Such a bridge was used to perform reproducibility tests of the QHE\cite{Schopfer2007}. Fig.\ref{fig:Fig-Wheatstone}a shows a Wheatstone bridge mounted from four GaAs/AlGaAs Hall bars using a triple connection. The unbalance current of the bridge $I_\mathrm{ub}$ is related to the relative deviation of the quantized resistance $\Delta R/R$ of one standard from the others according to $\Delta R/R=4I_\mathrm{ub}/I$, where $I$ is the biasing current of the bridge. Measuring $I_\mathrm{ub}$ using a sensitive CCC winding allowed the demonstration of the reproducibility of the quantized Hall resistance\cite{Schopfer2013} with a record relative uncertainty of $3\times10^{-11}$, as shown on fig.\ref{fig:Fig-Wheatstone}b.
\subsubsection{Graphene: towards a user-friendly standard}
\paragraph{Dirac Physics}
\begin{figure*}[!h]
\includegraphics[width=5.2in]{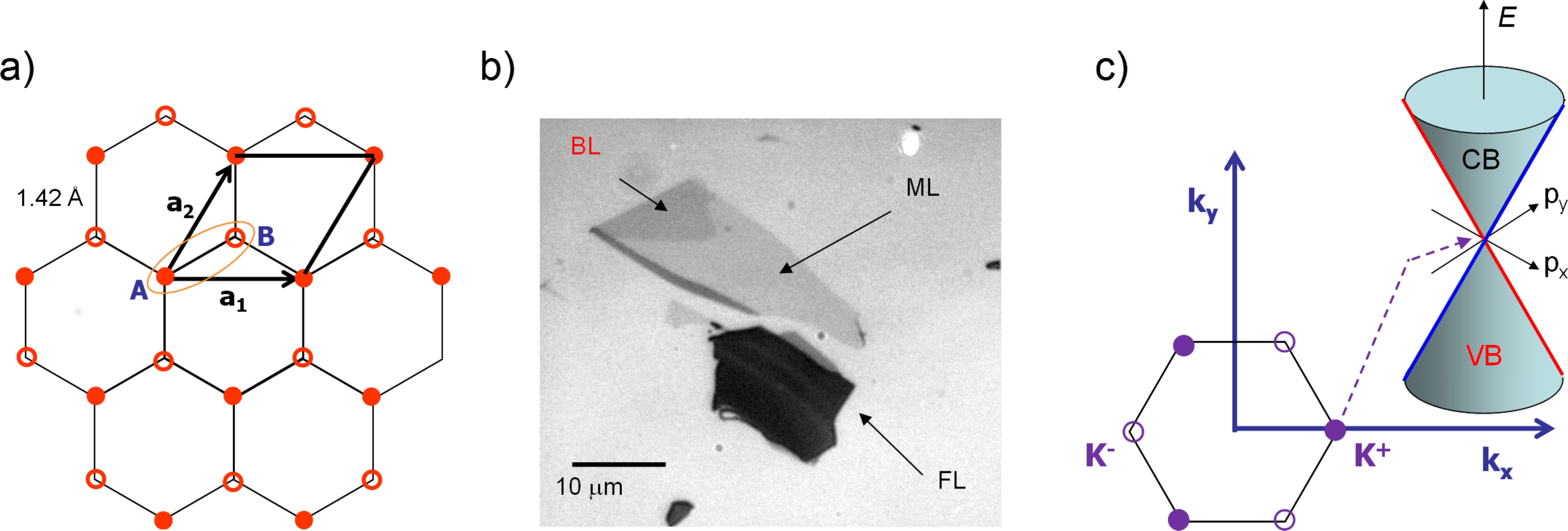}
\caption{a) Honeycomb lattice of graphene with two atoms A and B per cell ($a_1$ and $a_2$ are base vectors). b) Optical pictures of graphene flakes on top of a $\mathrm{SiO_2/Si}$ substrate: ML (monolayer), BL (bilayer), (FL) a few layers. c) First Brillouin zone with two independent vertices: $K^+$ and $K^-$ Dirac points. Conical energy spectrum around Dirac points.}\label{fig:Fig-Graphene}
\end{figure*}
Graphene is a monolayer of carbon atoms crystallized (fig.\ref{fig:Fig-Graphene}a) in a 2D honeycomb lattice. Fig.\ref{fig:Fig-Graphene}b shows optical pictures of graphene flakes of different numbers of layers. Its quantum electronic transport properties have been discovered\cite{Novoselov2004} by Geim and Novoselov in 2004. Since then, based on the exceptional properties of graphene, not only electrical, but also mechanical, optical, thermal and chemical, many works have been carried out for fundamental research\cite{Geim2007} and for industry applications\cite{Ferrari2015} as well. Graphene is a gapless semiconductor with two valleys corresponding to the two independent vertices, called Dirac points, of the hexagonal Brillouin zone (fig.\ref{fig:Fig-Graphene}c). At low energy around Dirac points, the energy spectrum\cite{Wallace1947} is conical and charge carriers behave as relativistic particles moving at Fermi velocity (fig.\ref{fig:Fig-Graphene}c). Dirac physics\cite{Castro2009,Sarma2011} manifests itself and determines many properties including electronic transport: Berry's phase $\pi$, chirality (helicity is a good quantum number and is preserved in elastic scattering process), cancellation of backscattering at normal incidence, anti-localization. In addition, the absence of gap between the conduction and valence bands makes this material ambipolar: charge carriers can be either electrons or holes.
\begin{figure*}[!h]
\includegraphics[width=5.2in]{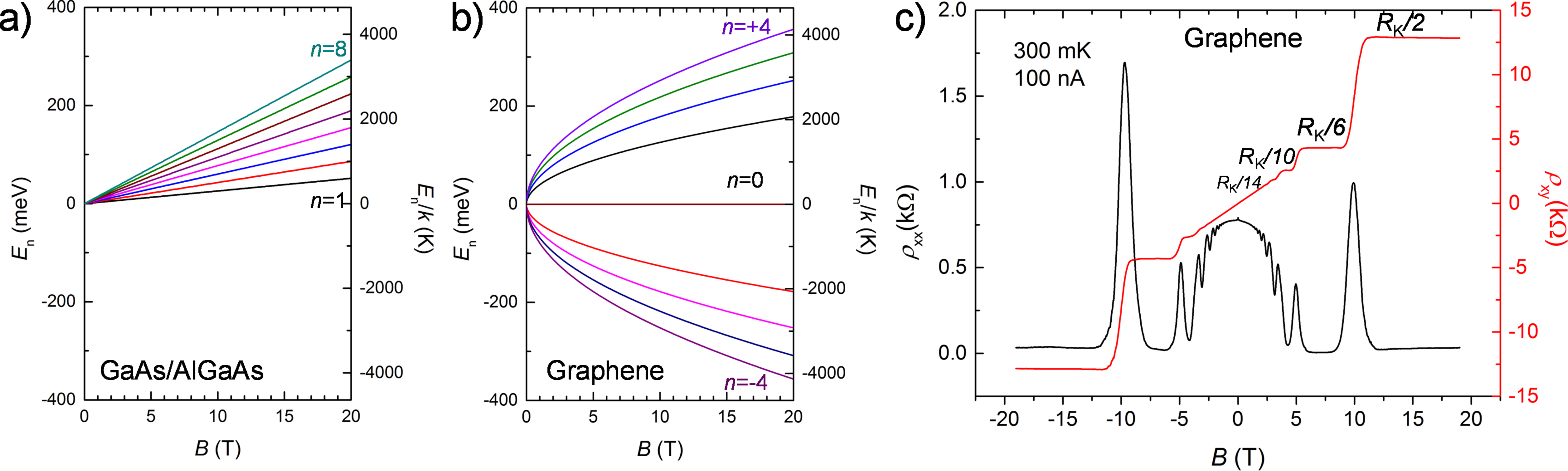}
\caption{Energy as a function of the magnetic induction $B$ of Landau levels with index $n$ in a) GaAs/AlGaAs ($n=1:8$) and in b) graphene ($n=0, \pm1, \pm2, \pm3,\pm 4$). c) Hall resistivity $\rho_{xy}$ and longitudinal resistivity $\rho_{xx}$ as a function of $B$, measured in graphene grown on SiC by thermal decomposition with an electronic density of $1\times 10^{12}\mathrm{cm^{-2}}$ and a mobility of $9 000~\mathrm{cm^{2}V^{-1}s^{-1}}$ \cite{Pallecchi2012}.}\label{fig:Fig-QHEGraphene}
\end{figure*}
One other emblematic property is a specific half-integer quantum Hall effect which was highlighted\cite{Novoselov2005,Zhang2005} right after the graphene discovery. The energy spectrum\cite{Castro2009,Goerbig2011} is quantized in Landau levels at energies given by:
\begin{equation}
\epsilon_n=\pm\sqrt{2e\hbar v_\mathrm{F}^2Bn},
\end{equation}
where $n$ is an integer value. The QHE in graphene differs from that in usual semiconductors by several peculiarities (fig.\ref{fig:Fig-QHEGraphene}a and fig.\ref{fig:Fig-QHEGraphene}b). The degeneracy of each Landau level is $4eB/h$ (spin and valley).The Landau level energy scales with $\sqrt{B}$ and energy gaps depend on $n$. Moreover, it exists a Landau level at zero energy and Hall plateaus occurs at unusual filling factors $\nu=\pm(2n+1)$, \emph{\emph{i.e.}} at resistance values $R_\mathrm{H}=\pm\frac{h}{e^2}\frac{1}{2(2n+1)}$, as can be observed in fig.\ref{fig:Fig-QHEGraphene}c. Lift of spin and valley degeneracy leads to the observation\cite{Zhang2006} of plateaus at others filling factors values such as $\nu=0,\pm1,\pm 4$. Coulomb interaction is responsible for the manifestation of the Fractional QHE\cite{Bolotin2009,Du2009} and also of ferromagnetic states\cite{Nomura2006,Goerbig2011,Kharitonov2012} that lift the n=0 Landau level degeneracy\cite{Zhang2006,Abanin2007,Young2012}.

\paragraph{Advantage for metrology}
\begin{figure*}[!h]
\includegraphics[width=5.2in]{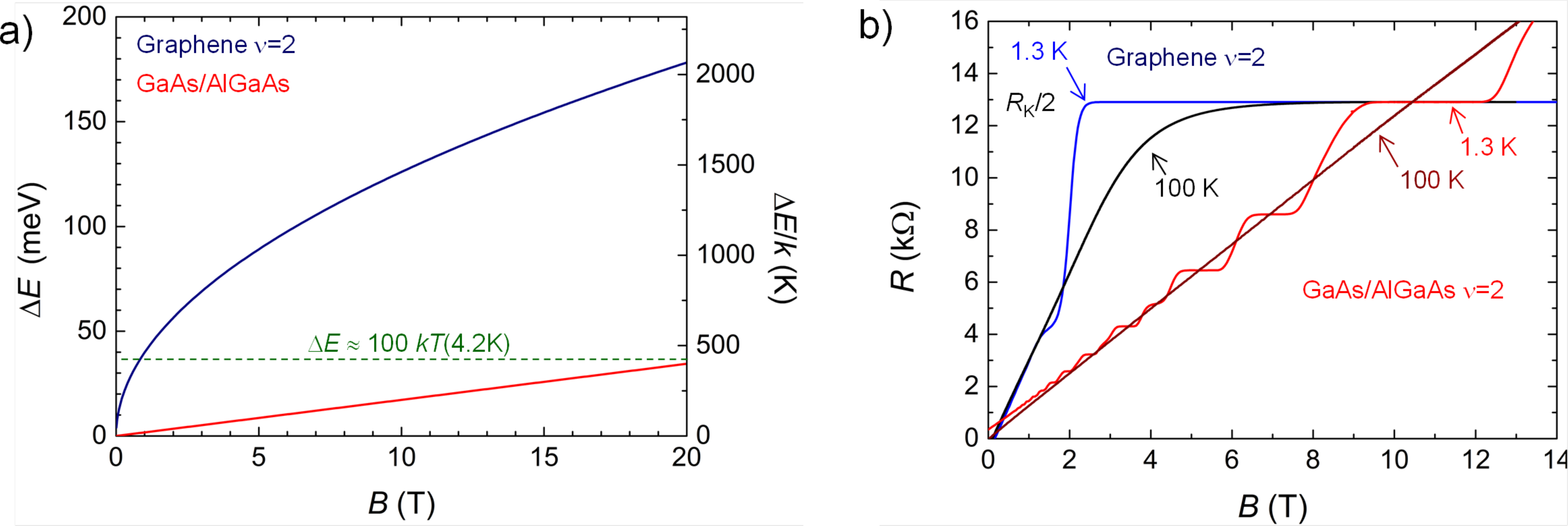}
\caption{a) Energy gap between the two first Landau levels protecting the Hall plateau at $\nu=2$ in graphene and between two nearest Landau levels in GaAs/AlGaAs (whatever their indexes), as a function of $B$. The energy level corresponding to 420 K is represented: it fixes the empirical minimum gap ensuring accurate Hall resistance quantization to within $10^{-9}$ at 4.2 K.  b) Hall resistance as a function of $B$, at two temperatures $1.3$ K and $100$ K, in a graphene device obtained by hydrogen/propane CVD on SiC \cite{Ribeiro2015} and in a typical GaAs/AlGaAs device used in metrology.}\label{fig:Fig-QHEMetroGraphene}
\end{figure*}
One specific property of the integer QHE, of strong interest for metrology, is that the energy gap $\Delta E$ in between the two first Landau levels is much larger in graphene than in GaAs/AlGaAs heterostructures for accessible magnetic fields, as shown in fig.\ref{fig:Fig-QHEGraphene}a. This explains the QHE robustness in graphene that has allowed the observation of the Hall resistance quantization at $\nu=2$ even at room temperature\cite{Novoselov2007}. Fig.\ref{fig:Fig-QHEMetroGraphene}b, which shows a $\nu=2$ Hall resistance plateau measured in a graphene-based Hall bar that remains much wider at $T=100$ K than the one usually measured in a GaAs/AlGaAs device at $T=1.3$ K, also highlights this robustness. More precisely, the empirical criterion, $\Delta E>100kT$, which is roughly valid for GaAs/AlGaAs material in conditioning the Hall resistance quantization, would indicate that $10^{-9}$-accuracy could be achieved in graphene at 4.2 K from only 0.8 T. These energy considerations have motivated\cite{Poirier2009,Poirier2010} the development of a graphene-based quantum resistance standard able to operate in more easy and accessible experimental conditions than its GaAs/AlGaAs counterpart. 

\begin{figure*}[!h]
\includegraphics[width=5.2in]{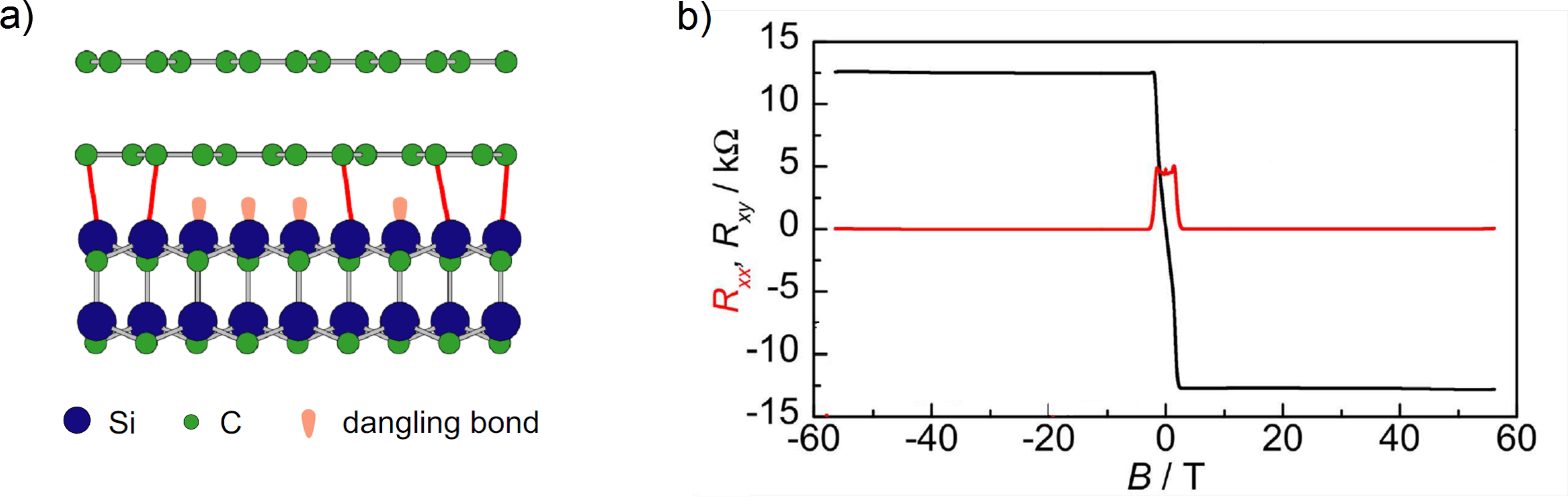}
\caption{a) Structural model of a monolayer graphene on SiC after growth on the Si-terminated face. The graphene is growing on top of the $(6\sqrt{3}\times 6\sqrt{3})R30^{\circ}$ reconstructed interface layer, also called buffer layer. Only one atom over three of this layer is bonded to the substrate. From ref.\cite{Riedl2010}. b) Measurements of the longitudinal resistance ($R_{\mathrm{xx}}$, red) and of the Hall resistance ($R_{\mathrm{xy}}$, black) in the in epitaxial graphene on SiC performed in pulsed magnetic field at $T=2$ K. An exceptionally wide Hall plateau quantized at $R_{\mathrm{K}}/2$ value is observed. From ref.\cite{Alexander-Webber2016}.}\label{fig:Fig-GraphenePlateau}
\end{figure*}

\paragraph{Development of the graphene-based quantum Hall resistance standard: state-of-the-art} Considering the great promise of graphene for developing a user-friendly quantum Hall resistance standard, research works started shortly after the first observations\cite{Novoselov2005,Zhang2005} of the QHE in graphene. The first precise measurements\cite{Schopfer2012,Janssen2013} of the quantized Hall resistance were carried out in 2008 by VSL, the Dutch metrology institute with samples from Geim and Novoselov's group, made of graphene obtained by exfoliation of graphite using the original "scotch tape" technique and then deposited on SiO$_{2}$/Si substrates\cite{Giesberg2009}. The accuracy was limited to $1.5\times 10^{-5}$, mainly because of the high resistance of the metallic contacts. Further works\cite{Guignard2012,Wosczczyna2012} in exfoliated graphene have shown that the small typical size of these devices and the extreme sensitivity of graphene electronic properties to the close environment may impede the measurement of the Hall resistance quantization with accuracy in graphene. Using graphene on SiC produced in Link$\mathrm{\ddot{o}}$ping university, in Sweden, by thermal decomposition of the substrate, NPL performed, in 2010, the first demonstration that the Hall resistance can be quantized in graphene with the same accuracy as in GaAs/AlGaAs\cite{Tzalenchuk2010}. The agreement between Hall resistance measurement performed in both materials to within $8.7\times 10^{-11}$ is worth one of the most precise QHE universality test\cite{Janssen2011,Janssen2012}. Nevertheless, the experimental conditions of magnetic induction and temperature required for the graphene device were not competitive with those of typical GaAs/AlGaAs devices. The carrier density was too high to get the quantization of the $\nu = 2$ Hall resistance plateau, which is expected to be the most robust, at lower magnetic induction. This high doping results from a charge transfer caused by the coupling of the graphene layer to the SiC substrate $via$ an interface layer, so-called buffer layer (fig.\ref{fig:Fig-GraphenePlateau}a).

This interface layer, which only exists in case of graphene grown on the Si-terminated face of the SiC substrate, is electrically inactive but can host a large density of localized donors. It acts as a charge carrier reservoir located very close to the graphene at a distance of about 0.3-0.4 nm. The charge transfer depends on the magnetic field. It exits magnetic field intervals where the carrier density in graphene increases linearly with the magnetic field which results in  the  pinning of the Landau level filling factor\cite{Tzalenchuk2011}, particularly at $\nu=2$. This pinning explains the broad magnetic field extension of the Hall resistance plateau observed in graphene on SiC (fig.\ref{fig:Fig-GraphenePlateau}b). Besides, with the objective to increase the size of QHE graphene devices, LNE developed collaborations with CNRS-Institut N\'eel to exploit graphene grown by chemical vapor deposition on copper, which has also the advantage to be a scalable production technique that can be transferred to industry. In this work, it was demonstrated that, in case of polycrystalline samples, the Hall resistance quantization is not accurate in accessible conditions, because of grain boundaries short-circuiting the quantum Hall edge states\cite{Lafont2014}. Generally speaking, the different attempts show how much the material quality is crucial to achieve the goal of a graphene-based quantum Hall resistance standard operating in more accessible conditions.
  \begin{figure*}[!h]
\includegraphics[width=5.2in]{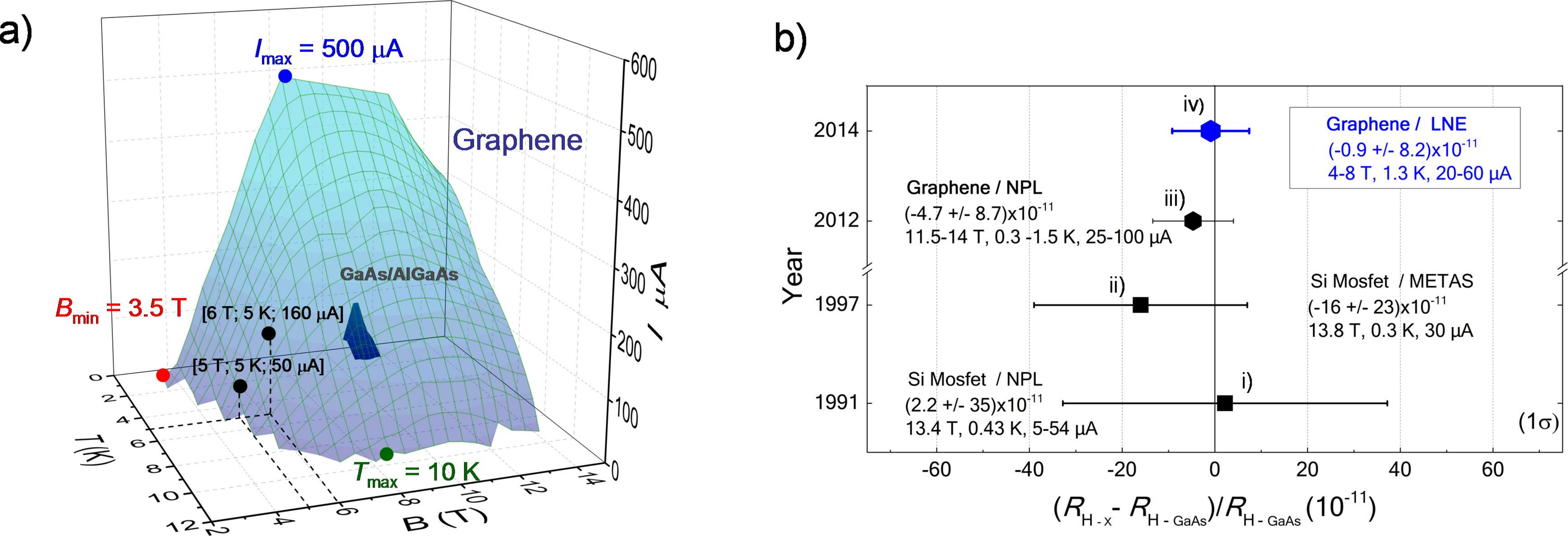}
\caption{a) Experimental conditions of magnetic induction $B$, temperature $T$ and current $I$, under which a graphene device obtained by hydrogen/propane CVD on SiC exhibits quantization of the Hall resistance with accuracy to within $1 \times 10^{-9}$ and below\cite{Ribeiro2015}. Two relaxed working points are pointed out. These conditions are compared with those typical (inner volume) that rules for a GaAs/AlGaAs device used as quantum Hall resistance standard. b) Results of the most precise QHE universality tests, based on comparisons of the quantized Hall resistance in different materials with that realized in GaAs/AlGaAs: i) A. Hartland \emph{et al.}\cite{Hartland1991}, ii) B. Jeckelmann \emph{et al.}\cite{Jeckelmann1997}, iii) T. J. B. M. Janssen \emph{et al.}\cite{Janssen2012}, and including the most recent and precise one realized with graphene iv) Ribeiro-Palau \emph{et al.}\cite{Ribeiro2015}. Experimental conditions of the Si-MOSFET and graphene devices compared to GaAs/AlGaAs devices are also indicated.}\label{fig:Fig-NatNanoLNEGraphene}
\end{figure*}

  One breakthrough\cite{Jabakhanji2014,Lafont2015} came from the use of samples made of graphene produced at CNRS-CRHEA by an hybrid technique\cite{Michon2010} of hydrogen/propane CVD on SiC and processed at CNRS-C2N. In one (fig.\ref{fig:Fig-QHEMetroGraphene}b) of these samples of moderate carrier density, $1.8\times 10^{11}\mathrm{cm^{-2}}$, and a relatively high mobility, $9 400~\mathrm{cm^{2}V^{-1}s^{-1}}$, the quantization of the Hall resistance at $\nu = 2$ was demonstrated by LNE with state-of-the-art accuracy below $1\times 10^{-9}$, at magnetic inductiosn from 14 T down to 3.5 T, temperatures up to 10 K, or currents up to 0.5 mA \cite{Ribeiro2015}. This extended and relaxed range of experimental conditions, enabled by graphene, largely surpasses the conditions required by GaAs/AlGaAs devices, as shown in fig.\ref{fig:Fig-NatNanoLNEGraphene}a. In addition, the studied graphene device demonstrates all the properties of a reliable primary quantum Hall resistance standard. Finally, the accuracy of the graphene device has been tested by comparison with a GaAs/AlGaAs device down to the record relative uncertainty of $8.2\times 10^{-11}$. This led to the most precise QHE universality test\cite{Ribeiro2015} as highlighted in fig.\ref{fig:Fig-NatNanoLNEGraphene}. After this demonstration, the efforts are now focused on improving the technology reliability: reproducibility, stability, control. One of the main issues is the control of the carrier density down to a low value, \emph{i.e.} $5\times 10^{10}\mathrm{cm^{-2}}$, that is required to get operation of the graphene-based QHRS at very low magnetic induction, \emph{i.e.} around 1 T, while keeping an excellent spatial homogeneity. This is a very big challenge, taking account the gapless character of graphene and its sensitivity to the environment. Another issue concerns the identification of the key control parameter for robust and accurate Hall resistance quantization. In graphene grown on SiC, the buffer layer certainly plays an important role for these two issues\cite{Brun-PicardCPEM2018}. Several current works address these points. On one hand, PTB and NIST, work to optimise the growth process to get high quality reconstruction of the SiC surface\cite{Kruskopf2016,Yanfei2017}. On the other hand RISE is developing a chemical gating technique to achieve low, homogeneous and stable charge carrier density\cite{He2018}. Considering the results demonstrated so far in graphene grown on SiC, research are presently mainly focused on this material, but other routes still deserve to be explored. For example, graphene embedded in hexagonal boron nitride (h-BN) offer a better control of the environment, higher mobility, easier control of doping by electrostatic gating. Nevertheless, samples have smaller sizes and remain difficult to fabricate up to now.

To end, several extensions of  graphene use in QHE metrology are also considered, where the material can provide the advantages of the low magnetic induction and high temperature operation and even further ones. The first one is the development of a quantum Hall resistance standard operating in ac for impedance measurement traceability\cite{Kalmbach2014}. The second one is the realization of series arrays with more compact and less risky design, exploiting PN junctions, $\emph{i.e.}$ the ambipolarity of the material\cite{Woszczyna2011}.

\section{The ampere realization from the elementary charge}
\label{SectionAmpere}
\subsection{Using new monoelectronic devices}
\label{Newmonoelectronicdevices}
\begin{figure*}[!h]
\includegraphics[width=5.2in]{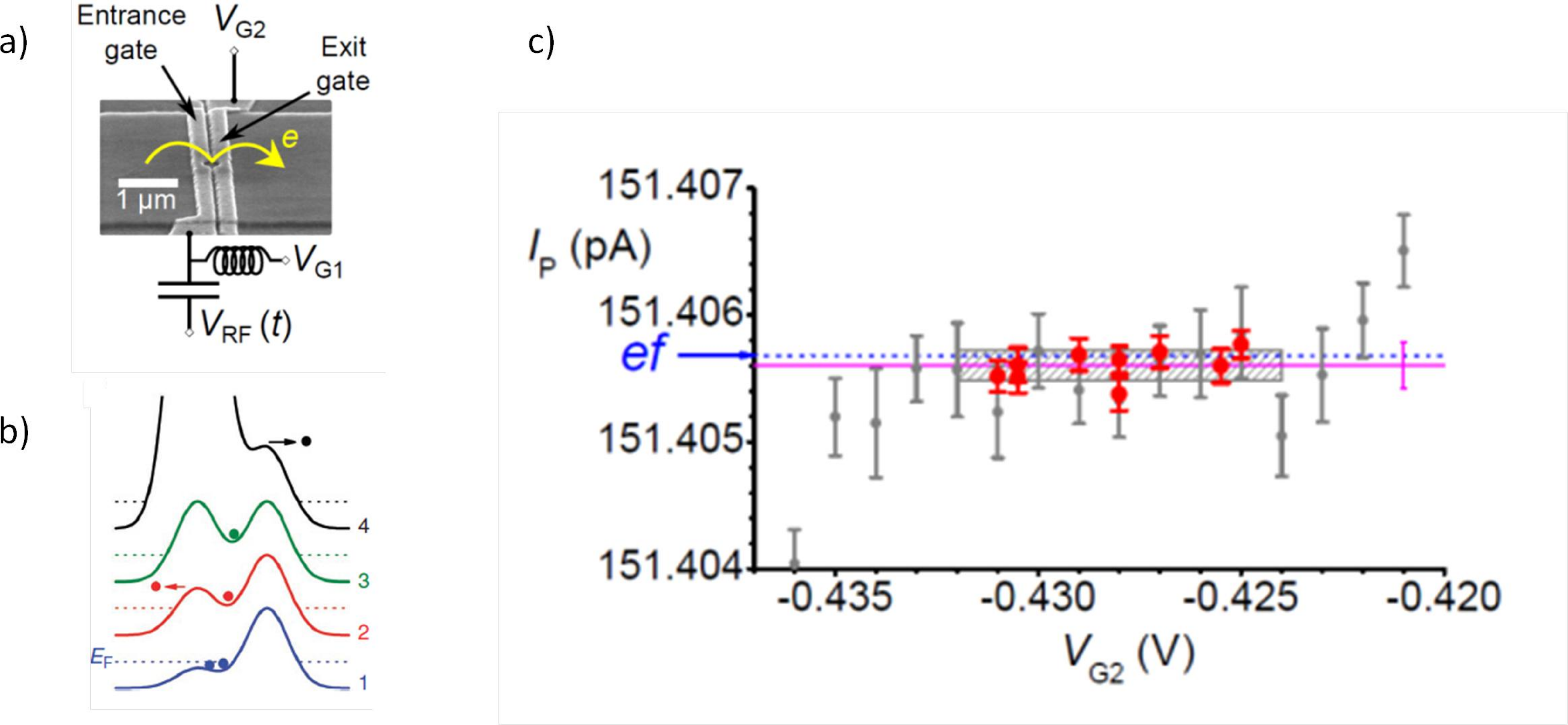}
\caption{a) SEM picture of a current pump based on a quasi-1D GaAs wire with two gates at potential $V_\mathrm{G1}$ and $V_\mathrm{G2}$. b) Electron transfer: four steps of modulation of $V_\mathrm{G1}$ potential at fixed $V_\mathrm{G2}$. c) Quantized current step obtained varying $V_\mathrm{G2}$. Adapted from ref.\cite{Giblin2012}}\label{fig:Fig-NewPump}
\end{figure*}
Metallic electron pumps with fixed insulating barrier described previously are in the strong Coulomb blockade regime, where the tunnel barriers are highly resistive to ensure localized states, and where the tunneling is treated as a perturbation\cite{Pekola2013,Grabert1991}. This regime is favorable to the precise transport of individual electrons. However, high tunnel barriers prevent the rapid loading of the electrons onto the metallic island, and thus, limit the operating frequency to preserve a low error rate due to missing electrons during the pumping cycle. As the corresponding output currents were too low (pA range) to realize a practical quantum current standard, several different systems have been investigated.

Among them is the hybrid superconducting-normal metal turnstile \cite{Pekola2008}, which shares the geometry of a single-electron transistor (SET), i.e. a mesoscopic conducting island connected through tunnel junctions to two bulk electrodes, but for which the source and drain electrodes are superconducting (S). A gate voltage source is coupled capacitively to the central island of this SINIS structure and a small bias voltage is applied over the SET in order to define a preferred direction for single-electron tunneling. Under this conditions, for a normal SET, a gate span between different charge states always crosses regions where the current freely flows without control. However, in the hybrid SET, thanks to the presence of the superconducting gap $\triangle$ in the leads, and for bias voltages below $2\triangle/e$, these regions are suppressed. Then, an accurate quantized current can be generated by driving the SET between two adjacent charge states using a single ac gate voltage. An electron is transferred at each cycle of the driving frequency. This technique allowed the parallelization of ten turnstiles with an increased current up to 100 pA \cite{Maisi2009}. However, the limitation of the current level to 10 pA for a single aluminium-based-SINIS turnstile \cite{Pekola2013}, in order to reduce the errors due to high-order processes, limits drastically the metrological applications.

Most of the recent research has been concentrated on single-electron sources based on semiconductor quantum dots \cite{Fujiwara2001,Fujiwara2004,Fujiwara2008,Ono2003,Blumenthal2007,Kaestner2008,Kaestner2009,Giblin2010,Giblin2012,Jehl2012,Jehl2013} (for review \cite{Pekola2013,Kaestner2015}). These devices allow similar manipulation of individual electrons, but provide also the possibility to tune the barriers defining the dot by applying gate voltages. The pioneering work of Kouwenhoven \emph{et al.} in 1991 demonstrated the transport of electrons through a quantum dot in GaAs/AlGaAs heterostructures by varying alternatively the two barriers height at a frequency $f$ \cite{Kouwenhoven1991}.
Very recent works\cite{Stein2015,Stein2017,Yamahata2016,Zhao2017} have improved the uncertainty in highly-controlled GaAs/AlGaAs and silicon quantum dots.
Fig.\ref{fig:Fig-NewPump}a shows a SEM picture of a state-of-the-art single-electron device based on a quasi-1D GaAs wire which was studied by Giblin \emph{et al.}\cite{Giblin2012}. Fig.\ref{fig:Fig-NewPump}b illustrates the pumping scheme: it shows the evolution of the electrostatic potential and the electron transfer during the pumping cycle. The left barrier alone is modulated, the right barrier is set well above the Fermi energy to prevent electrons from escaping. To increase the working frequency, during the loading phase the left barrier is completely opened such that few electrons can be loaded on the dot. Then the left barrier is raised to isolate the dot, while some electrons tunnel back to the reservoir, leaving a unique electron in the dot. The barrier is raised until the potential is much higher than the right barrier, so that the trapped electron is ejected to the reservoir on the right side. The decay-cascade model \cite{Kashcheyevs2010,Kaestner2015} describes the process of back tunneling, and gives the framework for the understanding of the pumping cycle in a tunable barrier quantum dot. The single electron transfer can be experimentally optimized by applying a non-sinusoidal signal to the gate $V_\mathrm{G1}$ \cite{Giblin2012} and a magnetic field $B$ to obtain better electron confinement \cite{Fletcher2012}. This non-adiabatic pumping cycle allows the control of the number of electrons pumped per cycle by varying the exit gate voltage $V_\mathrm{G2}$. A quantized current step has been obtained by varying this parameter as illustrated in fig.\ref{fig:Fig-NewPump}c. At $T\leq 0.3~\mathrm{K}$ and $B\geq 14~\mathrm{T}$, the quantization of the current was demonstrated with a relative measurement uncertainty of $1.2\times10^{-6}$ at 150 pA ($f_\mathrm{P}=945~\mathrm{MHz}$).

More recently, Stein and co-authors demonstrated the accuracy of a 96 pA current measured with an ultra-stable current amplifier. They reached a $1.6\times10^{-7}$ relative combined uncertainty ($k=1$) in a GaAs/AlGaAs device at $T=0.1~\mathrm{K}$ and $B\geq 9.2~\mathrm{T}$, ($f_\mathrm{P}=600~\mathrm{MHz}$)\cite{Stein2017}. The accuracy of single-electron pumping has also been improved recently in a metal-oxide-semiconductor silicon quantum dot driven by a 1-GHz sinusoidal wave
in the absence of magnetic field, where a relative combined uncertainty of $2.7\times10^{-7}$ ($k=1$)\cite{Zhao2017} was achieved.
\begin{figure*}[!h]
\includegraphics[width=5.2in]{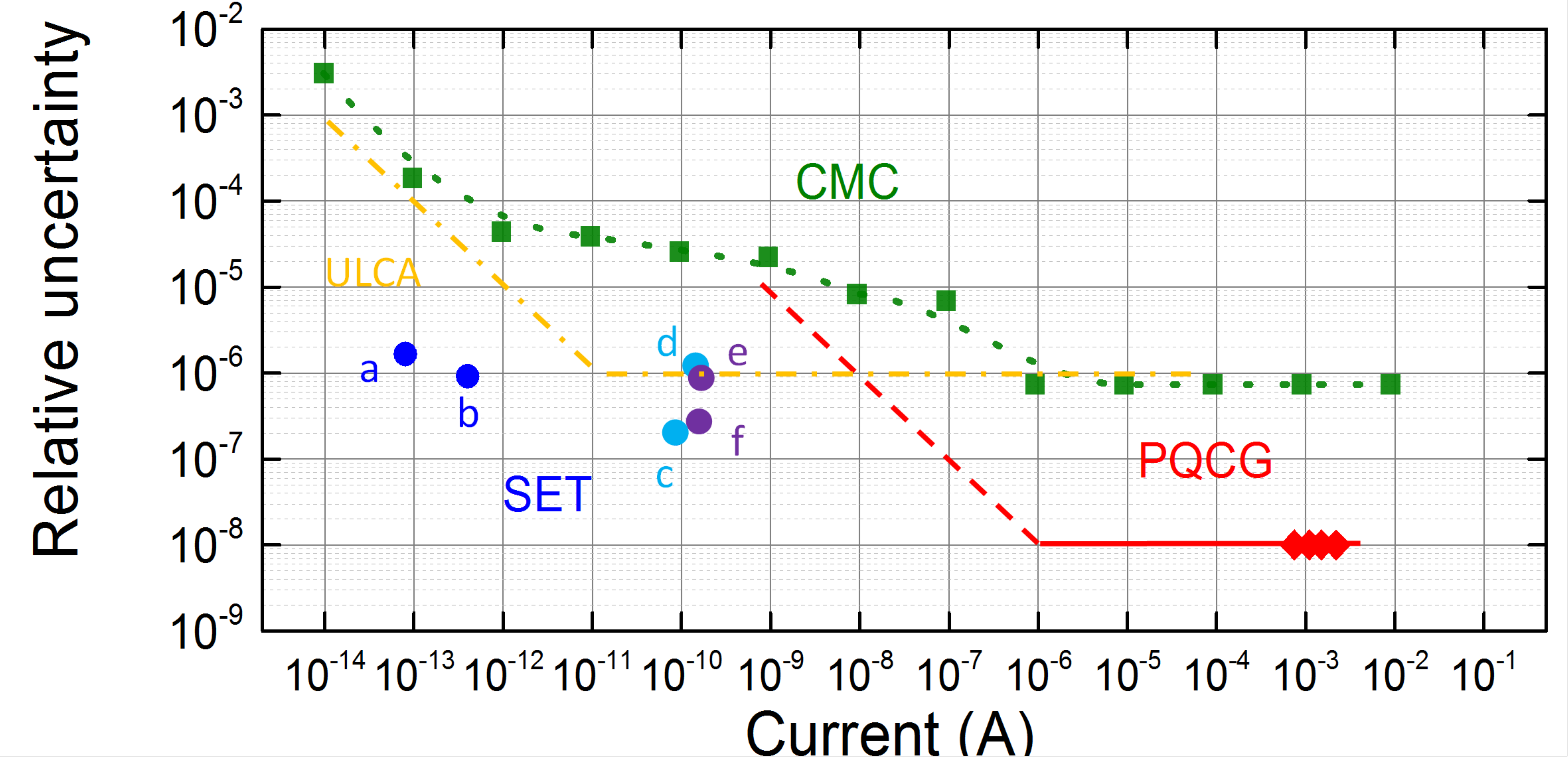}
\caption{\textbf{Relative uncertainty for ampere traceability (measurement/generation)}. Uncertainties of SET devices (blue dots). Metallic SET: a\cite{Camarota2012},b\cite{Keller2007}. GaAs SET: c\cite{Stein2015},d\cite{Giblin2012}. Silicon SET: e\cite{Yamahata2016},f\cite{Zhao2017}. Best calibration and measurement capabilities (CMCs) (green squares: from $10^{-14}$ up to $10^{-11}$ A by charging a capacitor (Physikalisch-Technische Bundesanstalt), from $10^{-10}$ up to $10^{-1}$ A by applying Ohm's law (Laboratoire national de m\'{e}trologie et d'essais)\cite{BestCMC}. Uncertainties of the ULCA on a quarter (yellow dashed-dotted line, from ref.\cite{Drung2017}. Uncertainty of the PQCG from $\mathrm{1~\mu A}$ up to 10 mA (red line) and uncertainties demonstrated through comparisons in the milliampere range (red diamonds), from ref.\cite{Brun-Picard2016}. Estimated uncertainty of the PQCG from 1 nA up to 1 $\mu$A (red dashed line). Uncertainties corresponds to one standard deviation (k=1).}\label{fig:Fig-CMCAmpere}
\end{figure*}
Tunable barriers semiconductor pumps have improved the uncertainties in the range below 1 nA as can be observed on fig.\ref{fig:Fig-CMCAmpere}. However, these new pumps do not yet constitute practical quantum current standards. A quantitative model of non-adiabatic effects in charge capture remains to be further developed \cite{Kaestner2015}. Moreover, compared to QHR or JE standards, the amount of data confirming the robustness against variations of the operating parameters is rare \cite{Stein2017,Zhao2017}. Quick characterizations that ensure that the quantification is at a certain level of uncertainty is still lacking.

To circumvent this difficulty, Fricke and co-authors have recently proposed self-referenced electron pumps equipped with a counting system of electron transfer errors based on additional quantum dots coupled to SET transistors\cite{Fricke2013,Fricke2014}. In the mean time, there are attempts to use SET pumps, which are very low noise current source, for fundamental research in the field of single-electron optics \cite{Johnson2017}.
\subsection{Applying Ohm's law or charging capacitor}
In NMIs, the traceability of current is realized by applying Ohm's law to secondary standards of voltage and resistance or charging a capacitor and calibrating the voltage at its terminals. Uncertainties claimed by NMIs in their best calibration and measurement capabilities (CMC), reported in fig.\ref{fig:Fig-CMCAmpere} are not better than $10^{-6}$ above 1 $\mu$A, and are higher at lower current values\cite{BestCMC}. Limitations come from the higher calibration uncertainties of secondary standards, although they are traceable to $R_\mathrm{K}$ and $K_\mathrm{J}$ constants, and the lack of sensitivity of measurement methods below 1 $\mu$A. As shown in fig.\ref{fig:Fig-CMCAmpere}, the traceability of low currents was recently improved by an ultra-low current amplifier (ULCA) based on a more stable voltage to current converter\cite{DrungRSI2015,Drung2015,Drung2017}. This device demonstrated a better relative reproducibility over time: $10^{-7}$ over a week, a quarterly stability of $10^{-6}$ and a stability of $5\times10^{-6}$ over a year. On the other hand, in the range of higher currents which covers the main calibration requests, no measurement improvement was expected until the development of a programmable quantum current generator.
\subsection{The programmable quantum current generator}
\begin{figure*}[!h]
\includegraphics[width=5.2in]{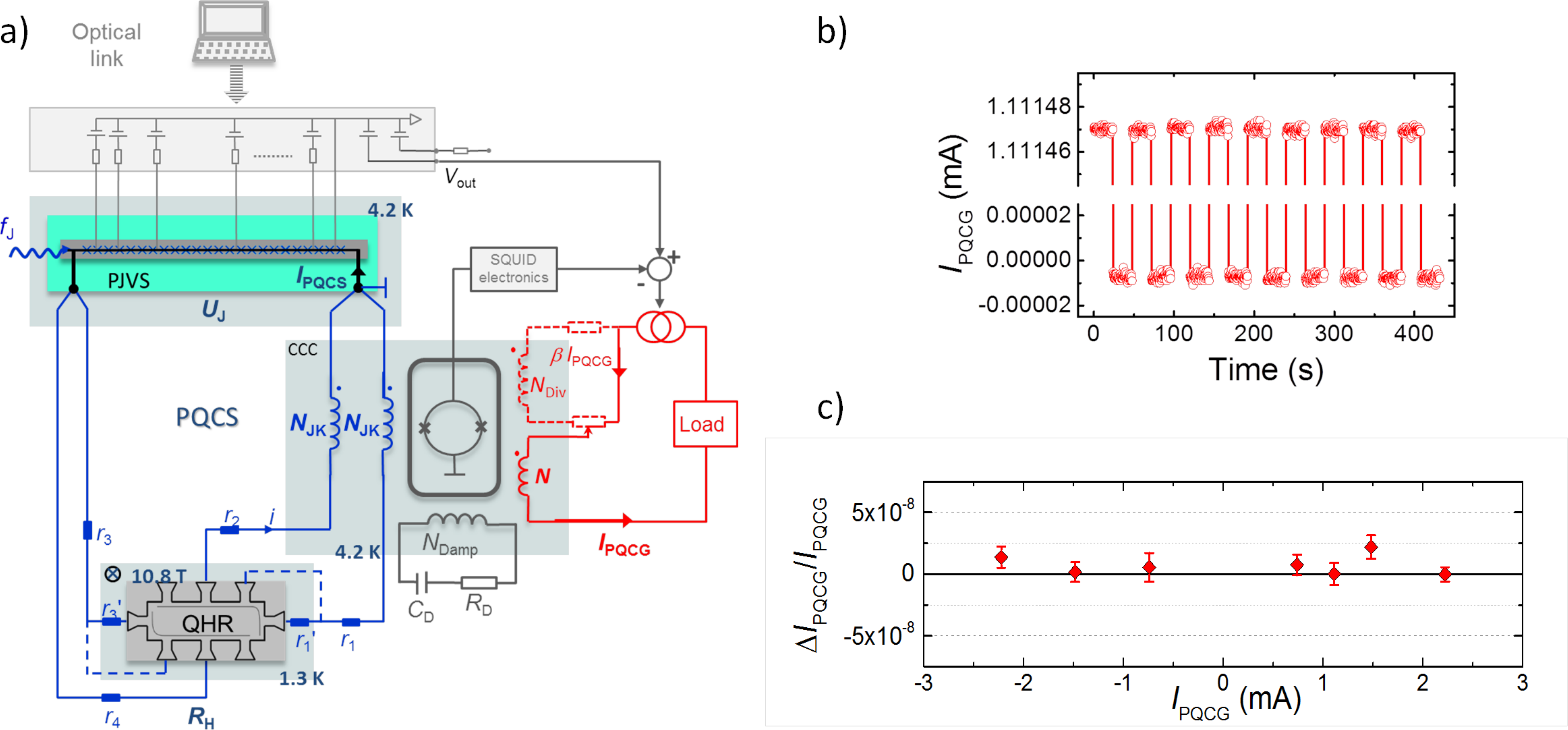}
\caption{a) Schematic of the programmable quantum current generator (PQCG) developed by LNE: the current $I_\mathrm{PQCG}$ of the generator is servo-controlled by a CCC that detect and amplify the current $I_\mathrm{PQCS}$ circulating through a QHR connected to a PJVS. b) Sequence of switching on/off of the current $I_\mathrm{PQCG}$ generated by the PQCG (@ 1.1 mA). c) Relative deviation $\Delta I_\mathrm{PQCG}/I_\mathrm{PQCG}$ of the current from its theoretical value in the milliampere range.}\label{fig:Fig-PQCG}
\end{figure*}
Very recently, a programmable quantum current generator (PQCG) based on the application of Ohm's law directly to the quantum voltage and resistance standards demonstrated quantized currents in terms of $ef_\mathrm{J}$ ($f_\mathrm{J}$ is a Josephson frequency) to within one part in $10^8$ in the range from 1 $\mu$A up to a few mA\cite{Poirier2014,Brun-Picard2016}. This performance relies on the use of a cryogenic current comparator to detect and then amplify the current $I_\mathrm{PQCS}$ flowing in a QHR multiply-connected to a PJVS, as shown in fig.\ref{fig:Fig-PQCG}a. The multiple (double in figure) connection is used to drastically reduce the correction to the quantized current caused by the wire resistance: $I_\mathrm{PQCS}=(U_\mathrm{J}/R_\mathrm{H})(1-\alpha)$, where $U_\mathrm{J}=n_\mathrm{J}(h/2e)f_\mathrm{J}$ is the Josephson voltage, $n_\mathrm{J}$ is the number of Josephson junction biased, $R_\mathrm{H}=h/2e^{2}$ is the Hall resistance and $\alpha\sim 2\times10^{-7}\pm2.5\times10^{-9}$. Due to the chirality of the edge states in the QHE regime, the main part of the current $I_\mathrm{PQCS}$ circulates through a wire, while a minor part flows through the second wire. The total current is detected by two CCC windings of same number of turns $N_\mathrm{JK}$ which are inserted in the two wires. A winding of number of turns $N$ is connected to an external battery-powered current source delivering a current $I_\mathrm{PQCG}$. The latter is servo-controlled by the CCC so that $N_\mathrm{JK}I_\mathrm{PQCS}-NI_\mathrm{PQCG}=0$. It results that the current of the generator is set to $I_\mathrm{PQCG}=(N_\mathrm{JK}/N)n_\mathrm{J}ef_\mathrm{J}(1-\alpha)$. Its amplitude can be modulated by varying either the gain $G=N_\mathrm{JK}/N$ by four orders of magnitude, or the number $n_\mathrm{J}$ between unity and several thousands or more finely the frequency value. An example of a sequence of on/off switching of the current (amplitude of $\sim 1.1$ mA) is given in fig.\ref{fig:Fig-PQCG}b. The accuracy of the PQCG was determined in the milliampere range by comparing the voltage drop at the terminals of a 100 $\Omega$ resistor feed by the quantized current with a reference Josephson voltage. As shown in fig.\ref{fig:Fig-PQCG}c, the PQCG current is quantized to its theoretical value to within one part in $10^{8}$ for current values between $\pm 2.2 mA$. The relative standard deviation of the results amounts to $8\times10^{-9}$ only. By principle, the current $I_\mathrm{PQCG}$ remains quantized with the same accuracy over the wide range of current values accessible by changing $G$ which is highly-accurate and can span two orders of magnitude above or below the unity gain. Moreover, the relative current density noise $S_I/I$ does not depend on $G$ at a given value $I_\mathrm{PQCS}$. Consequently, the PQCG can accurately generate currents with a combined relative measurement uncertainty of $10^{-8}$ in the whole range from $\mathrm{1~\mu A}$ up to 10 mA, as illustrated in fig.\ref{fig:Fig-CMCAmpere}. A linear increase of the uncertainty is expected at lower currents.

This device provides an accurate realization of the new ampere definition. More fundamentally, the PQCG works as a multi-electron current pump. At each cycle of the external radio-frequency signal, $n_\mathrm{J}$ electrons are indeed transferred through the QHR device. This comes from that a radio-frequency pulse irradiating the PJVS generates a quantized voltage pulse whose time-integral is equal to $n_\mathrm{J}h/2e$. This results in a quantized current pulse corresponding to a total charge $n_\mathrm{J}h/2e/(h/2e^2)=n_\mathrm{J}e$. As discussed further below, this calls for considering the development of an AC quantum current source in future. Besides, the PQCG relies on an instrumentation yet available in NMIs equipped with a CCC resistance bridge and a PJVS. Thus, there is no additional cost for its realization.

As reported in fig.\ref{fig:Fig-CMCAmpere}, The PQCG improves the current traceability by two orders of magnitude. This new quantum current generator, which can be quickly validated by checking quantization criteria, was successfully used to calibrate a digital ammeter in ranges from 1 $\mu$A to 10 mA with a record uncertainty ($\sim 2\times10^{-7}$) only limited by the performance of the device itself.

Further reduction of uncertainty (down to $10^{-9}$), extension of the current range and simplification are expected by implementing some improvements in the PQCG\cite{Brun-Picard2016}. Moreover, the use of a graphene-based quantum Hall resistance standard will simplify its experimental conditions of operation. The PQCG should constitute a key element of the quantum calibrator based on a single cryogen-free system, described in section \ref{Quantum calibrator}.
\subsection{Future of the metrological triangle}
Despite the adoption of the relationships $R_\mathrm{K}=h/e^2$, $K_\mathrm{J}=2e/h$ and $Q=e$ in the revised SI, the metrological triangle experiment remains of fundamental and practical interest. Any highlighting of a discrepancy to the equation $R_\mathrm{K}K_\mathrm{J}Q=2$ in the future would, of course, open a deep debate about quantum mechanics and would question the adoption of the individual relationships. Let us remember that tiny corrections (about $10^{-20}$ in relative value at 20 T) caused by a renormalization of the electron charge in presence of a magnetic field have been predicted both for $R_\mathrm{K}$ and $K_\mathrm{J}$ by A. A. Penin\cite{Penin2009,Penin2010} using quantum electrodynamics calculations. Regardless of such an hypothesis, this experiment keeps on being the best way to test the quantization of any single-electron pump.
\subsection{A new ampere metrology}
Current traceability is probably the field in electrical metrology where progress in accuracy will be the most important. As illustrated by fig. \ref{fig:Fig-CMCAmpere}, uncertainties of ampere realization have recently decreased by a factor of ten or one hundred in a range extending over more than 10 orders of magnitude. The PQCG offering highly-accurate current calibration, one can now consider the development of a new generation of stable transfer current source or ammeter. The ULCA\cite{Drung2017} is an example of such device. As shown in ref\cite{Brun-Picard2016}, commercial digital precision multimeters operating in current mode could also constitute good transfer ammeters. A simplification and improvement of the traceability chain for current is therefore expected. This would benefit to the end-users by a reduction of uncertainty and a possible cost reduction.
\begin{figure*}[!h]
\includegraphics[width=5.2in]{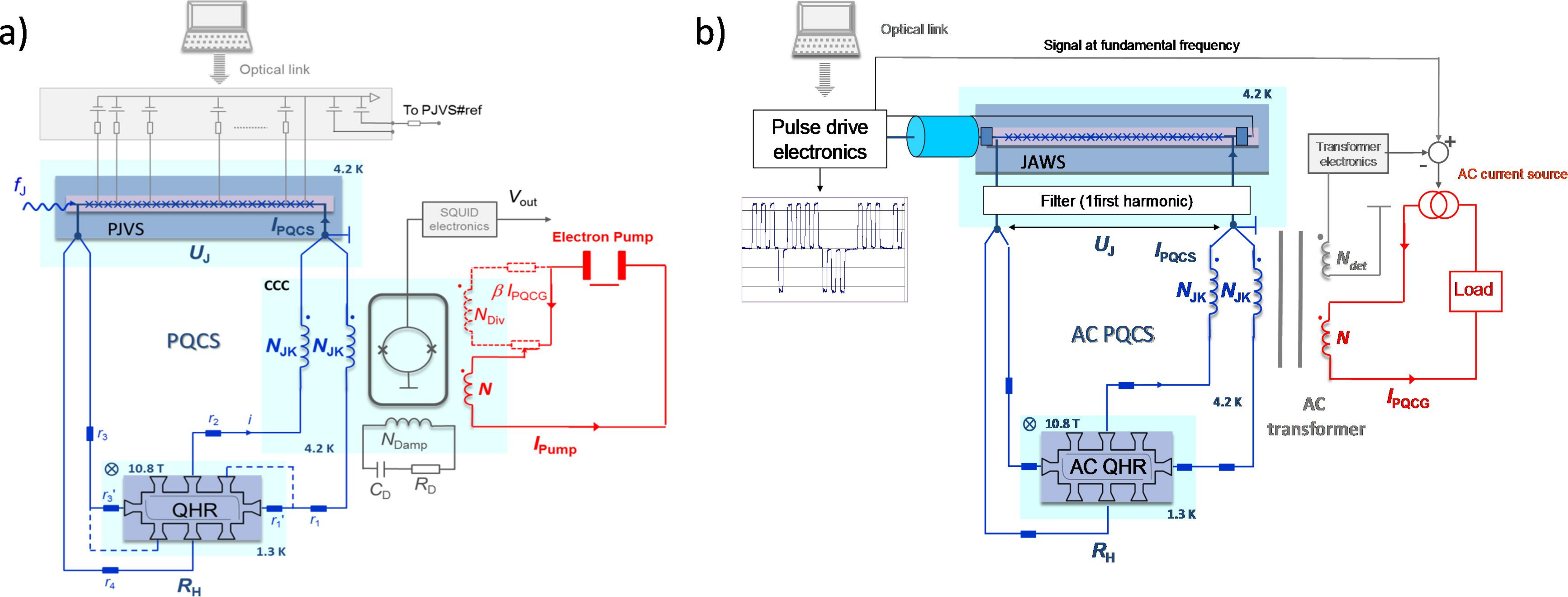}
\caption{a) Principle of calibration of a current source (here, an electron pump) using a quantum ammeter. b) Principle of a programmable quantum generator of alternating current.}\label{fig:Fig-PQCGPerspectives}
\end{figure*}

The PQCG relies on the accurate exploitation by a CCC of a reference current $I_\mathrm{PQCS}$. This principle is seminal and can be applied to others devices\cite{Poirier2014}. One can cite the quantum ammeter (fig.\ref{fig:Fig-PQCGPerspectives}a) based on the direct comparison of the current generated by a device under test (DUT) with the reference current. Ultra-accurate universality tests of the QHE can be performed by comparing, using the CCC, two reference currents obtained from the same Josephson voltage and two different quantum Hall resistors. Finally, the PQCG can be adapted to audio-frequency alternating current (fig.\ref{fig:Fig-PQCGPerspectives}b) by replacing the CCC by a magnetic transformer and the PJVS by a pulse-driven Josephson array. To conclude, the availability of a quantum current source opens the way to a renewed metrology of the ampere.
\section{New applications of quantum standards}
\subsection{The Kibble balance or the quantum kilogram}
\label{Kibble balance}
One emblematic application of the quantum electrical standards, which stimulated the revolution of the SI\cite{Mills2005}, is the realization of the kilogram from the Planck constant $h$ using a Kibble balance\cite{Kibble1976}. As yet mentioned, this experiment consists in measuring the mechanical power of a mass $m$ moving at a velocity $v$ under the gravitational acceleration $g$ in terms of an electric power in a coil calibrated from the Josephson constant $K_\mathrm{J}$ and the von Klitzing constant $R_\mathrm{K}$.
\begin{figure*}[!h]
\includegraphics[width=5.2in]{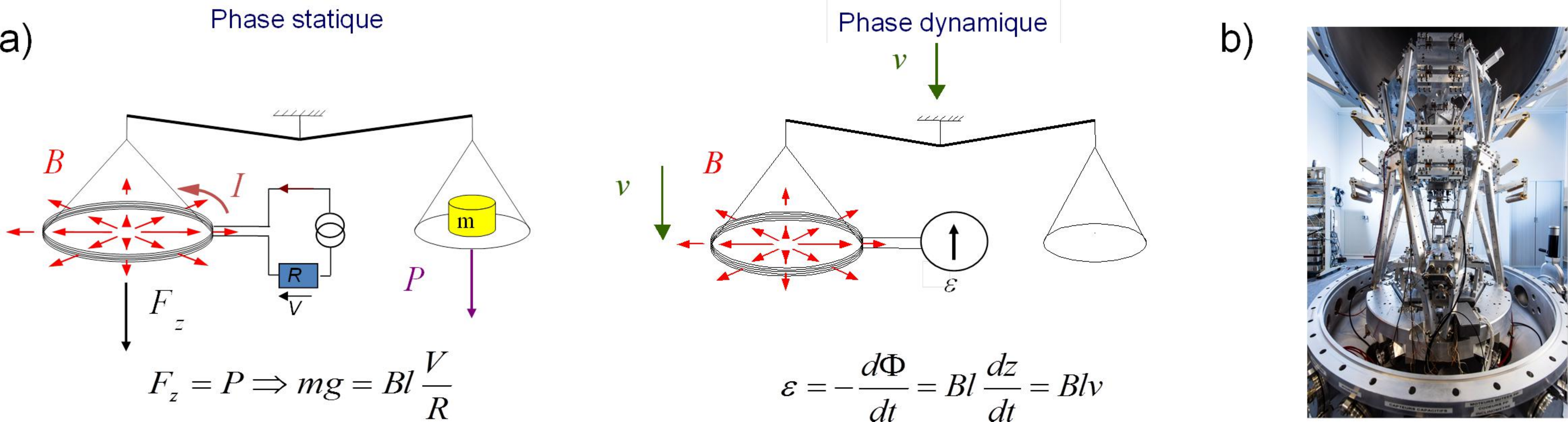}
\caption{a) Description of the two phases of the Kibble balance experiment. The mechanical power $mgv$ is determined from the electrical power $\epsilon V/R$. b) Picture of the Kibble balance at LNE.}\label{fig:Fig-BW}
\end{figure*}
In practice, this measurement has two phases (fig.\ref{fig:Fig-BW}a): the mechanical force is first balanced by the magnetic force resulting from the circulation of a current $I$ in a coil under a radial magnetic induction $B$ in the static phase, the voltage at the terminal of the coil moving at a velocity $v$ is then recorded in the dynamic phase. Measurements of both phases are combined to cancel the geometrical factor $l$ of the coil used, which leads to the power comparison. Adoption of the theoretical relationships $K_\mathrm{J}=2e/h$ and $R_\mathrm{K}=h/e^2$ in the SI provides a direct link between mass and Planck constant according to $m=h\frac{A}{4gv}$, where $A$ is a quantity involving Josephson frequencies, number of Josephson junctions and Hall plateau index. After participating in $h$ determinations\cite{Mohr2018}, it is now a question of using Kibble's balances, notably those having participated in the determination of $h$ from NIST\cite{Haddad2017}, NRC\cite{Wood2017} and LNE (fig.\ref{fig:Fig-BW}b)\cite{Thomas2017}, to calibrate mass standards with a $10^{-8}$ relative uncertainty from the Planck constant value\cite{Newell2018} $h=6.62607015\times10^{-34}$ J.s. This extension of the application of solid-state quantum effects beyond electrical metrology is to benefit from the user-friendly graphene-based quantum resistance standard and cryogen-free cooling techniques.
\subsection{Quantum impedance standard}
\label{QIS}
The Hall resistance being expected\cite{Kuchar1986,Viehweger1991} to remain quantized within one part in $10^9$ at frequencies in the kilohertz range, an important field of research has therefore been to realize a quantum resistance standard operating in alternating current (AC). Targeted applications are traceability of AC resistance and more generally of impedances.
\subsubsection{AC quantum Hall effect}
\begin{figure*}[!h]
\includegraphics[width=5.2in]{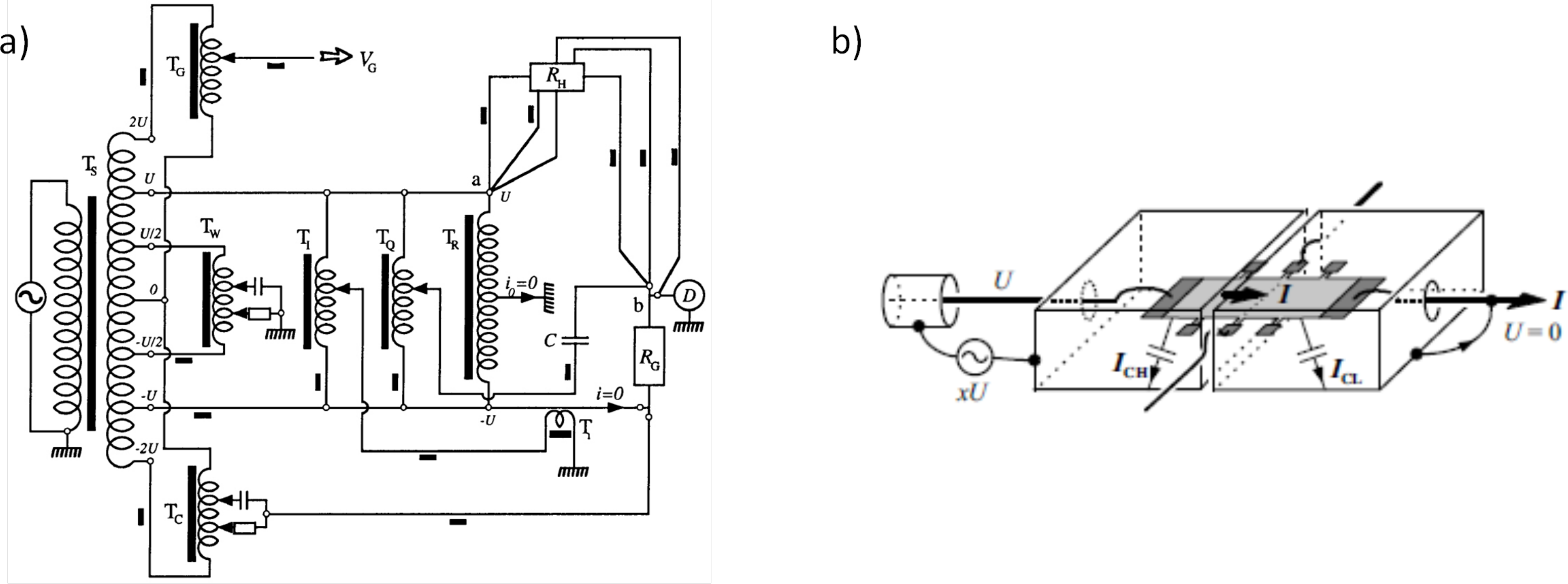}
\caption{a) Schematic of a resistance bridge used in AC regime. From ref.\cite{Delahaye1995}. b) Double shielding of a Hall bar for operation in AC regime. From ref.\cite{Kibble2008}.}\label{fig:Fig-ACQHE}
\end{figure*}
The study and implementation of the AC QHE relies on the development of terminal-pair resistance bridges and quadrature bridges that are based on the coaxial measurement techniques\cite{Awan2011Book}. As shown in fig.\ref{fig:Fig-ACQHE}a, specific techniques are used to preserve a perfect quantization of the Hall resistance. Firstly, the Hall bar is implemented using the multiple series connection technique. On one hand, this technique cancels large quadratic frequency dependencies due to series inductance, but also simplifies the bridge since zero current requirement in the voltage arm of the QHR is automatically ensured\cite{Delahaye1995,Overney2003}. Given the high impedance of the shielding conductors connecting the QHR at low temperature, active equalizers are used to ensure a good coaxiality. Despite these precautions, several works have highlighted a residual deviation of the Hall resistance from its quantized value that linearly increases with frequency and measurement current\cite{Ahlers2009}. This discrepancy, which can amount to about a few $10^{-8}$ at 1 kHz, is linearly coupled to the longitudinal resistance, as often observed in DC regime. Its origin is attributed to losses of AC charging current in internal capacitances of the Hall bar and in external capacitances of coupling with ground. However, it is possible to cancel its impact on the Hall resistance to within about one part in $10^9$ per kilohertz by using a double-shielding technique\cite{Kibble2008} of the Hall bar, as described in fig.\ref{fig:Fig-ACQHE}b.

Given the robustness of the QHE in graphene, it was attractive to perform studies in graphene-based Hall bars. In 2014, Kalmbach and co-authors showed that large quantum Hall plateaus measured with alternating current were flat within one part in $10^7$. Moreover, they measured a intrinsic frequency dependence similar in magnitude to that of GaAs devices\cite{Kalmbach2014}. Owing to graphene, a more user-friendly quantum standard for both resistance and impedance is therefore expected.
\subsubsection{Impedance calibration}
\begin{figure*}[!h]
\includegraphics[width=5.2in]{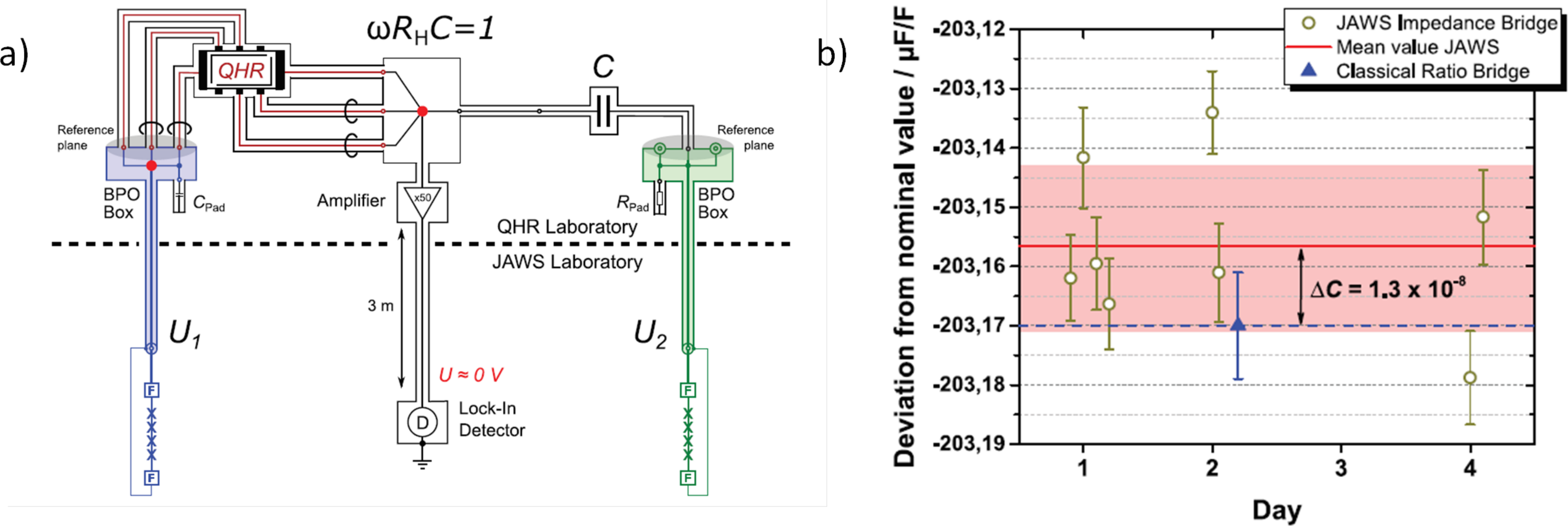}
\caption{a) Schematic overview of the quadrature bridge for measuring a 10 nF capacitance standard from the quantum Hall resistance. The two reference voltages $U_1$ and $U_2$ defining the standard ratio of the bridge are provided by two Pulse-driven Josephson voltage standards. b) Comparison of the results obtained with the pulse-driven Josephson bridge (open circle) and with a classical impedance bridge (filled triangle). Error bars correspond to one standard deviation (k=1). From ref.\cite{Bauer2017}.}\label{fig:Fig-QHEImpedance}
\end{figure*}
Operation of the QHE in AC regime at frequencies of a few kilohertz has opened up the way towards a quantum standard of impedance\cite{Overney2006,Schurr2007} linked to $R_\mathrm{K}$ constant. In 2009, J. Schurr and B. Kibble\cite{Schurr2009} demonstrated a new way to realize the unit of farad by calibrating a capacitance from two quantum Hall resistances used in a quadrature bridge with a relative measurement uncertainty of $6\times10^{-9}$. This method provides a direct realization of the farad from $R_\mathrm{K}$ which avoids an additional calibration step relying on calculable resistors\cite{HaddadPhd1969,Gibbings1963}. However, the calibration of impedance requires very precise comparison bridges which are usually based on inductive voltage dividers (IVD)\cite{Awan2011Book,Cutkosky1970}. Even if they can reach uncertainties of a few parts in $10^9$, their measurement capabilities are limited to pure impedances. Furthermore the ratio of the measured impedances is restricted to a few fixed nominal ratios like 1:1 and 10:1. Finally, these impedance bridge require realizing a long and tedious calibration of the IVDs at each ratio and frequency used. The recent development of pulse-driven Josephson voltage standards\cite{Lee2010,Overney2018,Palafox2016} able to generate sine-waves with high spectral purity, now makes possible the comparison of arbitrary impedances at the same level of uncertainty as the IVD's bridges over a wider frequency range. The ratio of the bridge is then defined by two pulse-driven Josephson series arrays adjustable in magnitude and phase offering very high accuracy, and the possibility to compare any kind of impedance at frequencies up to 40 kHz\cite{Overney2016}. Using such a technology, S. Bauer and co-authors\cite{Bauer2017} performed the calibration of a capacitance from $R_\mathrm{K}$: two pulse-driven Josephson voltage standards and one quantum Hall resistor were involved in the experiment (fig.\ref{fig:Fig-QHEImpedance}a). They demonstrated an agreement of the measurements with those performed using classical bridges within about 1.3 parts in $10^{8}$ (fig.\ref{fig:Fig-QHEImpedance}b). Further progress in the experimental conditions of operation of graphene-based QHE devices, \emph{i.e.} reduction of the magnetic induction below 2 T, should allow in a near future that Josephson voltage standard and QHE devices could operate in the same cryogen-free cryostat. This would support the development of a user-friendly quantum standard ensuring impedance calibration in the whole complex plane.
\subsection{Quantum electronic kelvin}
\label{QVNS}
\begin{figure*}[!h]
\begin{center}
\includegraphics[width=2.5in]{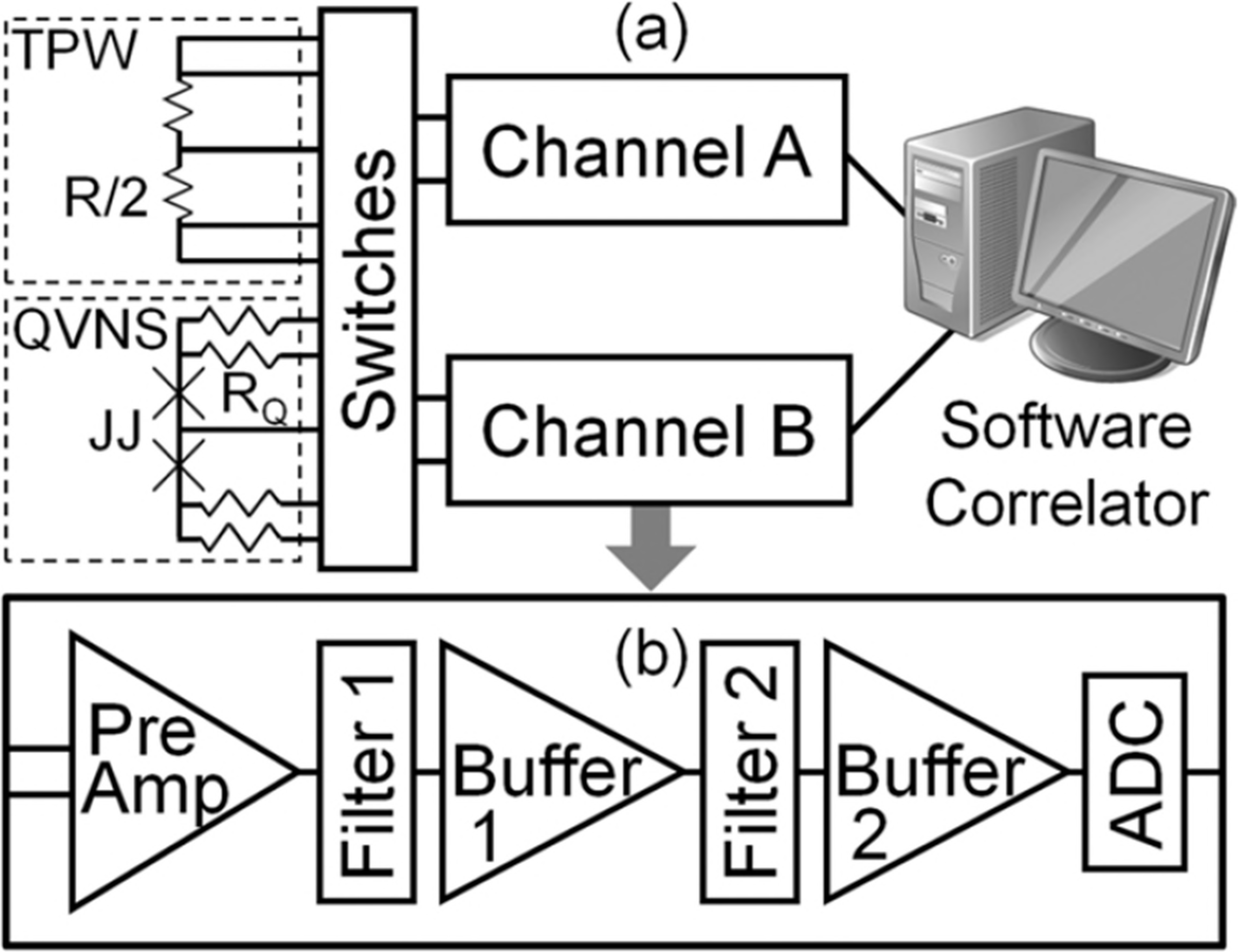}
\caption{Schematic of the QVNS-JNT cross-correlation electronics: a) Two channels
(A and B) of the correlator that simultaneously measure one of the two voltage sources. The switching network alternates between the two input signals. b) Each channel consists of a series of amplifiers and filters, followed by an analogue-to-digital converter (ADC). The digitized signals from each channel are optically transmitted to the computer that performs the correlation analysis. From ref.\cite{Benz2011}.}\label{fig:Fig-QVNS}
\end{center}
\end{figure*}

The Johnson noise thermometry (JNT) \cite{White1996} is a primary thermometry based on the Johnson-Nyquist noise of a resistor $R$ \cite{Johnson1927,Johnson1928}. This noise results from the fluctuation-dissipation theorem, which predicts a relation between the resistance and the thermal voltage fluctuations in a conductor due to the random thermal motion of the electrons. For a temperature $T$, the mean-squared voltage noise is given by Nyquist equation :
\begin{equation}\label{Johnson-noise}
  \overline{V_{T}^{2}}=4 k T R \Delta f
\end{equation}
where $k$ is the Boltzmann's constant and $\Delta f$ is the bandwidth of the measurements.

Experimentally, the JNT is used to infer a temperature by comparing the mean-square noise voltage measured at the terminals of a first resistor at the unknown temperature and of a second resistor at a reference temperature. By this way, the calibration of the measurement chain, notably the bandwidth $\Delta f$, is circumvented. Due to the extremely small voltages, of only 1.2 nV/$\sqrt{\mathrm{Hz}}$ for a resistance of 100 $\Omega$ at 273.16 K (triple point of water), cross-correlation techniques are used. Cross correlation is needed because the small noise voltage is comparable to the noise of the low-noise amplifier.  In 2003, Benz \emph{et al.} \cite{Benz2003,Nam2003} proposed to replace the reference calibrated resistor by a JAWS (fig.\ref{fig:Fig-QVNS}). The requirements for the JAWS system in JNT experiment contrast with those in AC voltage metrology where the challenge is to increase the output amplitude. Here, the JAWS system have been designed to produce multi-tone pseudo-noise waveforms with small (1$\mu$V peak) voltage amplitudes (few JJs) \cite{Benz2011}  approximately matched to the expected Johnson noise. The quantum voltage noise source (QVNS) is a comb of harmonic tones (MHz range), equally spaced in frequency, of identical amplitudes and random relative phases \cite{White2008}. The different improvements of the QVNS-JNT lead to high precision temperature measurements, which were used to determine the Boltzmann's constant \cite{Benz2011,Flowers-Jacobs2017,Urano2017,Qu2017}. The lowest uncertainty with this technique was achieved by Qu et al. \cite{Qu2017} with a relative uncertainty of 2.7 parts in $10^{6}$.
\subsection{The quantum calibrator}
\label{Quantum calibrator}
\begin{figure*}[!h]
\includegraphics[width=5.2in]{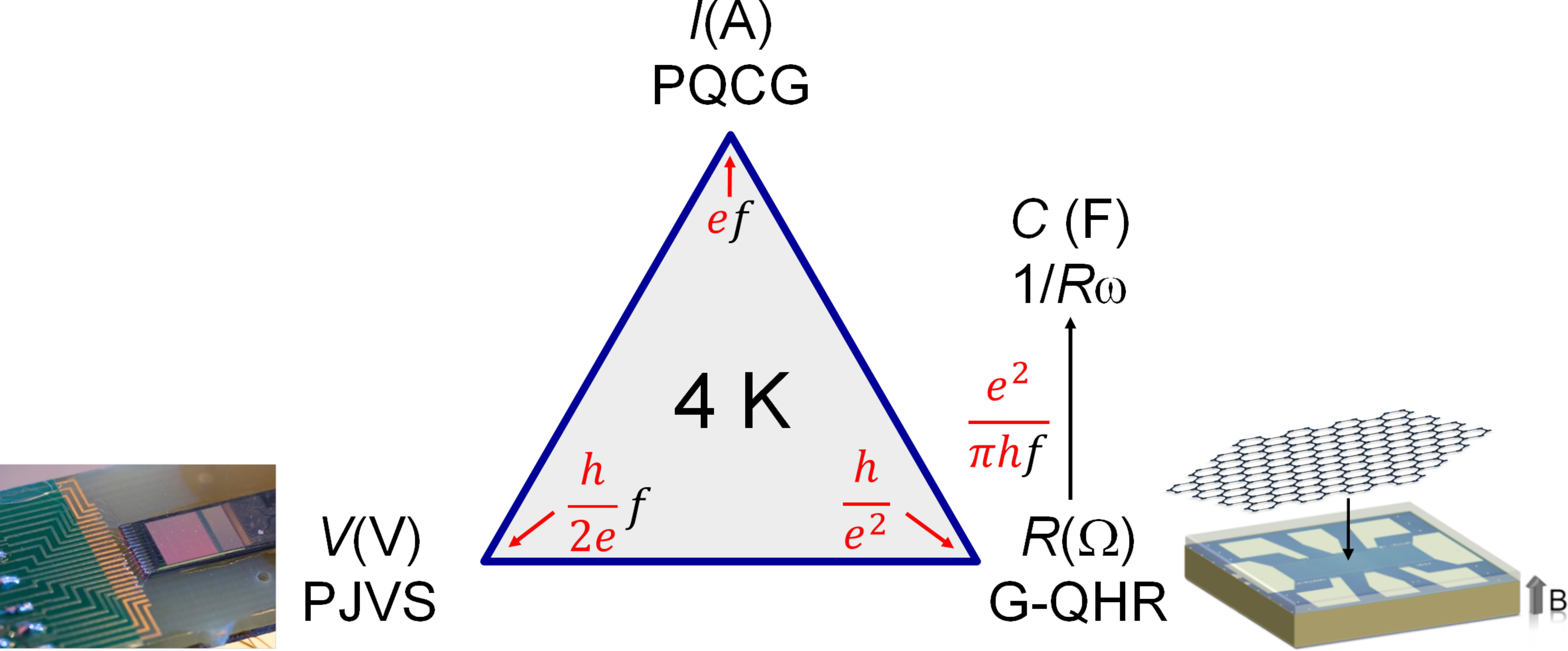}
\caption{Illustration of a quantum calibrator realizing the electrical units (A, V, $\Omega$, F) from $h$, $e$ and the frequency $f$. It is based on a programmable Josephson voltage standard (PJVS) and a graphene-based quantum Hall resistance standard (G-QHR). The ampere is realized using the programmable quantum current generator (PQCG). The farad is realized from the G-QHR using a quadrature bridge.}\label{fig:Fig-Calibrator}
\end{figure*}
The quantum calibrator consists in a user-friendly device realizing accurately the main electrical units, \emph{i.e.} the ampere, the volt, the ohm and the farad from the Planck constant $h$ and the elementary charge $e$ only. Until recently, this ambitious idea was facing several difficulties. The first one was that the relationships of the fundamental constants $Q$, $K_\mathrm{J}$ and $R_\mathrm{K}$ with $h$ and $e$ were spoiled by large uncertainties. The revised SI solves this problem. The second critical problem was that the three quantum devices used to realize the units were operating in very different experimental conditions which forbade their implementation in a single cryostat: $T=4.2$ K and $B=0$ T for the Josephson array, $T<2$ K and $B\sim 10$ T for the quantum Hall bar and $T<0.3$ K and $B\sim 10$ T for the recent SET devices. After the many works carried out in the last ten years, this difficulty is being solved. At first, it was demonstrated that a graphene-based quantum resistance standard can operate at a temperature $T\geq 4.2$ K and a magnetic induction $B\simeq 3.5$ T. These conditions are now closer from those required by a Josephson array. Second, the PQCG now offers a $10^{-8}$-accurate realization of the ampere with adapted current values to calibrations centers which is only based on the Josephson and quantum Hall resistance standards. This avoids the supplementary experimental constraints imposed by SET devices, which besides have not yet managed to realize the ampere with the required accuracy and reproducibility. These results give hope that further progress, concerning the graphene growth for achieving an even lower operational magnetic field (hopefully less than 1 T) and some engineering works to screen small magnetic field, should allow the development of a device based on a Josephson array and a quantum Hall resistance standard only. The third difficulty was to realize the farad from the QHE operating in ac without referring to calculable coaxial resistors. Calibration in such way of a capacitance was demonstrated with a relative uncertainty of a few $10^{-9}$ using an adapted quadrature bridge. Finally, the availability of cryogen-free cryostats with a base temperature lower than 4.2 K makes easier and less-costly the operation of quantum devices. Thus, the quantum calibrator is no longer just an idea but now becomes a project in many NMIs to support a high-accuracy dissemination of the electrical units towards the end-users.
\section{Further perspectives}
Advances in metrology have always closely followed the scientific and technological discoveries. Let us evoke recent works that could be promising for electrical metrology.

Mooij and Nazarov \cite{Mooij2006} suggested to use quantum phase slips in disordered superconducting nanowires to realize a quantized current source, which could produce larger currents. Phase slip events occur in low dimension superconductors where thermodynamic fluctuations of the order parameter become significant. When a phase slip occurs at some point in the wire, the superconducting order parameter vanishes locally. The phase difference changes by $2\pi$ over the wire and this gives rise to a quantized voltage pulse. If the phase slips happen frequently, they produce a finite dc voltage or a finite resistance. Well below $T_c$, the phase of a homogeneous superconducting wire can slip by $2\pi$ due to quantum tunnelling, a process analogous to Cooper-pair tunnelling in Josephson junctions. The quantum-phase-slip junction is formally the exact dual of the Josephson junction with respect to the exchange of the canonically conjugated quantum variables, phase and charge. Hence, equation (\ref{RCSJ_equation}) can be rewritten for the charge instead of the phase by considering a dual circuit, where the capacitance and the resistance shunting the JJ are replaced by an inductance and a resistance in series with the quantum-phase-slip junction. Mapping the problem of the Josephson junction, dual-Shapiro steps have been predicted in ultrathin superconducting nanowire with sufficiently high series resistance submitted to microwave irradiation. Although, very promising, these dual Shapiro steps have not been observed yet \cite{Webster2013}. However, coherent quantum phase slips have been unambiguously observed by spectroscopy, in narrow nanowires of strongly disordered superconductors near the superconductor/insulator transition, integrated in a superconducting loop coupled to a coplanar resonator \cite{Astafiev2012}. More recently, the dual of the SQUID has been demonstrated in a device that integrates several coherent quantum phase slip junctions \cite{DeGraaf2018}.

The observation of gigahertz quantized charge pumping in graphene quantum dots\cite{Connolly2013} and the recent discovery of superconductivity in magic-angle graphene superlattices\cite{Cao2018} could enable the use of graphene as a common material platform for developing not only the QHR but also voltage and current quantum standards. The ideal goal would even be to integrate several quantum standards on a single graphene chip, although this requires a magnetic field of operation of the QHE low enough to preserve the superconductivity in Josephson devices.
\begin{figure*}[!h]
\includegraphics[width=5.2in]{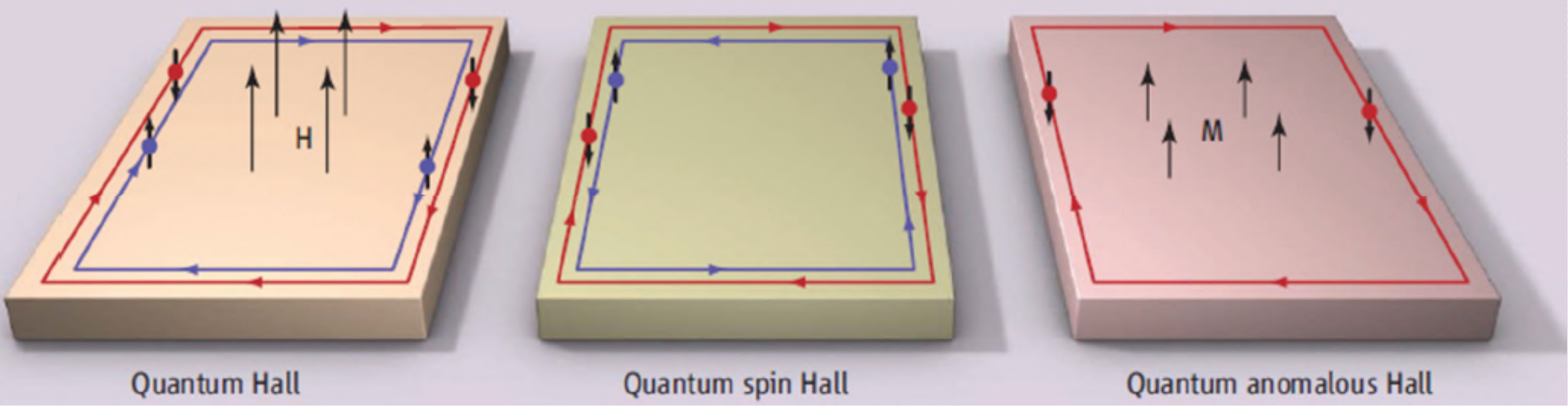}\hfill
\caption{Edge states in the three quantum Hall effects. The usual quantum Hall effect (left): electrons with opposite spin move in the same direction. The spin quantum Hall effect (center): electrons with opposite spin move in opposite directions. The anomalous quantum Hall effect (right): electrons moving alon an edge have a defined spin direction (here spin down). From ref.\cite{Oh2013}.}
\label{fig:Fig-AllQHE}
\end{figure*}

The discovery of the quantum anomalous Hall Effect (QAHE) which manifests itself by the Hall resistance quantization at zero magnetic field\cite{Oh2013} opens another way to get in a single cryostat both the QHR and the Josephson array. Generally, the QHE relies on the existence of chiral edge states having opposite momentum on both sides of a sample which suppresses electron backscattering, \emph{i.e.} dissipation (fig.\ref{fig:Fig-AllQHE}-left). One fundamental question was to know whether dissipation-less states can exist without magnetic field. Following some theoretical predictions supporting this hypothesis\cite{Haldane1988}, the discovery of the spin quantum Hall effect brought the demonstration of dissipation-less edge states where electrons of opposite spin directions are counter-propagating as a result of a strong spin-orbit coupling\cite{Konig2007} (fig.\ref{fig:Fig-AllQHE}-center).
\begin{figure*}[!h]
\includegraphics[width=5.2in]{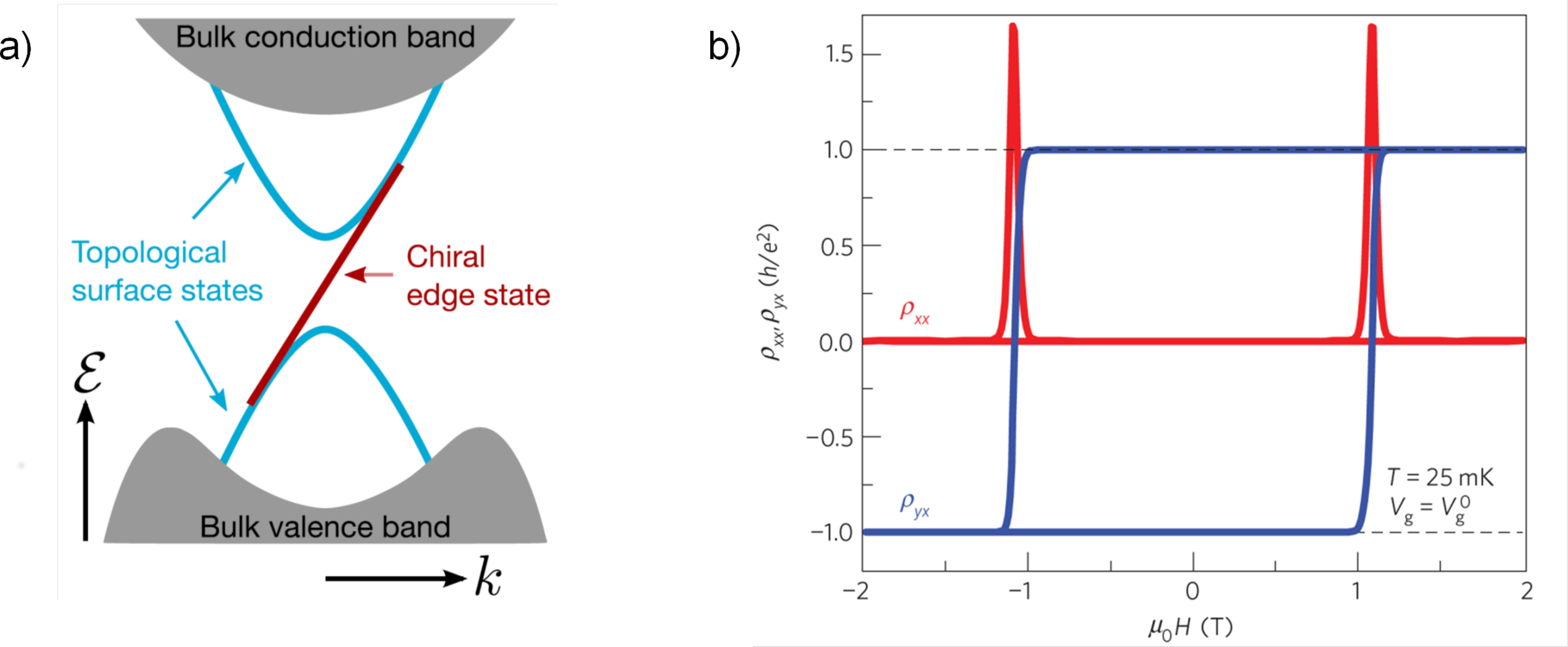}\hfill
\caption{a) Schematic energy spectrum $\epsilon(k)$ of a FTI in the QAHE state. From ref.\cite{Fox2018}. b) Observation of the QHAE in a 4-Quintuple Layer of $\mathrm{(Bi_{0.29}Sb_{0.71})_{1.89}V_{0.11}Te_3}$ at 25 mK. Longitudinal resistivity $\rho_\mathrm{xx}$ and transverse resistivity $\rho_\mathrm{xy}$ \emph{versus} $B$ at the charge neutrality point $V_g=V^{0}_{g}$. From ref.\cite{Chang2015}.}\label{fig:Fig-SpectrumChang}
\end{figure*}
The QAHE corresponds to the fundamental state where only one spin direction edge-state is kept (fig.\ref{fig:Fig-AllQHE}-right) which can be achieved by introducing ferromagnetism. As described in fig.\ref{fig:Fig-SpectrumChang}a from ref.\cite{Fox2018}, the gapless chiral edge state is hosted in the exchange-induced gap in the Dirac spectrum of the topological surface states inside the 3D bulk gap. The Hall resistance is quantized for a Fermi level in the surface-state gap. This new quantum Hall effect was first observed in 2013 in a thin ferromagnetic topological insulator (FTI)\cite{Chang2013}. Fig.\ref{fig:Fig-SpectrumChang}b reports on the dependence of the Hall and longitudinal resistances as a function of the magnetic induction in a 4-quintuple layer of $\mathrm{(Bi_{0.29}Sb_{0.71})_{1.89}V_{0.11}Te_3}$ at a temperature $T=25$ mK. It shows a hysteresis cycle and flat Hall plateaus centered around $B=0$ T. The quantization of the Hall resistance was demonstrated to within a few $10^{-4}$ at $B=0$ T. A similar accuracy was obtained in ref.\cite{Bestwick2015}.

Accurate comparisons of the Hall resistances measured both in a FTI at zero magnetic field and in a GaAs-based reference standard were recently performed using a CCC-based resistance bridge. Actually, Fox and co-authors\cite{Fox2018} have demonstrated the quantization of the Hall resistance with a relative uncertainty of about one part in $10^{6}$ at a temperature of 21 mK for a measurement current of 100 nA in a top-gated 100 $\mu$m wide Hall bar made of 6-quintuple-layer sample of $\mathrm{Cr_{0.12}(Bi_{0.26}Sb_{0.62})_2Te_3}$ grown on a GaAs substrate by molecular beam epitaxy (MBE). Limitations in temperature and current are explained by an effective energy gap much lower than expected and strong electron heating in bulk current flow respectively. G\"{o}tz and co-authors determined a relative discrepancy of $(0.17\pm0.25)\times10^{-6}$ between the von Klitzing constant $R_\mathrm{K}$ and the quantized resistance measured, at a temperature of 20 mK and for currents lower than 10 nA, in a top-gated 200 $\mu$m wide Hall bar made of a 9 nm thick film of the ferromagnetic topological insulator $\mathrm{V_{0.1}(Bi_{0.21}Sb_{0.79})_{1.9}Te_3}$ grown by MBE on a hydrogen
passivated Si(111) substrate\cite{Gotz2018}. These recent results support the topological robustness of the QAHE and motivate further works to investigate its metrological application. Beyond, non-dissipative edge states at zero magnetic field in the QHAE instigate others interests, notably applications in low consumption electronics.

Generally, electrical metrology looks after quantum effects, with a very rich physics, which are of interest not only for fundamental research but also for applications. It turns out that the integration of quantum physics in the SI occurs simultaneously with an ambition of exploiting new quantum technologies in industry. It is about using individual particles, superposed coherent states or entangled states as a basis for a quantum computer, protected communications and more sensitive detectors. In this context, new measurement methods will be needed. Their development can rely on the know-how of NMIs in the field of quantum effects.  As an example of new quantum technology, one can cite single-electron interferometers as local sensitive electromagnetic field detectors\cite{SEQUOIA} for which expertise of NMIs in SET and QHE physics should constitute a clear support to their development and characterization.
\section{Conclusions}
The quantum voltage and resistance standards have greatly progressed since the discovery of the Josephson effect and the quantum Hall effect. They have become pillars of the electrical metrology allowing a traceability improvement for all basic electrical units, \emph{i.e.} the volt, the ohm, the ampere and the farad, but also the kilogram. This comes from the richness and universality of these two quantum phenomena. Pulse-driven Josephson arrays and graphene-based standards are recent examples of this fruitfulness for which short-term issues are the achievement in a reproducible way of 1 V and even 10 V pulse-driven voltage standards and of a stable low-magnetic field QHR respectively. Quantum standards having reached a certain level of maturity, one of the future issues will be to combine them together to develop new applications. The development of the quantum standards of impedance and current based on Ohm's law illustrates this new research direction. The quantum calibrator probably constitutes an emblematic challenge because it is the key for disseminating electrical units closer to end-users. We have evoked some recent scientific discoveries able to support the development of this device. Beyond, national metrology institutes, as experts of measurements, should be associated to this ambition of developing new quantum technologies for industry. As always, success in this new metrological challenge would rely on a close collaboration between NMIs, academic laboratories and relevant high-technology industries.
\section*{Acknowledgments}
This work was supported by the Joint Research Projects ‘e‐SI‐Amp’ (15SIB08) and 'Sequoia' (17FUN04). It received funding from the European Metrology Programme for Innovation and Research (EMPIR), co-financed by the Participating States, and from the European Union's Horizon 2020 research and innovation programme. This work was supported by the French Agence Nationale de la Recherche (ANR-16-CE09-0016).



\bibliographystyle{elsarticle-num}


\providecommand{\noopsort}[1]{}\providecommand{\singleletter}[1]{#1}%



\end{document}